\documentclass[onecolumn,showpacs,preprintnumbers,showkeys]{revtex4}
\usepackage{amsmath, amsthm, amscd, amsfonts, amssymb, graphicx, hhline}
\usepackage{dcolumn}
\usepackage{bm}
\usepackage[english]{babel}
\usepackage{footnpag}
\usepackage{nccfoots}
\usepackage{multirow}
\usepackage[symbol*]{footmisc}

\begin{document}

\title{Extended time-dependent Ginzburg-Landau theory}
\author{Konstantin V. Grigorishin}

\email{konst.phys@gmail.com} \affiliation{Bogolyubov Institute for Theoretical Physics of the National Academy of Sciences of
Ukraine, 14-b Metrolohichna str. Kiev, 03143, Ukraine.}
%\date{\today}

\begin{abstract}
We formulate the gauge invariant Lorentz covariant Ginzburg-Landau theory which describes nonstationary regimes: relaxation of a superconducting system accompanied by eigen oscillations of internal degrees of freedom (Higgs mode and Goldstone mode), and also forced oscillations under the action of an external gauge field. The theory describes Lorentz covariant electrodynamics of superconductors where Anderson-Higgs mechanism occurs, at the same time the dynamics of conduction electrons remains non-relativistic. It is demonstrated that Goldstone oscillations cannot be accompanied by oscillations of charge density and they generate the transverse field only. In addition, we consider Goldstone modes and features of Anderson-Higgs mechanism in two-band superconductors. We study dissipative processes, which are caused by movement of the normal component of electron liquid and violate the Lorentz covariance, on the examples of the damped oscillations of the order parameter and the skin-effect for electromagnetic waves. An experimental consequence of the extended time-dependent Ginzburg-Landau theory regarding the penetration of the electromagnetic field into a superconductor is proposed.
\end{abstract}


 \maketitle

\section{Introduction}\label{intr}

\begin{figure}[h]
\includegraphics[width=10cm]{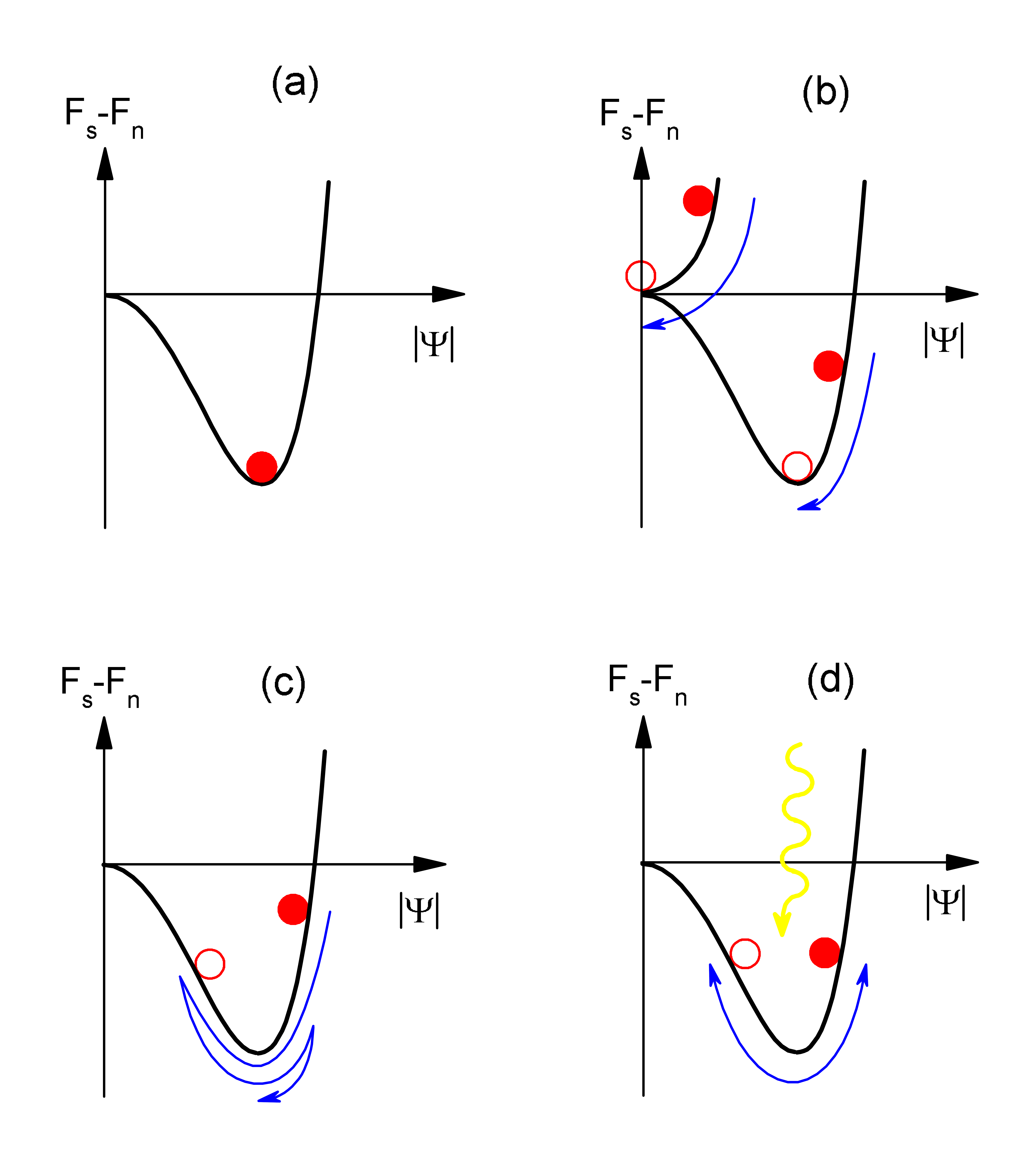}
\caption{Variation of Landau free energy $F_{s}-F_{n}$ (for example, SC state $s$ versus normal state $n$) with the order
parameter $\Psi$. (a) - equilibrium state of the field, (b) - monotonous relaxation to the equilibrium, (c) - relaxation to the
equilibrium with damped oscillation, (d) - forced oscillation under the action of an external field.} \label{Fig1}
\end{figure}

The system undergoing the second-order phase transition (on example of superconductivity) is described with a Landau functional
\begin{eqnarray}\label{1.1}
    F=\int d^{3}r\mathcal{F}=\int d^{3}r\left[\frac{\hbar^{2}}{4m}\left(\nabla\Psi\right)\left(\nabla\Psi^{+}\right)
    +a\left|\Psi\right|^{2}+\frac{b}{2}\left|\Psi\right|^{4}\right],
\end{eqnarray}
where $\mathcal{F}$ is density of free energy, $a=\alpha(T-T_{\mathrm{c}})$, $\Psi=\left|\Psi(\mathbf{r})\right|e^{i\theta(\mathbf{r})}$ is a two-component order parameter (the wave function of condensate of Cooper pairs) so that $n_{\mathrm{s}}=2\left|\Psi\right|^{2}$ is density of superconducting (SC) electrons, a term $\frac{\hbar^{2}}{4m}\left|\nabla\Psi\right|^{2}$ can be understood as density of kinetic energy of Cooper pairs of mass $2m$ each. Configuration of the field $\Psi(\textbf{r})$ which minimizes the functional (\ref{1.1}) is obtained from equation:
\begin{eqnarray}\label{1.2}
    \frac{\delta F}{\delta\Psi^{+}}=\frac{\partial\mathcal{F}}{\partial\Psi^{+}}-\nabla\frac{\partial\mathcal{F}}{\partial(\nabla\Psi^{+})}=0\Rightarrow
    \frac{\hbar^{2}}{4m}\Delta\Psi-a\Psi-b\left|\Psi\right|^{2}\Psi=0.
\end{eqnarray}
This configuration is shown in Fig.\ref{Fig1}a: they say that the field is a string laying in a valley of the potential
$a\left|\Psi\right|^{2}+\frac{b}{2}\left|\Psi\right|^{4}$.

At the same time, out the equilibrium the field $\Psi$ depends on time. For small deviations from the equilibrium it is natural to
assume \cite{lark,kop}, that the time derivative $\partial\Psi/\partial t$ is proportional to the
variational derivative of the free energy functional $\delta F/\delta\Psi^{+}$ which is equal to zero at the equilibrium. Thus,
one can write the time-dependent Ginzburg-Landau equation \cite{lark,gor,kop,watts,tinh,als,lars,art,rieg,bind} (TDGL equation) in a form:
\begin{eqnarray}\label{1.3}
\frac{\hbar^{2}}{4mD}\frac{\partial\Psi}{\partial t}=-\frac{\delta F}{\delta\Psi^{+}}\Rightarrow\tau\frac{\partial\psi}{\partial t}=\xi^{2}\Delta\psi+\psi-|\psi|^{2}\psi,
\end{eqnarray}
where $\psi=\Psi/\Psi_{0}$ is the dimensionless order parameter, $\Psi_{0}=\sqrt{-a/b}$ is an equilibrium value of the spatial homogeneous order parameter, $D$ is a diffusion coefficient of electrons in the normal state. The temperature-dependent coherence length $\xi(T)=\sqrt{\frac{\hbar^{2}}{4m|a(T)|}}$ and the temperature-dependent relaxation time $\tau(T)=\frac{\hbar^{2}}{4mD|a(T)|}$ can be found from the microscopic theory for the case of a gapless SC alloy containing a high concentration of paramagnetic impurities \cite{gor,rieg} and for dirty superconductors in the Ginzburg-Landau (GL) regime $|T_{c}-T|<<T_{c}$ \cite{kop,watts}. However, in the general case, we have to treat these parameters phenomenologically. For $T>T_{c}$ the TDGL equation has a form $\tau_{0}\frac{\partial\Psi}{\partial t}=\xi^{2}\Delta\Psi-\Psi$, where $\tau_{0}=\frac{\pi\hbar}{8(T-T_{c})}$ is the relaxation time for a homogeneous mode (here and further $k_{B}=1$) \cite{tinh}. In this case the equilibrium value is $\left\langle\Psi\right\rangle=0$ but $\left\langle\Psi^{2}\right\rangle\neq0$. The corresponding relaxation processes are illustrated in Fig.\ref{Fig1}b.

Thus, Eq.(\ref{1.3}) describes \emph{relaxation} of the order parameter at small deviations from the equilibrium. At the same time, in the system undergoing the second-order phase transition the collective excitations can exist which are resonant oscillations. When a continuous symmetry is spontaneously broken, there emerge two types of collective modes in general: massive Higgs mode, which is oscillation of modulus $|\Psi|$ of the order parameter, and Goldstone mode, which is oscillation of the phase $\theta$. As it has been demonstrated in \cite{volkov,varma,vad} in collisionless approximation the Higgs mode in SC system can exist as perturbation of amplitude of the gap $\Delta$ which takes the form of oscillations having a frequency $\sim|\Delta|$. At the same time, the oscillations of order parameter damp in some time $\tau$. Thus, the system is out from the equilibrium then the relaxation process depends on relation between period of the eigen oscillations $\sim 1/\omega$ and the damping time: if $1/\omega\gg\tau$ then aperiodic relaxation occurs as shown in Fig.\ref{Fig1}b and it is described with TDGL equation (\ref{1.3}). In general case the parameters $1/\omega$ and $\tau$ can be in arbitrary relation, hence the relaxation process can have more complicated form, for example, if $1/\omega\ll\tau$ then an oscillatory process with small damping occurs as shown in Fig.\ref{Fig1}c. Moreover, an external field can swing the system, that is the undamped oscillations occur, while heat is released (for example, electromagnetic wave, falling on a superconductor, induces Foucault currents both normal $\textbf{j}_{n}$ and superconducting $\textbf{j}_{s}$). Such situation is shown in Fig.\ref{Fig1}d.

In presence of electric $\varphi$ and magnetic $\textbf{A}$ potentials the replacements
\begin{equation}\label{1.4}
\frac{\partial}{\partial t}\rightarrow\frac{\partial}{\partial t}+\frac{i2e}{\hbar}\varphi,\quad\nabla\rightarrow\nabla-\frac{i2e}{c\hbar}\textbf{A}
\end{equation}
must be done in free energy functional (\ref{1.1}) and in Eqs.(\ref{1.2},\ref{1.3}) for a gauge invariance. Moreover, the total current is a sum of normal current and supercurrent: $\mathbf{j}=\sigma\left(-\nabla\varphi-\frac{1}{c}\frac{\partial \mathbf{A}}{\partial t}\right)+\mathbf{j}_{s}$, where $\sigma$ is conductivity, $\mathbf{j}_{s}=-\frac{ie\hbar}{2m}\left(\Psi^{+}\nabla\Psi-\Psi\nabla\Psi^{+}\right)-\frac{2e^{2}}{mc}\left|\Psi\right|^{2}\mathbf{A}$. Thus, the TDGL equations determine dynamics of both order parameter $\psi$ and electromagnetic field $A^{\mu}\equiv\left(\varphi,\textbf{A}\right)$. On the other hand, equations for electromagnetic field should be Lorentz covariant both in vacuum and within any medium, despite dynamics of particles (medium) is non-relativistic, for example, Maxwell equations in dielectrics. However the light speed in the medium is less than the speed in vacuum $c$ and is determined with dynamic properties of the system. At the same time, the dissipative mechanisms give terms which violate Lorentz covariance since the dissipation distinguishes a time direction, i.e., violates the time symmetry $t\leftrightarrow -t$ which is symmetry of the Lorentz boost. For example, Maxwell equation in medium with conductivity $\sigma$
has a form:
\begin{equation}\label{1.4b}
    \textrm{curl}\mathbf{H}=\frac{1}{c}\frac{\partial\mathbf{D}}{\partial t}+\frac{4\pi}{c}\sigma\mathbf{E}.
\end{equation}
Here, the first two terms are an equation from Lorentz covariant Maxwell equations, the last term $\frac{4\pi}{c}\sigma\mathbf{E}$ is a dissipative part. Depending on material and processes occurring in it, some terms in the equations can be neglected. So, in metals the dissipative term dominates, for example, the strength of electrostatic field must be $\mathbf{E}=0$ inside metal: the nonzero strength would lead to a current, meanwhile the propagation of the current is associated with energy dissipation according to Joule-Lenz law $Q=\mathbf{j}^{2}/\sigma=\sigma\mathbf{E}^{2}$ and, therefore, it cannot be supported in an equilibrium state by itself \cite{landau}. For not very large frequencies of electromagnetic waves (less than plasma frequency) a condition $\frac{\sigma}{\omega}\gg\varepsilon(\omega)$ is satisfied for good metals, then the conduction current $-\sigma\frac{1}{c}\frac{\partial\mathbf{A}}{\partial t}$ gives main contribution in electromagnetic response that gives the skin-effect, at the same time the displacement current $\frac{\varepsilon}{4\pi}\frac{\partial\mathbf{E}}{\partial t}$ can be neglected \cite{landau,tilley}. On the contrary, in dielectrics the dissipative processes can be neglected in the first approximation (since conductivity is zero); hence, Lorentz covariant electrodynamics remains only. In bad conductors (semiconductors, weak electrolytes, weakly ionized plasma) both conduction and displacement currents can be important. Analogously to dielectrics and metals electrodynamics of superconductors must be composed of both Lorentz covariant part and dissipative part (due to friction of normal electrons, damping of collective excitations and breaking of Cooper pairs). Which term is dominant is determined with physical conditions and processes. Unlike the normal metal phase, in the SC phase the conductivity essentially depends on temperature due to density of normal electrons $n_{\mathrm{n}}$: $\sigma(T)=\frac{\tau_{ph}}{m}e^{2}n_{\mathrm{n}}(T)\Rightarrow\sigma(T\rightarrow 0)\rightarrow 0$ (here $\tau_{ph}$ is the mean free time of electron caused by electron-phonon interaction), that is at low temperatures the contribution of the dissipative part decreases. It should be noted that superconductivity is a thermodynamic effect and is not electrodynamic one (superconductor is not ideal conductor). Hence equations for electromagnetic field in superconductor are result of variation of some functional of action (or free energy functional for equilibrium case): $\frac{\delta S}{\delta A_{\mu}}=0$. The action $S\left[\Psi(\mathbf{r},t),A_{\mu}(\mathbf{r},t)\right]$ must be Lorentz invariant in order to ensure Lorentz covariant electrodynamics of superfluid component. The  dissipative terms are introduced in the equations of motion by means of Rayleigh dissipation function. At the same time, the Lorentz invariance of the action should have consequences for dynamics of the scalar field $\Psi$: the collective pseudo-relativistic excitations (Higgs mode and Goldstone mode) occur. In \cite{kop,watts} the general equations for the dynamic behavior of dirty superconductors in GL regime $|T_{c}-T|<<T_{c}$ are derived from microscopic theory. The local equilibrium approximation leads to a simple generalized TDGL equation describing relaxation of the order parameter. As indicated above, in this nonequilibrium regime the dissipative processes dominant, hence the Lorentz covariation can be neglected and a relaxation equation of type TDGL (\ref{1.3}) is valid in mean field approximation.

The dynamic extension of GL theory has been proposed in \cite{aitch} as a time-dependent nonlinear Schr\"odinger Lagrangian:
\begin{equation}\label{1.4a}
\mathcal{L}=i\hbar\Psi^{+}\frac{\partial\Psi}{\partial t}-\frac{\hbar^{2}}{4m}\nabla\Psi^{+}\nabla\Psi-V\left(|\Psi|\right),
\end{equation}
which describes the low-frequency, long-wavelength dynamics of the pair field $\Psi(\mathbf{r},t)$ for a BCS-type s-wave superconductor at $T=0$, $V$ is potential leading to spontaneously broken $U(1)$ symmetry and is assumed to be a function of $|\Psi|$ only. We can see that Lagrangian (\ref{1.4a}) is Galilean invariant. In this Lagrangian the electromagnetic field $A^{\mu}=(\varphi,\mathbf{A})$ must be included by replacement (\ref{1.4}) based on gauge invariance, and Lagrangian of electromagnetic field $\left(\mathbf{E}^{2}-\mathbf{H}^{2}\right)/8\pi$ should be added. Varying the corresponding action: $\frac{\delta S\left(\Psi, A^{\mu}\right)}{\delta A^{\mu}}=0$ we will find equations for the electromagnetic field $A^{\mu}(\mathbf{r},t)$ in SC medium. Obviously, these equations will not be Lorentz covariant, since initial Lagrangian (\ref{1.4a}) is not Lorentz invariant. However, as mentioned above, equations for electromagnetic field in nondissipative medium must be Lorentz covariant like the Maxwell equations in vacuum.

It is believed that superconductors cannot contain macroscopic electric fields in static configurations. This fact is directly based on the first of the London equations:
\begin{eqnarray}
  &\frac{d\mathbf{j}_{s}}{dt}=\frac{n_{s}e^{2}}{m}\mathbf{E}\label{1.5a} \\
  &\nonumber \\
  &\mathrm{curl}\mathbf{j}_{s}=-\frac{n_{\mathrm{s}}e^{2}}{mc}\mathbf{H}\Rightarrow\Delta\mathbf{H}=\frac{1}{\lambda^{2}}\mathbf{H},\label{1.5b}
\end{eqnarray}
where $\lambda=\sqrt{\frac{mc^{2}}{4\pi n_{\mathrm{s}}e^{2}}}$ is the London penetration depth, $n_{\mathrm{s}}$ is density of SC electrons, $\mathbf{j}_{s}=n_{\mathrm{s}}e\mathbf{v}$ is supercurrent density, and $\mathbf{E}$, $\mathbf{H}$ are electric and magnetic fields respectively. Indeed, unlike normal metals, in superconductors due to zero resistance to support the direct current $\mathbf{j}$ the presence of the electric field $\mathbf{E}$ is not necessary. This means that in the stationary regime $\frac{d\mathbf{j}}{dt}=0$ we have $\textbf{E}=0$ inside superconductor. However, it should be noted, that the second London equation (\ref{1.5b}) is a result of minimization of free energy of the superconductor: $\frac{1}{8\pi}\int\left[\mathbf{H}^{2}+\lambda^{2}(\mathrm{curl}\mathbf{H})^{2}\right]dV$, where the first term is energy of the magnetic field, the second term is kinetic energy of the supercurrent. At the same time, the first London equation is not result of minimization of the free energy of superconductor:  it is suggested that motion of SC electrons is not accompanied with friction hence they are accelerated by electric field $\textbf{E}$, i.e., it is just the second Newton law. This means that \emph{the first London equation} (\ref{1.5a}) \emph{is equation of an ideal conductor} that discussed in Appendix \ref{london}. Superconductivity is the thermodynamically steady state: configuration of both the electric field $\mathbf{E}$ and the magnetic field $\mathbf{H}$ must be found from minimum of some free energy functional $F(\Psi,\Psi^{+},\varphi,\mathbf{A})$.

From Eqs.(\ref{1.5a},\ref{1.5b}) we can see that \emph{the coupling of the supercurrent density to the electromagnetic fields through the London and GL equations is not space-time covariant}: under Lorentz boost, the supercurrent density $\mathbf{j}$ in the presence of the electric field $\mathbf{E}$ ought to transform as the space components of a 4-vector whose time component would then play role of some supercharge density, while the electric $\mathbf{E}$ and magnetic $\mathbf{H}$ fields transform as components of the two index antisymmetric field strength tensor $F_{\mu\nu}$. Thus, it would seem, the Lorentz covariance requires the possibility of electric fields on the same footing as magnetic ones within superconductors, with an electric penetration depth equal to the familiar magnetic one. In works \cite{bert1,bert2} it has been proposed the natural covariant extension of the GL free energy (\ref{1.1}) in terms of the Higgs Lagrangian for spontaneous $U(1)$ gauge symmetry breaking in the vacuum (in SI units):
\begin{eqnarray}\label{1.6}
\mathcal{L}=\frac{1}{2}\varepsilon_{0}c^{2}\left(\frac{\hbar}{2e\lambda}\right)^{2}
\left\{\left|\left(\partial_{\mu}+i\frac{2e}{\hbar}A_{\mu}\right)\psi\right|^{2}-\frac{1}{2\xi^{2}}(|\psi|^{2}-1)^{2}\right\}
-\frac{1}{4}\varepsilon_{0}c^{2}F_{\mu\nu}F^{\mu\nu},
\end{eqnarray}
where the order parameter $\psi$ is normalized to the density of electron pairs in a bulk sample $\psi(x)=\Psi(x)/\Psi_{0}$ ($\Psi_{0}=\sqrt{-a/b}$) in the absence of any electromagnetic field $A_{\mu}=(\varphi/c,-\mathbf{A})$, the tensor $F_{\mu\nu}=\partial_{\mu}A_{\nu}-\partial_{\nu}A_{\mu}$ is the electromagnetic field strength with $\partial_{\mu}=\partial/\partial x^{\mu}$ being the space-time gradients for the coordinates $x^{\mu}=(ct,\mathbf{r})$, $\lambda$ is the magnetic penetration depth. The dynamics is determined from the Lorentz invariant action $S=\int\mathcal{L}d^{4}x$ through the variational principle provides the equations of motion for the scalar field $\psi$ and the vector field $A_{\mu}$ respectively. Main result of this model is the effect of electrostatic field on a superconductor: the field $\mathbf{E}$ as well as the field $\mathbf{B}$ can destroy the SC state. Namely, the critical values of the fields satisfy the relation:
\begin{equation}\label{1.7}
    \left(\frac{B}{B_{cm}}\right)^{2}+\left(\frac{E/c}{B_{cm}}\right)\simeq 1,
\end{equation}
where $B_{cm}$ is a thermodynamical magnetic field (at $B>B_{cm}$ a type-I superconductor goes into the normal state by the first-order phase transition). Thus, we can see that at electric field $E\simeq cB_{cm}$ superconductivity should be destroyed at absence of magnetic field. \emph{The analysis of experimental data presented in} \cite{bert3} \emph{suggests that an external electric field does not affect significantly the superconducting state}. This conclusion is in total contradiction with the expected behavior based on the covariant Lagrangian (\ref{1.6}). It has been suggested that such negative experimental result can be explained with that some non-paired normal electrons could play a crucial role against the electric field, whereas such contributions are not included in the model.

In the work \cite{hirs1} relativistically covariant London equations have been proposed, and they can be understood as arising from the "rigidity" of the superfluid wave function in a relativistically covariant microscopic theory. They predict that for slowly varying electric fields, both longitudinal and transverse, and assuming no charge density in the superconductor:
\begin{equation}\label{1.8}
    \nabla^{2}\mathbf{E}=\frac{1}{\lambda^{2}}\mathbf{E}.
\end{equation}
which implies that an electric field penetrates a distance $\lambda$, as a magnetic field does. Thus, the screening length of the electric field in the SC state is equal to the London penetration depth, not the Thomas-Fermi screening length $\lambda_{TF}$. The associated longitudinal dielectric function is obtained as
\begin{equation}\label{1.9}
    \varepsilon(q,\omega\rightarrow 0)=1+\frac{1}{\lambda^{2}q^{2}}
\end{equation}
unlike the dielectric function for normal metal $\varepsilon(q,\omega\rightarrow 0)=1+\frac{1}{\lambda^{2}_{\mathrm{TF}}q^{2}}$. This model remains debatable \cite{koy,hirs2} and requires experimental verification \cite{hirs3,pero}.

In the Lorentz covariant models the Higgs mode (spectrum with energy gap) and the Goldstone mode (acoustic spectrum) should arise. However, unlike the Higgs mode, the Goldstone mode is unobservable that can be explained with the Anderson-Higgs mechanism: oscillations of the phase $\theta$ are absorbed into the gauge field $A_{\mu}$. At the same time, the above-mentioned models essentially differ from the results of the theory of gauge-invariant response of superconductors to external electromagnetic field presented in \cite{ars}. In this model the Coulomb interaction "pushes" the frequency of the acoustic oscillations to the plasma frequency $\omega_{p}$. Thus, Goldstone mode becomes unobservable in itself since it turns to plasma oscillations. In addition, in such model the electric field is screened with Thomas-Fermi length $\lambda_{TF}$ like in the normal metal phase. However, it should be noted, that the phase oscillations $\theta(\textbf{r},t)$ are oscillations of the order parameter $|\Psi|e^{i\theta}$, i.e., they are specific for the SC state. At the same time, the plasma oscillations exist unchanged both in SC phase and in the normal metal phase, i.e., they are unrelated to the SC ordering and are not specific for the SC state. Thus, if the plasma oscillations were the phase oscillations of the order parameter, then this would be affected them at the transition point $T_{c}$ necessarily; however, the spectrum of the plasma oscillations does not depend on temperature.

Proceeding from aforesaid, we are aimed to generalize the GL theory for the nonstationary regimes Fig.\ref{Fig1}(b-d): the theory should describe the relaxation of the system with accounting of eigen oscillations of internal degrees of freedom, and also forced oscillations under the action of external field. Thus, this theory includes the GL theory and the TDGL equation as special cases, and we will call it as the \emph{extended TDGL theory}. Moreover, this theory must be Lorentz covariant without accounting of dissipative processes, since it includes Lorentz covariant electrodynamics, at the same time the dynamics of conduction electrons remains non-relativistic. Accounting of dissipative mechanisms should violate the Lorentz covariance of the theory and should lead to relaxation processes in SC system. Thus, our paper is organized by the following way. In Sect.\ref{normal} we generalize GL free energy functional to relativistic-like action by phenomenological approach. Using this action we study the possible eigen oscillations of the order parameter $\Psi(t,\mathbf{r})$: Higgs mode and Goldstone mode. Moreover, we   obtain this relativistic-like GL functional by microscopical approach also. In Sect.\ref{electro} using the gauge symmetry we formulate electrodynamics of superconductors in the sense of Lorentz covariant equations for 4D electromagnetic potential $A^{\mu}\equiv(\varphi,\mathbf{A})$. The equations describe propagate of electromagnetic field in SC medium where Anderson-Higgs mechanism (absorption of the Goldstone bosons into the gauge field $A^{\mu}$ and the gaining of mass by the gauge field) takes place. In Sect.\ref{two} we consider features of Anderson-Higgs mechanism in two-band superconductors and occurrence of Leggett's mode. In Sect.\ref{relax} we study influence of the motion of normal component of electron liquid which causes the damping of oscillations both order parameter and electromagnetic field. As example we consider the wave skin-effect, eigen electromagnetic oscillations and relaxation of fluctuation of the order parameter. We propose an experimental consequence of the extended TDGL theory regarding the penetration of the electromagnetic field in a superconductor. Besides we demonstrate that the London electrodynamics and the TDGL equation (\ref{1.3}) are limit cases of the extended TDGL theory.

\section{Normal modes and pseudo-relativistic collective excitations}\label{normal}

\subsection{Phenomenological approach}\label{phen}

In general case the SC order parameter $\Psi$ is both spatially inhomogeneous and it can change over time:
$\Psi=\Psi(\textbf{r},t)$. The order parameter is a complex scalar field which is equivalent to two real fields: modulus $\left|\Psi(\textbf{r},t)\right|$ and phase $\theta(\textbf{r},t)$ (the modulus-phase representation):
\begin{equation}\label{2.2}
    \Psi(\textbf{r},t)=\left|\Psi(\textbf{r},t)\right|e^{i\theta(\textbf{r},t)}.
\end{equation}
For stationary case $\Psi=\Psi(\textbf{r})$ the free energy functional (\ref{1.1}) exists and the steady configuration of the field $\Psi(\textbf{r})$ minimizes this functional. However for the nonstationary case $\Psi(\textbf{r},t)$ the minimizing procedure loses any sense. Thus, it is necessary to find an equation determining evolution of the order parameter in time. Our method for solving this problem is as follows. The parameter $t$ - the time can be turned into a coordinate $t\rightarrow\upsilon t$ in some 4D Minkowski space $\{\upsilon t,\textbf{r}\}$, where $\upsilon$ is an parameter of dimension of speed (like the light speed) which \emph{must be determined with dynamical properties of the system}. At the same time, the dynamics of conduction electrons remains non-relativistic. Then the two-component scalar field $\Psi(\textbf{r},t)$ minimizes some action $S$ (like in the relativistic field theory \cite{sad}) in the Minkowski space:
\begin{equation}\label{2.5}
    S=\frac{1}{\upsilon}\int\mathcal{L}(\Psi,\Psi^{+})d\Omega,
\end{equation}
where $\mathcal{L}$ is some Lagrangian (density of Lagrange function $L=\int\mathcal{L}d^{3}r$), $d\Omega\equiv\upsilon dtd^{3}r$ is an element of the 4D Minkowski space. The Lagrangian can be built by generalizing the density of free energy $\mathcal{F}$ in Eq.(\ref{1.1}) to "relativistic" invariant form by substitution of covariant and contravariant differential operators
\begin{equation}\label{2.4}
\widetilde{\partial}_{\mu}\equiv\left(\frac{1}{\upsilon}\frac{\partial}{\partial t},\nabla\right),\quad\widetilde{\partial}^{\mu}\equiv\left(\frac{1}{\upsilon}\frac{\partial}{\partial
t},-\nabla\right)
\end{equation}
instead the gradient operators:
$\nabla\Psi\rightarrow\widetilde{\partial}_{\mu}\Psi,\quad\nabla\Psi^{+}\rightarrow\widetilde{\partial}^{\mu}\Psi^{+}$. Then the required Lagrangian takes a form:
\begin{equation}\label{2.6}
    \mathcal{L}=\frac{\hbar^{2}}{4m}\left(\widetilde{\partial}_{\mu}\Psi\right)\left(\widetilde{\partial}^{\mu}\Psi^{+}\right)
    -a\left|\Psi\right|^{2}-\frac{b}{2}\left|\Psi\right|^{4}=\frac{\hbar^{2}}{4m}\frac{1}{\upsilon^{2}}\left(\frac{\partial\Psi}{\partial t}\right)\left(\frac{\partial\Psi^{+}}{\partial t}\right)-\frac{\hbar^{2}}{4m}\left(\nabla\Psi\right)\left(\nabla\Psi^{+}\right)
    -a\left|\Psi\right|^{2}-\frac{b}{2}\left|\Psi\right|^{4}\equiv\mathcal{T}-\mathcal{F}.
\end{equation}
Here, as we can see from Eq.(\ref{1.1}), the spatial terms of the Lagrangian (free energy $\mathcal{F}$) play role of "potential" energy, and the time term plays role of "kinetic" energy $\mathcal{T}$. Lagrange equation for functional (\ref{2.5}) is
\begin{eqnarray}\label{2.7}
    \widetilde{\partial}^{\mu}\frac{\partial\mathcal{L}}{\partial(\widetilde{\partial}^{\mu}\Psi^{+})}-\frac{\partial\mathcal{L}}{\partial\Psi^{+}}=0\Rightarrow
    \frac{\hbar^{2}}{4m}\widetilde{\partial}^{\mu}\widetilde{\partial}_{\mu}\Psi+a\Psi+b\left|\Psi\right|^{2}\Psi=0,
\end{eqnarray}
where
\begin{equation}\label{2.8}
    \widetilde{\partial}^{\mu}\widetilde{\partial}_{\mu}=\widetilde{\partial}_{\mu}\widetilde{\partial}^{\mu}=\frac{1}{\upsilon^{2}}\frac{\partial^{2}}{\partial
    t^{2}}-\Delta.
\end{equation}
Since the field $\Psi$ is complex and Lagrangian (\ref{2.6}) has $U(1)$ symmetry then the conserving charge takes place. Using Eq.(\ref{2.7}) and its complex conjugate equation we can obtain the continuity condition in a form:
\begin{equation}\label{2.9}
    \widetilde{\partial}^{\mu}j_{\mu}=\frac{\partial\rho}{\partial t}+\mathrm{div}\mathbf{j}_{s}=0,
\end{equation}
where $j_{\mu}$ is a 4D supercurrent:
\begin{eqnarray}\label{2.10}
    j_{\mu}&\equiv&(\upsilon\rho_{s},-\mathbf{j}_{s})=
    \frac{ie\hbar}{2m}\left(\Psi^{+}\widetilde{\partial}_{\mu}\Psi-\Psi\widetilde{\partial}_{\mu}\Psi^{+}\right)\nonumber\\
    &=&\left(\upsilon\frac{ie\hbar}{2m\upsilon^{2}}\left(\Psi^{+}\frac{\partial\Psi}{\partial t}-\Psi\frac{\partial\Psi^{+}}{\partial t}\right),
    \frac{ie\hbar}{2m}\left(\Psi^{+}\nabla\Psi-\Psi\nabla\Psi^{+}\right)\right)\nonumber\\
    &=&\left(-\upsilon\frac{e\hbar}{m\upsilon^{2}}|\Psi|^{2}\frac{\partial\theta}{\partial t},-\frac{e\hbar}{m}|\Psi|^{2}\nabla\theta\right).
\end{eqnarray}
It should be noted that the correct definition of the current $j_{\mu}$ is possible only with using of gauge invariance, which is done in Sect.\ref{electro}.

\begin{figure}[h]
\includegraphics[width=8cm]{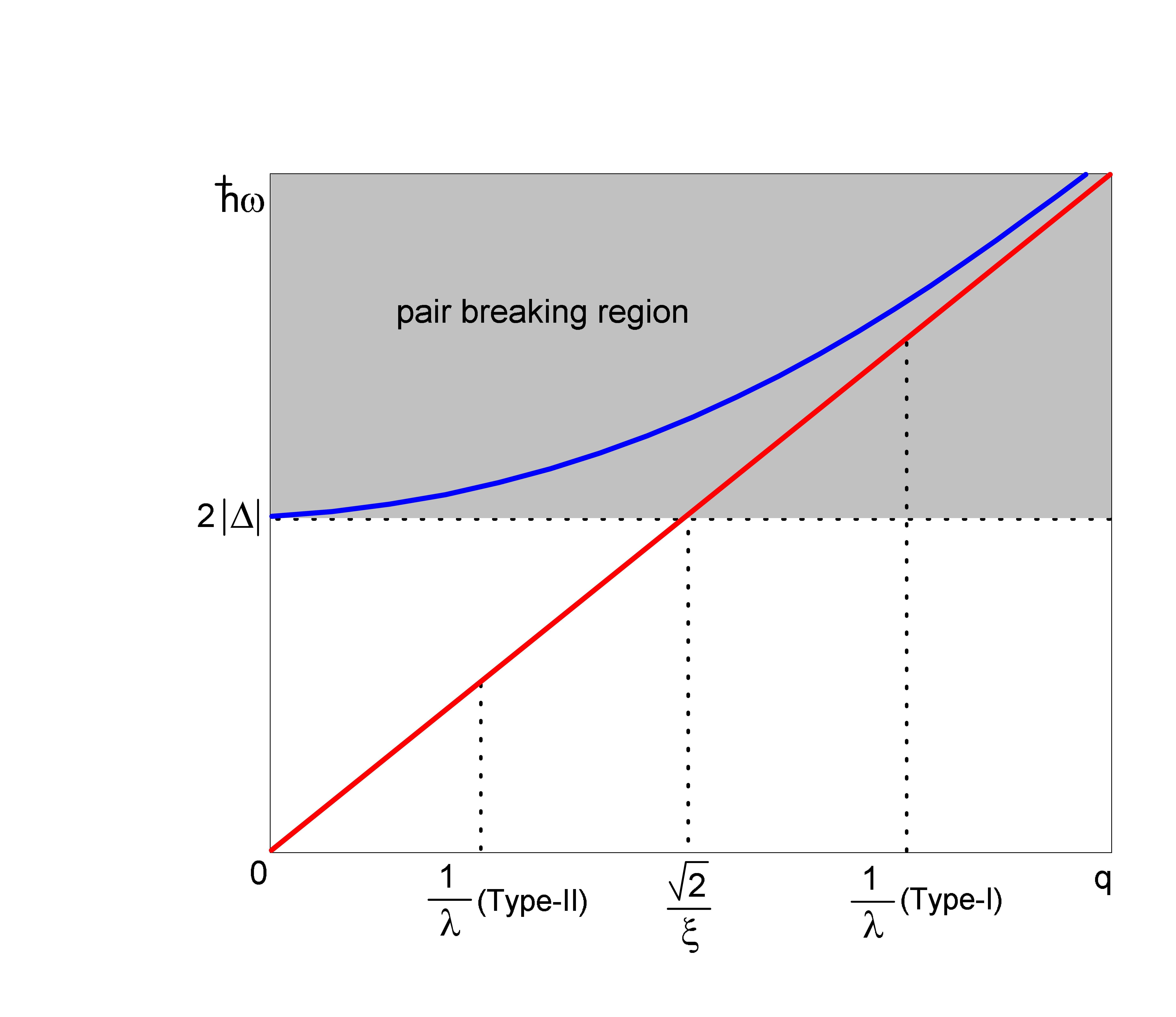}
\caption{Higgs oscillations with spectrum (\ref{2.15}) - blue line, and Goldstone oscillations with spectrum (\ref{2.16}) - red line. For $q=\frac{\sqrt{2}}{\xi}$ we have $\hbar\omega=2|\Delta|$ in the Goldstone mode. If to set $q=\frac{1}{\lambda}$, then $\hbar\omega(q)>2|\Delta|$ for type-I superconductors and $\hbar\omega(q)<2|\Delta|$ for type-II superconductors. The region where the pair breaking occurs, because $E>2|\Delta|$, is shaded in gray. The free Higgs mode lies entirely in this region, hence this mode is unstable due to decay to the above-condensate quasiparticles.} \label{Fig2}
\end{figure}

\begin{figure}[h]
\includegraphics[width=8cm]{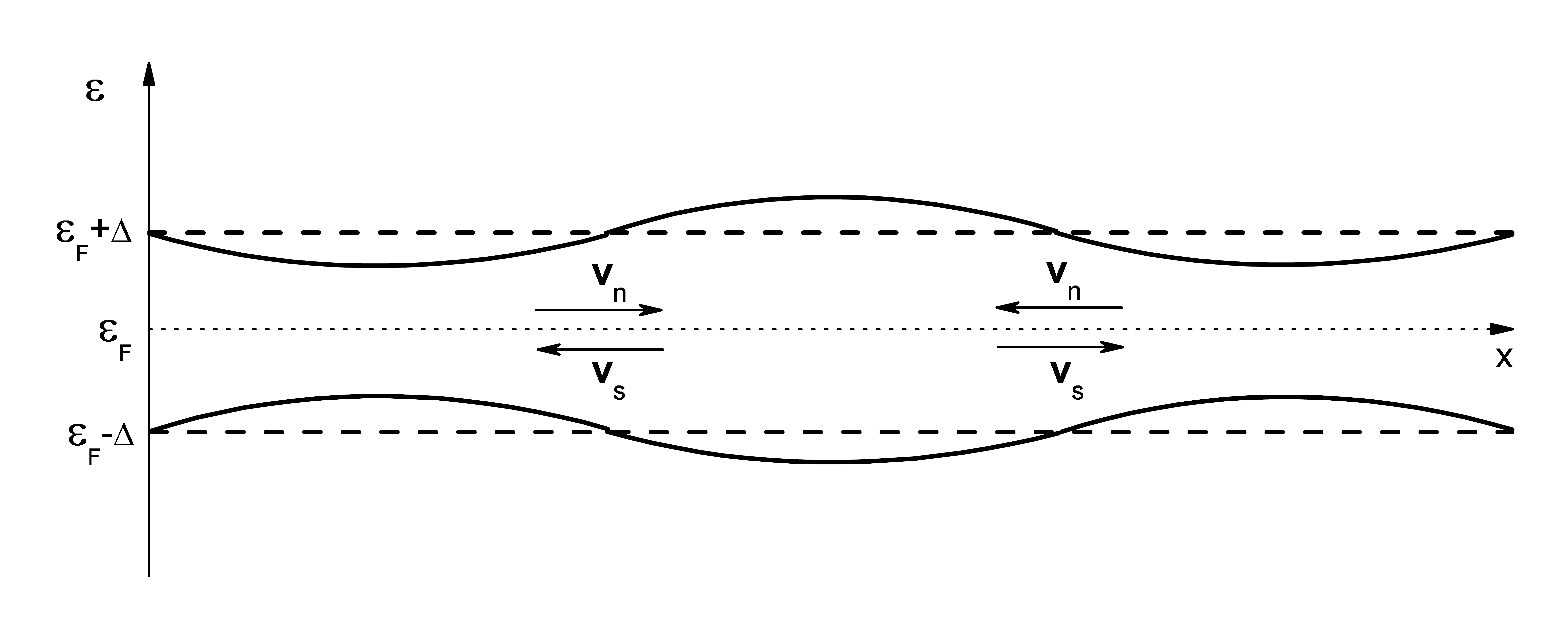}
\caption{Higgs oscillations with spectrum (\ref{2.15}) at $q\neq 0$. This mode is oscillations of modulus of order parameter $|\Psi|$, the oscillations are accompanied by changes of density of SC electrons $n_{\mathrm{s}}=2\left|\Psi\right|^{2}$. At the same time, the total electron density must be $n=n_{\mathrm{s}}+n_{\mathrm{n}}=\mathrm{const}$, hence the oscillations can be presented as counterflows of SC and normal components so that $n_{\mathrm{s}}\textbf{v}_{\mathrm{s}}+n_{\mathrm{n}}\textbf{v}_{\mathrm{n}}=0$.}
\label{Fig3}
\end{figure}

Substituting representation (\ref{2.2}) in the Lagrangian (\ref{2.6}) we obtain:
\begin{equation}\label{2.11}
    \mathcal{L}=\frac{\hbar^{2}}{4m}\widetilde{\partial}_{\mu}|\Psi|\widetilde{\partial}^{\mu}|\Psi|    +\frac{\hbar^{2}}{4m}|\Psi|^{2}\widetilde{\partial}_{\mu}\theta\widetilde{\partial}^{\mu}\theta
    -a\left|\Psi\right|^{2}-\frac{b}{2}\left|\Psi\right|^{4}.
\end{equation}
Let us consider small variations of modulus of order parameter from its equilibrium value: $|\Psi|=\sqrt{-\frac{a}{b}}+\phi\equiv\Psi_{0}+\phi$, where $|\phi|\ll\Psi_{0}$. Then at $T<T_{c}$ the Lagrangian takes a form:
\begin{equation}\label{2.12}
    \mathcal{L}=\frac{\hbar^{2}}{4m}\widetilde{\partial}_{\mu}\phi\widetilde{\partial}^{\mu}\phi-2|a|\phi^{2}
    +\frac{\hbar^{2}}{4m}\Psi_{0}^{2}\widetilde{\partial}_{\mu}\theta\widetilde{\partial}^{\mu}\theta+\frac{1}{2}\frac{a^{2}}{b}.
\end{equation}
Corresponding Lagrange equations are
\begin{eqnarray}
&&\frac{\hbar^{2}}{4m}\widetilde{\partial}_{\mu}\widetilde{\partial}^{\mu}\phi+2|a|\phi\equiv\frac{\hbar^{2}}{4m}
\left(\frac{1}{\upsilon^{2}}\frac{\partial^{2}\phi}{\partial t^{2}}-\Delta\phi\right)+2|a|\phi=0,\label{2.13}\\
&&\widetilde{\partial}_{\mu}\widetilde{\partial}^{\mu}\theta\equiv\frac{1}{\upsilon^{2}}\frac{\partial^{2}\theta}{\partial t^{2}}-\Delta\theta=0.\label{2.14}
\end{eqnarray}
We can see that modulus and phase variables are separated. Thus, the field coordinates $|\Psi(\mathbf{r},t)|$ and $\theta(\mathbf{r},t)$ are normal coordinates, and their small oscillations are normal oscillations. From Eqs.(\ref{2.13},\ref{2.14}) we obtain dispersion relations for harmonic oscillations  $\phi(\mathbf{r},t)=\phi_{0}e^{i(\mathbf{qr}-\omega t)}$ and $\theta(\mathbf{r},t)=\theta_{0}e^{i(\mathbf{qr}-\omega t)}$ accordingly:
\begin{eqnarray}
&&(\hbar\omega)^{2}=8|a|m\upsilon^{2}+(\hbar q)^{2}\upsilon^{2}\quad-\quad\texttt{Higgs mode},\label{2.15}\\
&&(\hbar\omega)^{2}=(\hbar q)^{2}\upsilon^{2}\quad-\quad\texttt{Goldstone mode}.\label{2.16}
\end{eqnarray}
Thus, we have two types of collective excitations: with an energy gap $\sqrt{8|a|m\upsilon^{2}}$ - Higgs mode and with acoustic spectrum $\omega=\upsilon q$ - Goldstone mode illustrated in Fig.\ref{Fig2}. We can see that for these dispersion relations $\min\left[\frac{\omega(q)}{q}\right]=\upsilon>0$, that is the Landau criterion for superfluidity is satisfied. Since the Higgs mode is oscillations of modulus of the order parameter $|\Psi|$ which determines the density of SC electrons as $n_{\mathrm{s}}=2\left|\Psi\right|^{2}$ at $T\rightarrow T_{c}$, then these oscillations are accompanied by changes of SC density. At the same time, the total electron density must be $n=n_{\mathrm{s}}+n_{\mathrm{n}}=\mathrm{const}$, because otherwise the oscillations of charge (plasmons) should take place (the plasmons exist in normal metal too, hence these oscillations are not specific for SC state). Thus, the oscillation of $n_{\mathrm{s}}$ when $n=\mathrm{const}$ and $q\neq 0$ can be presented as counterflows of SC and normal components so that $n_{\mathrm{s}}\mathbf{v}_{s}+n_{\mathrm{n}}\mathbf{v}_{n}=0$ - Fig.\ref{Fig3}. Hence the Higgs mode can be considered as sound in the gas of above-condensate quasiparticles (with spectrum $\sqrt{|\Delta|^{2}+v_{F}^{2}(p-p_{F})^{2}}$) - the second sound. It should be noted that since the normal component $n_{\mathrm{n}}=n-2\left|\Psi\right|^{2}$ is gas of excitations above condensate of the Cooper pairs (moreover, the size of a pair is much more than average distance between electrons), then separation of electrons into SC and normal is some conditionality: in reality each electron makes SC and normal movements simultaneously.

The speed $\upsilon$ can be found with the following way. Let us consider the long-wave limit $q\rightarrow 0$ for Higgs mode, that is the whole system oscillates in a phase. To change the SC density and, hence, the normal density, one Cooper pair must be broken as minimum. For this the energy $2|\Delta|$ must be spent as minimum (SC energy gap). In turn, from Eq.(\ref{2.15}) it can see that minimal energy to excite one Higgs boson is $\sqrt{8|a|m\upsilon^{2}}$, then
\begin{equation}\label{2.17}
    \sqrt{8|a|m\upsilon^{2}}=2|\Delta|\Rightarrow\upsilon=\frac{v_{F}}{\sqrt{3}},
\end{equation}
where $v_{F}$ is Fermi velocity, we have used $|\Delta|=T_{c}\left(\frac{8\pi^{2}}{7\zeta(3)}\right)^{1/2}\left(1-\frac{T}{T_{c}}\right)^{1/2}$, $a=\frac{6\pi^{2}T_{c}^{2}}{7\zeta(3)\varepsilon_{F}}\left(\frac{T}{T_{c}}-1\right)$ for pure superconductor \cite{sad2}. Thus, the speed $\upsilon$ does not depend on temperature (at $T\rightarrow T_{c}$) and is $\sim 10^{6}\mathrm{m}/\mathrm{s}\ll c$. Physical sense of this speed will be defined in the next section. Using Eq.(\ref{2.17}) we can see that a value $\hbar\omega=2|\Delta|$ is achieved in the Goldstone mode (\ref{2.16}) at $q=\frac{\sqrt{2}}{\xi}$, where $\xi=\sqrt{\frac{\hbar^{2}}{4m|a|}}$ is a temperature-dependent coherence length. Noteworthy that, if we set $q=\frac{1}{\lambda}$, then $\hbar\omega(q)>2|\Delta|$ be for type-I superconductors, and $\hbar\omega(q)<2|\Delta|$ be for type-II superconductors - Fig.\ref{Fig2}, since $\kappa\equiv\frac{\lambda}{\xi}<\frac{1}{\sqrt{2}}$ for the type-I and $\kappa>\frac{1}{\sqrt{2}}$ for the type-II.


The dispersion relation (\ref{2.15}) can be rewritten in the following relativistic-like form:
\begin{equation}\label{2.18}
    E^{2}=\widetilde{m}^{2}\upsilon^{4}+p^{2}\upsilon^{2},
\end{equation}
where
\begin{equation}\label{2.19}
    \widetilde{m}\equiv\frac{\sqrt{8|a|m}}{\upsilon}=\frac{\sqrt{2}\hbar}{\xi\upsilon}\propto\left(T_{c}-T\right)^{1/2}
\end{equation}
is the mass of a Higgs boson. The mass $\widetilde{m}$ determines the scale of spatial inhomogeneities in a superconductor: let $E=0$ (that is we consider a stationary configuration: $\Psi(t)=\mathrm{const}$) then $p^{2}=-\widetilde{m}^{2}\upsilon^{2}\Rightarrow p=i\hbar\frac{\sqrt{2}}{\xi}$, hence $\Psi=\Psi_{0}-\phi_{0}e^{i\frac{p}{\hbar} x}=\Psi_{0}-\phi_{0}e^{-\frac{\sqrt{2}}{\xi}x}$ (for example, proximity effect, where a superconductor occupies half-space $x>0$), that is the order parameter is recovered on length $\frac{\xi}{\sqrt{2}}=\frac{\hbar}{\widetilde{m}\upsilon}$. It should be noticed since in the normal phase (that is at $T>T_{c}$ or $H>H_{c}$ where $n_{\mathrm{s}}=0$, $n_{\mathrm{n}}=n$) equilibrium value of the order parameter is $\Psi=0$ then the Goldstone and Higgs oscillations loss any sense. In the normal phase the relaxation of the nonequilibrium order parameter $\Psi\neq 0$ to the equilibrium one $\Psi=0$ takes place only. In the normal phase the speed $\upsilon$ losses physical sense and the correct limit transition to the normal state in expressions, which depend on the factor $c/\upsilon$, will be formulated in Sect.\ref{electro}. As seen from Eqs.(\ref{2.15},\ref{2.17}) the energy of Higgs boson is $E\geq 2|\Delta|$, that is this mode exists in the free quasiparticle continuum. Hence the free Higgs oscillation decays to quasiparticles with energy $\sqrt{|\Delta|^{2}+v_{F}^{2}(p-p_{F})^{2}}$ each. Thus, \emph{the free Higgs mode in a pure superconductor is unstable}, hence its observation is problematical. So, investigations of resonant excitations in cuprate superconductors using THz pulse in \cite{cea} exhibits that in this nonlinear optical process the light-induced excitation of Cooper pairs is dominated, while the collective amplitude (Higgs) fluctuations of the SC order parameter give, in general, a negligible contribution. At the same time, Higgs mode is a scalar excitation of the order parameter, distinct from charge or spin fluctuations, and thus does not couple to electromagnetic fields linearly. It makes possible to study the Higgs mode through the nonlinear light–Higgs coupling which manifests itself, for example, in the third harmonic generation and pump-probe spectroscopy mediated by the Higgs mode \cite{shim}. It should be noted that the impurity scattering drastically enhances the light–Higgs coupling: in the pure limit the Higgs-mode contribution is subleading to quasiparticles, whereas in the dirty regime it becomes comparable with or even larger than the quasiparticle contribution. The precise ratio between the Higgs and quasiparticle contributions may depend on details of the system.

The Goldstone mode (\ref{2.14}) is eddy (Foucault) currents: alternating current $\mathbf{j}=\frac{e\hbar}{m}|\Psi|^{2}\nabla\theta$ (since $\theta\propto e^{i\omega t}$) generates alternating magnetic field $\mathrm{curl}\mathbf{H}(t)=\frac{4\pi}{c}\mathbf{j}(t)$, hence the electric field $\mathrm{curl} \mathbf{E}=\frac{1}{c}\frac{\partial\mathbf{H}}{\partial t}$ is induced, which drives the eddy currents in turn. In the long wave limit ($q\rightarrow 0$) the energy of the Goldstone mode is $E=0$, that means the passing of nondissipative direct current. These processes will be detail considered in Sects.\ref{electro},\ref{relax}. Substituting the supercurrent (\ref{2.10}) in the continuity condition (\ref{2.9}) we obtain the equation for Goldstone mode (\ref{2.14}). Thus, \emph{the equation for Goldstone mode} (\ref{2.14}) \emph{is the continuity condition for the corresponding eddy currents}. In the Sect.\ref{electro} we will reveal that Goldstone oscillations are not accompanied by the charge oscillations, i.e., $\frac{\partial\rho_{s}}{\partial t}=0$, hence these supercurrents are closured $\mathrm{div}\mathbf{j}_{s}=0$.

At the same time, in this Section we neglected by the normal component $n_{\mathrm{n}}=n-2|\Psi|^{2}$ with corresponding friction $-\frac{m}{\tau_{ph}}\mathbf{v}$, where $\tau_{ph}$ is the mean free path time of electrons caused by electron-phonon interaction (or by scattering on lattice defects). The friction causes damping of oscillations of the order parameter and can lead to the overdamped regime when monotonic relaxation of a fluctuation occurs. These processes are considered in Sect.\ref{relax}.

\subsection{Microscopic derivation}\label{micro}

\begin{figure}[h]
\includegraphics[width=16cm]{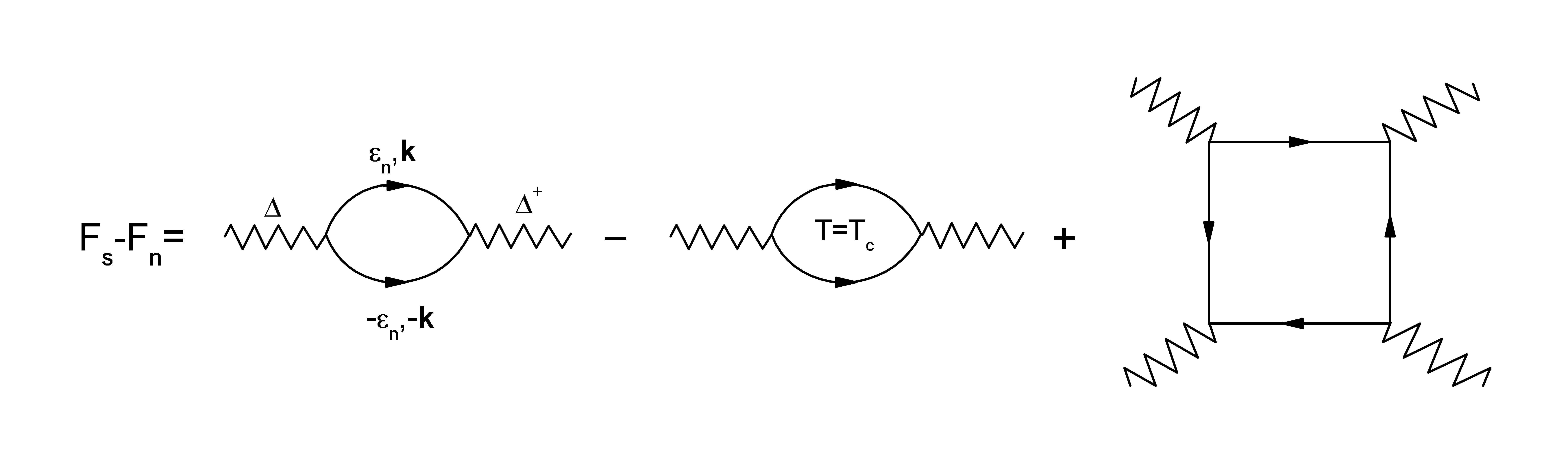}
\caption{Diagrammatic representation of Ginzburg-Landau expansion from \cite{sad1}.}\label{Fig4}
\end{figure}

Following \cite{sad1} let us consider electrons in a normal metal $(T>T_{c})$ propagating in a random "field" of thermodynamic fluctuations of the SC order parameter $\Delta$ (however $\langle\Delta\rangle=0$) which are static and "smooth" enough in space. Then we can write down the following Hamiltonian for interaction of an electron with these fluctuations:
\begin{eqnarray}\label{2.24}
    \widehat{H}_{int}&=&\sum_{\mathbf{k}}\left[\Delta a_{\mathbf{k}\uparrow}^{+}a_{-\mathbf{k}\downarrow}^{+}+\Delta a_{-\mathbf{k}\downarrow}a_{\mathbf{k}\uparrow}\right].
\end{eqnarray}
Correction to thermodynamic potential (free energy) due to (\ref{2.24}) is
\begin{eqnarray}\label{2.25}
  F_{s}-F_{n}&=&-|\Delta|^{2}T\sum_{\textbf{k}}\sum_{n}G(\mathbf{k},\varepsilon_{n})G(-\mathbf{k},-\varepsilon_{n})
  +|\Delta|^{2}T_{c}\sum_{\mathbf{k}}\sum_{n}G(\textbf{k},\varepsilon_{n})G(-\textbf{k},-\varepsilon_{n})|_{T=T_{c}} \nonumber\\
  &+&\frac{T}{2}|\Delta|^{4}\sum_{\mathbf{k}}\sum_{n}G^{2}(\mathbf{k},\varepsilon_{n})G^{2}(-\mathbf{k},-\varepsilon_{n})+\ldots,
\end{eqnarray}
where $G=\frac{1}{i\varepsilon_{n}-\xi}$ is an electron propagator for normal metal, $\varepsilon_{n}=\pi T(2n+1)$, $\xi=\hbar\upsilon_{F}(k-k_{F})$ is energy of an electron near Fermi surface. The first term (which is $\propto |\Delta|^{2}T$) diverges, but it is compensated by the second term (which is $\propto |\Delta|^{2}T_{c}$), the third term is finite and it can be taken at $T=T_{c}$ near $T_{c}$.  The terms from Eq.(\ref{2.25}) can be represented by diagrams shown in Fig.\ref{Fig4}. Calculation of the coefficients at $|\Delta|^{2}$, $|\Delta|^{4}$ gives:
\begin{equation}\label{2.26}
   F_{s}-F_{n}=V\nu_{F}\frac{T-T_{c}}{T_{c}}|\Delta|^{2}+V\nu_{F}\frac{7\zeta(3)}{16\pi^{2}T_{c}^{2}}|\Delta|^{4}+\ldots,
\end{equation}
where $\nu_{F}=\frac{mk_{F}}{2\pi^{2}\hbar^{2}}$ is density of state on Fermi surface per spin, $V$ is volume of the system, $\sum_{\textbf{k}}\rightarrow\frac{V}{(2\pi)^{2}}\int d^{3}k$. Thus, we have microscopic method of derivation of GL expansion of the free energy functional which will be generalized as follows.

\begin{figure}[h]
\includegraphics[width=8cm]{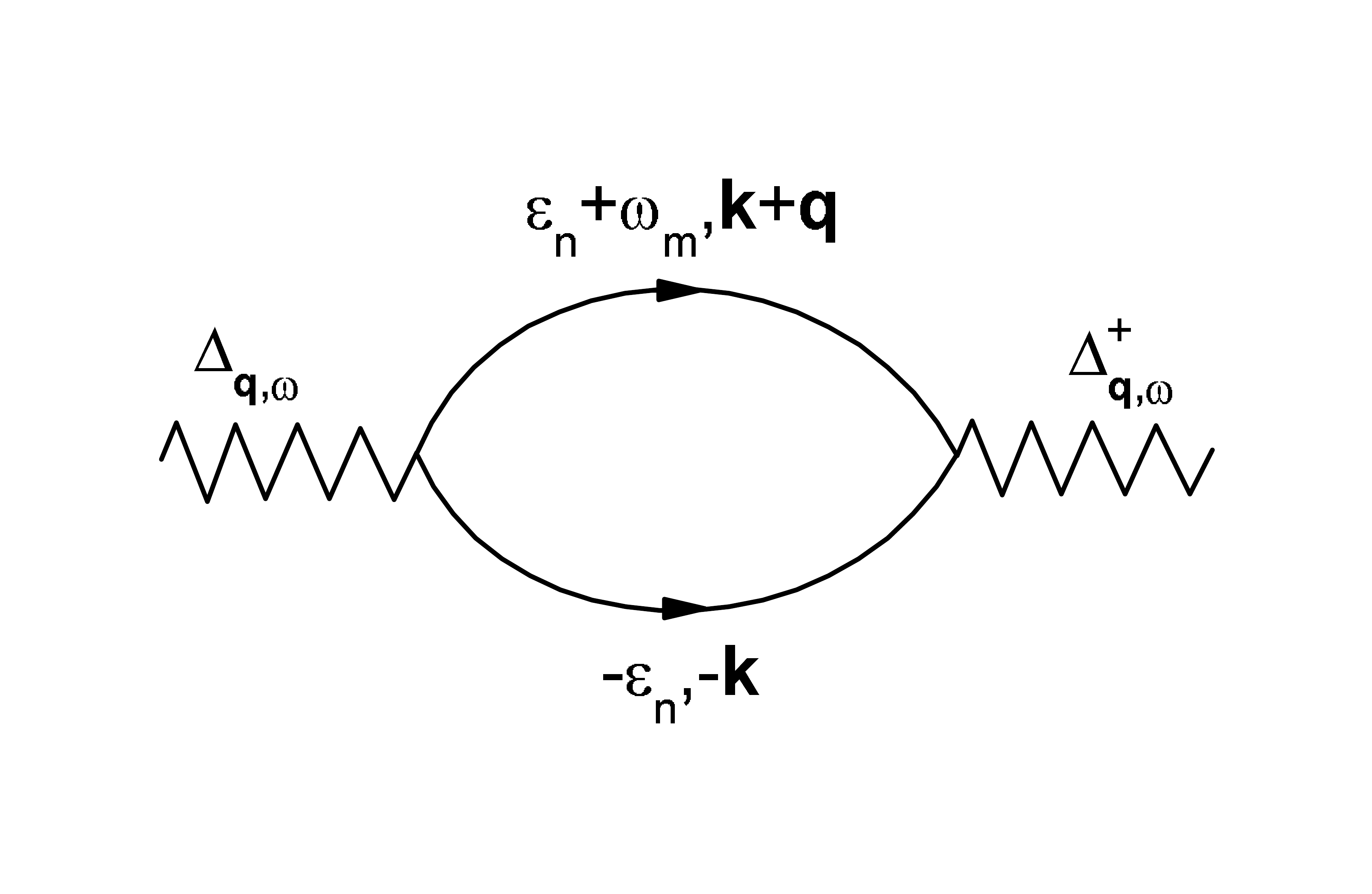}
\caption{The first diagram from Fig.\ref{Fig4} which accounts spatial and time inhomogeneities of the order parameter.}\label{Fig5}
\end{figure}

Let us consider the first diagram in Fig.\ref{Fig4} and let us give an additional momentum $\textbf{q}$ and an additional energy parameter $\omega_{m}=\pi T(2m+1)$ to one side of the loop as illustrated in Fig.\ref{Fig5}. This means that the order parameter will be function of these parameters $\Delta(\mathbf{q},\omega_{m})$, and note $\xi_{\mathbf{k}+\mathbf{q}}\approx\xi_{\mathbf{k}}+\hbar^{2}\frac{\mathbf{kq}}{m^{2}}$ where $|\mathbf{k}|=k_{F}$. Then instead the first term in the expansion (\ref{2.25}) we have:
\begin{eqnarray}\label{2.27}
  &&-|\Delta|^{2}T\sum_{\mathbf{k}}\sum_{n}G(\mathbf{k}+\mathbf{q},\varepsilon_{n}+\omega_{m})G(-\mathbf{k},-\varepsilon_{n})
  =-|\Delta|^{2}T\sum_{\mathbf{k}}\sum_{n}\frac{1}{i\varepsilon_{n}-\xi+\left(i\omega_{m}-\hbar^{2}\frac{\mathbf{kq}}{m}\right)}
  \frac{1}{-i\varepsilon_{n}-\xi}\nonumber\\
  &&\approx -|\Delta|^{2}TV\nu_{F}\int_{-\infty}^{+\infty}d\xi\sum_{n}\frac{1}{\varepsilon_{n}^{2}+\xi^{2}}
  +|\Delta|^{2}T_{c}V\frac{\nu_{F}}{2}\int_{-\infty}^{+\infty}d\xi\int_{-1}^{1}d(\cos\theta)  \sum_{n}\frac{\varepsilon_{n}^{2}-\xi^{2}}{\left(\varepsilon_{n}^{2}+\xi^{2}\right)^{3}}  \left[(i\omega_{m})^{2}+\hbar^{4}\frac{k_{F}^{2}}{m^{2}}\mathbf{q}^{2}\cos^{2}\theta \right]\nonumber\\
  &&=-|\Delta|^{2}TV\nu_{F}\int_{-\infty}^{+\infty}d\xi\sum_{n}\frac{1}{\varepsilon_{n}^{2}+\xi^{2}}
  +|\Delta|^{2}V\nu_{F}\frac{1}{4\pi^{2}T_{c}^{2}}\frac{7\zeta(3)}{4}(i\omega_{m})^{2}
  +|\Delta|^{2}V\nu_{F}\frac{\hbar^{2}\upsilon_{F}^{2}}{12\pi^{2}T_{c}^{2}}\frac{7\zeta(3)}{4}\mathbf{q}^{2}
\end{eqnarray}
up to the second-order terms in the additional parameters $\textbf{q}$ and $\omega_{m}$.


At first let us consider the expansion in the parameter $\mathbf{q}$ only:
\begin{equation}\label{2.28}
   F_{\mathbf{q}}=V\nu_{F}\frac{\hbar^{2}\upsilon_{F}^{2}}{12\pi^{2}T_{c}^{2}}\frac{7\zeta(3)}{4}\mathbf{q}^{2}|\Delta|^{2}
   +V\nu_{F}\frac{T-T_{c}}{T_{c}}|\Delta|^{2}+V\nu_{F}\frac{7\zeta(3)}{16\pi^{2}T_{c}^{2}}|\Delta|^{4}.
\end{equation}
Let us introduce "wave function" of Cooper pairs which determines density of SC component $n_{s}=2\left|\Psi\right|^{2}$ \cite{sad1,levit} so that to obtain the term $\frac{\hbar^{2}}{4m}\mathbf{q}^{2}\left|\Psi\right|^{2}$ ("kinetic energy" of Cooper pairs) instead the first term in Eq.(\ref{2.28}):
\begin{equation}\label{2.29}
   \Psi=\frac{\left(14\zeta(3)n\right)^{1/2}}{4\pi T_{c}}\Delta,
\end{equation}
where $n=\frac{k_{F}^{3}}{3\pi^{2}}$ is the total electron density. Then
\begin{equation}\label{2.30}
   F_{\mathbf{q}}=V\frac{\hbar^{2}}{4m}\mathbf{q}^{2}\left|\Psi\right|^{2}+Va\left|\Psi\right|^{2}+V\frac{b}{2}\left|\Psi\right|^{4},
\end{equation}
where
\begin{equation}\label{2.31}
   a=\nu_{F}\frac{(4\pi T_{c})^{2}}{14\zeta(3)n}\frac{T-T_{c}}{T_{c}},\quad b=\nu_{F}\frac{(4\pi T_{c})^{2}}{14\zeta(3)n^{2}},
\end{equation}
so that $|\Psi_{q=0}|^{2}=-\frac{a}{b}=n\frac{T_{c}-T}{T_{c}}\Rightarrow \frac{n_{\mathrm{s}}}{n}=2\left(1-\frac{T}{T_{c}}\right)$. The function $F_{\mathbf{q}}$ should be understood as free energy per a wave-vector $\mathbf{q}$. Corresponding free energy functional is
\begin{eqnarray}\label{2.32}
   F=\sum_{\mathbf{q}}F_{\mathbf{q}}=V\sum_{\mathbf{q}}\left[\frac{\hbar^{2}}{4m}\mathbf{q}^{2}\left|\Psi_{\mathbf{q}}\right|^{2}
   +a\left|\Psi_{\mathbf{q}}\right|^{2}+\frac{b}{2}\left|\Psi_{\mathbf{q}}\right|^{4}\right]
   \approx\int\left[\frac{\hbar^{2}}{4m}\left|\nabla\Psi(\mathbf{r})\right|^{2}   +a\left|\Psi(\mathbf{r})\right|^{2}+\frac{b}{2}\left|\Psi(\mathbf{r})\right|^{4}\right]d^{3}r,
\end{eqnarray}
where
\begin{equation}\label{2.33}
  \Psi(\mathbf{q})=\frac{1}{V}\int\Psi(\mathbf{r})e^{-i\mathbf{qr}}d^{3}r,\quad
  \Psi(\mathbf{r})=\sum_{\mathbf{q}}\Psi(\mathbf{q})e^{i\mathbf{qr}}=V\int\Psi(\mathbf{q})e^{i\mathbf{qr}}\frac{d^{3}q}{(2\pi)^{3}}.
\end{equation}
Thus, we have standard GL expansion.

Now let us consider the expansion with a term with $(i\omega_{m})^{2}$ in Eq.(\ref{2.27}). Then we have an expression:
\begin{equation}\label{2.34}
   \widetilde{F}_{\mathbf{q},\omega_{m}}=V\frac{1}{4m}\frac{(i\omega_{m})^{2}}{(\upsilon_{F}/\sqrt{3})^{2}}\left|\Psi_{\mathbf{q},\omega_{m}}\right|^{2}
   +V\frac{\hbar^{2}}{4m}\mathbf{q}^{2}\left|\Psi_{\mathbf{q},\omega_{m}}\right|^{2}
   +Va\left|\Psi_{\mathbf{q},\omega_{m}}\right|^{2}+V\frac{b}{2}\left|\Psi_{\mathbf{q},\omega_{m}}\right|^{4}.
\end{equation}
To perform analytic continuation of this expression to the real axis we have to make a substitution $i\omega_{m}\rightarrow\hbar\omega+i\delta$ where $\delta\rightarrow 0$ \cite{sad1}. Then we obtain:
\begin{equation}\label{2.35}
   \widetilde{F}_{\mathbf{q},\omega}=V\frac{\hbar^{2}}{4m\upsilon^{2}}\omega^{2}\left|\Psi_{\mathbf{q},\omega}\right|^{2}
   +V\frac{\hbar^{2}}{4m}\mathbf{q}^{2}\left|\Psi_{\mathbf{q},\omega}\right|^{2}
   +Va\left|\Psi_{\mathbf{q},\omega}\right|^{2}+V\frac{b}{2}\left|\Psi_{\mathbf{q},\omega}\right|^{4}\equiv T_{\mathbf{q}}+F_{\mathbf{q}}.
\end{equation}
Here, we have noted:
\begin{equation}\label{2.36}
   \upsilon\equiv\frac{\upsilon_{F}}{\sqrt{3}},
\end{equation}
that coincides with a "light" speed (\ref{2.17}). Presence of the frequency $\omega$ means dependence of the order parameter on time since
\begin{equation}\label{2.37}
  \Psi(t)=\mathfrak{T}\int\Psi(\omega)e^{-i\omega t}\frac{d\omega}{2\pi},\quad \Psi(\omega)=\frac{1}{\mathfrak{T}}\int\Psi(t)e^{i\omega t}dt,
\end{equation}
where $\mathfrak{T}\sim\sqrt[3]{V}/\upsilon$ is some time interval introduced for the same dimensions of $\Psi(\omega)$ and $\Psi(t)$. The function $\widetilde{F}_{\mathbf{q},\omega}$ does not have sense of free energy. From Eq.(\ref{2.35}) we can see that the first term plays role of "kinetic" energy $T_{\mathbf{q}}$ and the remaining terms $F_{\mathbf{q}}$ plays role of "potential" energy. Therefore corresponding Lagrangian has a form:
\begin{equation}\label{2.38}
   L_{\mathbf{q},\omega}=T_{\mathbf{q}}-F_{\mathbf{q}}=V\frac{\hbar^{2}}{4m\upsilon^{2}}\omega^{2}\left|\Psi_{\mathbf{q},\omega}\right|^{2}
   -V\frac{\hbar^{2}}{4m}\mathbf{q}^{2}\left|\Psi_{\mathbf{q},\omega}\right|^{2}
   -Va\left|\Psi_{\mathbf{q},\omega}\right|^{2}-V\frac{b}{2}\left|\Psi_{\mathbf{q},\omega}\right|^{4}.
\end{equation}
Then an action takes place (like a free energy functional (\ref{2.32})):
\begin{equation}\label{2.39}
   S=\mathfrak{T}^{2}\sum_{\mathbf{q}}\int\frac{d\omega}{2\pi}L_{\mathbf{q},\omega}=\int \mathcal{L}\left[\Psi(\mathbf{r},t)\right]d^{3}rdt=\frac{1}{\upsilon}\int \mathcal{L}\left[\Psi(\mathbf{r},t)\right]d\Omega,%\quad (d\Omega=\upsilon dtd^{3}r),
\end{equation}
where the Lagrangian is
\begin{eqnarray}\label{2.40}
    \mathcal{L}\left[\Psi(\mathbf{r},t)\right]&\approx&\frac{\hbar^{2}}{4m}\frac{1}{\upsilon^{2}}\left(\frac{\partial\Psi}{\partial t}\right)\left(\frac{\partial\Psi^{+}}{\partial t}\right)
    -\frac{\hbar^{2}}{4m}\left(\nabla\Psi\right)\left(\nabla\Psi^{+}\right)-a\left|\Psi\right|^{2}-\frac{b}{2}\left|\Psi\right|^{4}\nonumber\\
    &\equiv&\frac{\hbar^{2}}{4m}\left(\widetilde{\partial}_{\mu}\Psi\right)\left(\widetilde{\partial}^{\mu}\Psi^{+}\right)
    -a\left|\Psi\right|^{2}-\frac{b}{2}\left|\Psi\right|^{4},
\end{eqnarray}
which coincides with phenomenological Lagrangian (\ref{2.6}) in the extended TDGL theory. Here, $\widetilde{\partial}_{\mu}$ and $\widetilde{\partial}^{\mu}$ are covariant and contravariant differential operators (\ref{2.4}) accordingly. We can see that Lagrangian (\ref{2.40}) is Lorentz invariant in some 4D Minkowski space $\{\upsilon t,\textbf{r}\}$, where $\upsilon$ is an parameter of dimension of speed (like the light speed) which is determined by dynamical properties of the system. At the same time, the dynamics of conduction electrons remains non-relativistic.

\section{Electrodynamics}\label{electro}

\subsection{Gauge invariance}\label{electro1}

Let a superconductor be in electromagnetic field $A_{\mu}=(\varphi,-\textbf{A})$ (or contravariant vector $A^{\mu}=(\varphi,\textbf{A})$). In order to ensure the gauge invariance the differential operation $\partial_{\mu}\Psi$ must be changed as follows:
\begin{equation}\label{3.1}
    \partial_{\mu}\Psi\rightarrow\left(\partial_{\mu}+\frac{i2e}{c\hbar}A_{\mu}\right)\Psi
    =e^{i\theta}\left[\partial_{\mu}|\Psi|+i|\Psi|\left(\partial_{\mu}\theta+\frac{2e}{c\hbar}A_{\mu}\right)\right].
\end{equation}
Indeed, making a gauge transformation
\begin{equation}\label{3.2}
\theta=\theta'+\frac{2e}{c\hbar}\chi,\quad A_{\mu}=A_{\mu}'-\partial_{\mu}\chi=\left\{\begin{array}{c}
  \varphi=\varphi'-\frac{1}{c}\frac{\partial\chi}{\partial t} \\
  \mathbf{A}=\mathbf{A}'+\nabla\chi \\
\end{array}\right\},
\end{equation}
we have:
\begin{equation}\label{3.3}
    \partial_{\mu}\theta+\frac{2e}{c\hbar}A_{\mu}=\partial_{\mu}\theta'+\frac{2e}{c\hbar}A_{\mu}'.
\end{equation}
However in Lagrangian (\ref{2.6}) the differential operators (\ref{2.4}) take place:
\begin{equation}\label{3.4}
    \widetilde{\partial}_{\mu}\equiv\left(\frac{1}{\upsilon}\frac{\partial}{\partial t},\nabla\right)\neq
\partial_{\mu}\equiv\left(\frac{1}{c}\frac{\partial}{\partial t},\nabla\right).
\end{equation}
At the same time, in order to ensure the gauge invariance of Maxwell equations the field $A_{\mu}$ must be transformed with transformation (\ref{3.2}) only. Hence the equality (\ref{3.3}) with the differential operator $\widetilde{\partial}_{\mu}$ cannot be satisfied:
\begin{equation}\label{3.3a}
    \widetilde{\partial}_{\mu}\theta+\frac{2e}{c\hbar}A_{\mu}\neq\widetilde{\partial}_{\mu}\theta'+\frac{2e}{c\hbar}A_{\mu}',
\end{equation}
therefore the gauge invariance of the Lagrangian is violated. This fact is consequence of that the fields $\Psi$ and $A_{\mu}$ move in different Minkowski spaces: with the limit speeds $\upsilon$ and $c$ accordingly.

In order to ensure the gauge invariance of the Lagrangian (\ref{2.6}) with electromagnetic field we should consider a field
\begin{equation}\label{3.5}
    \widetilde{A}_{\mu}=\left(\frac{c}{\upsilon}\varphi,-\mathbf{A}\right)\equiv(\widetilde{\varphi},-\mathbf{A}).
\end{equation}
Then from the transformation (\ref{3.2}) we obtain the gauge transformations for the field $\widetilde{A}_{\mu}$:
\begin{equation}\label{3.6}
\theta=\theta'+\frac{2e}{c\hbar}\chi,\quad \widetilde{A}_{\mu}=\widetilde{A}_{\mu}'-\widetilde{\partial}_{\mu}\chi=\left\{\begin{array}{c}
  \widetilde{\varphi}=\widetilde{\varphi}'-\frac{1}{\upsilon}\frac{\partial\chi}{\partial t} \\
  \mathbf{A}=\mathbf{A}'+\nabla\chi \\
\end{array}\right\}.
\end{equation}
As a result we have
\begin{equation}\label{3.7}
    \widetilde{\partial}_{\mu}\theta+\frac{2e}{c\hbar}\widetilde{A}_{\mu}=\widetilde{\partial}_{\mu}\theta'+\frac{2e}{c\hbar}\widetilde{A}_{\mu}'.
\end{equation}
It should be noticed the following property:
\begin{equation}\label{3.8}
    \widetilde{\partial}_{\mu}\widetilde{A}^{\mu}\equiv \left(\frac{1}{\upsilon}\frac{\partial\widetilde{\varphi}}{\partial t},\nabla\mathbf{A}\right)=\left(\frac{1}{c}\frac{\partial\varphi}{\partial t},\nabla\mathbf{A}\right)\equiv\partial_{\mu}A^{\mu}.
\end{equation}

Interaction of a charge $e$ with electromagnetic field $A_{\mu}$ is described with an action
\begin{equation}\label{3.9}
    S_{int}=-\int\frac{e}{c}A_{\mu}dx^{\mu}.
\end{equation}
Taking into account $A_{\mu}dx^{\mu}=\varphi cdt-\mathbf{A}d\mathbf{r}=\widetilde{\varphi}\upsilon dt-\mathbf{A}d\mathbf{r}=\widetilde{A}_{\mu}d\widetilde{x}^{\mu}$ and defining the new charge $\widetilde{e}$ as
$\frac{e}{c}=\frac{\widetilde{e}}{\upsilon}$ we have
\begin{equation}\label{3.10}
    S_{int}=-\int\frac{e}{c}A_{\mu}dx^{\mu}=-\int\frac{\widetilde{e}}{\upsilon}\widetilde{A}_{\mu}d\widetilde{x}^{\mu}.
\end{equation}
Thus, interaction of the new charges with the new fields does not change, for example $e\varphi=\widetilde{e}\widetilde{\varphi}$. Moreover, the magnetic flux quantum does not change also: $\Phi_{0}=\frac{\pi\hbar c}{e}=\frac{\pi\hbar \upsilon}{\widetilde{e}}$. Then the action for a charged particle in the field is
\begin{equation}\label{3.11}
    S_{p+int}=\int\left(\frac{m\mathrm{v}^{2}}{2}+\frac{\widetilde{e}}{\upsilon}\mathbf{A}\mathbf{v}-\widetilde{e}\widetilde{\varphi}\right)dt.
\end{equation}
Therefore equation of motion is
\begin{eqnarray}\label{3.12}
    m\frac{d\mathbf{v}}{dt}&=&-\frac{\widetilde{e}}{\upsilon}\frac{\partial\mathbf{A}}{\partial t}  -\widetilde{e}\nabla\widetilde{\varphi}+\frac{\widetilde{e}}{\upsilon}\mathbf{v}\times\textrm{curl}\mathbf{A}
    \equiv\widetilde{e}\widetilde{\mathbf{E}}+\frac{\widetilde{e}}{\upsilon}\mathbf{v}\times\mathbf{H}=e\mathbf{E}+\frac{e}{c}\mathbf{v}\times\mathbf{H}.
\end{eqnarray}
Here, the electric field $\widetilde{\mathbf{E}}$ is such that
\begin{equation}\label{3.13}
    \widetilde{\mathbf{E}}=\frac{c}{\upsilon}\mathbf{E},\quad\widetilde{e}\widetilde{\mathbf{E}}=e\mathbf{E}.
\end{equation}
From the definition of $\widetilde{\mathbf{E}}$ and $\mathbf{H}$ in Eq.(\ref{3.12}) the first pair of Maxwell equations follows:
\begin{eqnarray}
  \textrm{curl}\widetilde{\mathbf{E}} &=& -\frac{1}{\upsilon}\frac{\partial\textbf{H}}{\partial t}, \label{3.14}\\
  \textrm{div}\textbf{H} &=& 0.\label{3.15}
\end{eqnarray}
Then corresponding action for the electromagnetic field $\widetilde{A}_{\mu}$ should be
\begin{equation}\label{3.16}
    S_{f}=-\frac{1}{16\pi\upsilon}\int\widetilde{F}_{\mu\nu}\widetilde{F}^{\mu\nu}d\Omega=
    \frac{1}{8\pi}\int\left(\widetilde{\mathbf{E}}^{2}-\mathbf{H}^{2}\right)dVdt,
\end{equation}
where $d\Omega=\upsilon dtdV$ is an element of the 4D Minkowski space, $\widetilde{F}_{\mu\nu}=\widetilde{\partial}_{\mu}\widetilde{A}_{\nu}-\widetilde{\partial}_{\nu}\widetilde{A}_{\mu}\equiv\left(\widetilde{\mathbf{E}},\mathbf{H}\right)$ is Faraday tensor. The action (\ref{3.10}) can be written as
\begin{equation}\label{3.17}
    S_{int}=-\frac{1}{\upsilon}\int\widetilde{\rho}\widetilde{A}_{\mu}d\widetilde{x}^{\mu}dV=    -\frac{1}{\upsilon^{2}}\int\widetilde{A}_{\mu}\widetilde{j}^{\mu}d\Omega,
\end{equation}
where the 4D current is\enlargethispage{\baselineskip}
\begin{equation}\label{3.18}
    \widetilde{j}^{\mu}=\widetilde{\rho}\frac{d\widetilde{x}^{\mu}}{dt}=
    \left(\upsilon\widetilde{\rho},\widetilde{\rho}\mathbf{v}\right)=\left(\upsilon\widetilde{\rho},\widetilde{\mathbf{j}}\right),
\end{equation}
here the new charge and current are
\begin{equation}\label{3.19}
    \widetilde{\rho}=\frac{\upsilon}{c}\rho,\quad\widetilde{\mathbf{j}}=\frac{\upsilon}{c}\mathbf{j}.
\end{equation}
Then the action describing the electromagnetic field and interaction of the current with the field is
\begin{equation}\label{3.20}
    S_{int+f}=\int\left[-\frac{1}{\upsilon^{2}}\widetilde{A}_{\mu}\widetilde{j}^{\mu}  -\frac{1}{16\pi\upsilon}\widetilde{F}_{\mu\nu}\widetilde{F}^{\mu\nu}\right]d\Omega.
\end{equation}
Variation of the action $\delta S_{int+f}$ gives the second pair of Maxwell equations:
\begin{eqnarray}
  \textrm{curl}\mathbf{H} &=& \frac{1}{\upsilon}\frac{\partial\widetilde{\mathbf{E}}}{\partial t}+
  \frac{4\pi}{\upsilon}\widetilde{\mathbf{j}}, \label{3.22}\\
  \textrm{div}\widetilde{\mathbf{E}} &=& 4\pi\widetilde{\rho},\label{3.23}
\end{eqnarray}
from where we can obtain conservation of charges both $\widetilde{\rho}$ and $\rho$ in a form:
\begin{equation}\label{3.24}
\widetilde{\partial}_{\mu}\widetilde{j}^{\mu}=0\Rightarrow\frac{\partial\widetilde{\rho}}{\partial t}+\textrm{div}
\widetilde{\mathbf{j}}=0\Rightarrow\frac{\partial\rho}{\partial t}+\textrm{div}\mathbf{j}=0.
\end{equation}
In order to figure out the physical sense of field $\widetilde{\mathbf{E}}$, charge $\widetilde{\rho}$ and speed $\upsilon$ let us consider Maxwell equations for electromagnetic field in a dielectric:
\begin{equation}\label{3.25}
    \begin{array}{c}
      \textrm{curl}\mathbf{E} = -\frac{1}{c}\frac{\partial\mathbf{H}}{\partial t} \\
      \textrm{div}\mathbf{H} = 0 \\
      \textrm{curl}\mathbf{H}=\frac{1}{c}\frac{\partial\mathbf{D}}{\partial t}+\frac{4\pi}{c}\mathbf{j} \\
      \textrm{div}\mathbf{D} = 4\pi\rho_{\mathrm{f}} \\
    \end{array},
\end{equation}
here $\mathbf{D}=\varepsilon\mathbf{E}$ is electric displacement, $\varepsilon$ is electric permittivity, $\rho_{\mathrm{f}}$ is density of free charges, $\mathbf{j}=\rho_{\mathrm{f}}\mathbf{v}$ is conduction current. Then for a case $\varepsilon=\mathrm{const}$ we can write:
\begin{equation}\label{3.26}
    \begin{array}{c}
      \textrm{curl}\mathbf{E} = -\frac{1}{c}\frac{\partial\mathbf{H}}{\partial t} \\
      \textrm{div}\mathbf{H} = 0 \\
      \textrm{curl}\mathbf{H}=\frac{\varepsilon}{c}\frac{\partial\mathbf{E}}{\partial t}+\frac{4\pi}{c}\mathbf{j} \\
      \textrm{div}\mathbf{E} = 4\pi\frac{\rho_{\mathrm{f}}}{\varepsilon} \\
    \end{array}\equiv
\begin{array}{c}
      \textrm{curl}\left(\sqrt{\varepsilon}\mathbf{E}\right) = -\frac{\sqrt{\varepsilon}}{c}\frac{\partial\mathbf{H}}{\partial t} \\
      \textrm{div}\mathbf{H} = 0 \\
      \textrm{curl}\mathbf{H}=\frac{\sqrt{\varepsilon}}{c}\frac{\partial\left(\sqrt{\varepsilon}\mathbf{E}\right)}{\partial t}+\frac{4\pi\sqrt{\varepsilon}}{c}\frac{\mathbf{j}}{\sqrt{\varepsilon}} \\
      \textrm{div}\left(\sqrt{\varepsilon}\mathbf{E}\right) = 4\pi\frac{\rho_{\mathrm{f}}}{\sqrt{\varepsilon}} \\
    \end{array}\equiv
    \begin{array}{c}
      \textrm{curl}\widetilde{\mathbf{E}} = -\frac{1}{\upsilon}\frac{\partial\mathbf{H}}{\partial t} \\
      \textrm{div}\mathbf{H} = 0 \\
      \textrm{curl}\mathbf{H}=\frac{1}{\upsilon}\frac{\partial\widetilde{\mathbf{E}}}{\partial t}+\frac{4\pi}{\upsilon}\widetilde{\mathbf{j}} \\
      \textrm{div}\widetilde{\mathbf{E}} = 4\pi\widetilde{\rho}_{\mathrm{f}} \\
    \end{array},
\end{equation}
where we have defined
$\widetilde{\mathbf{E}}\equiv\sqrt{\varepsilon}\mathbf{E}$, $\widetilde{\rho}_{\mathrm{f}}\equiv\rho_{\mathrm{f}}/\sqrt{\varepsilon}$,
$\widetilde{\mathbf{j}}\equiv\mathbf{j}/\sqrt{\varepsilon}$, $\upsilon\equiv c/\sqrt{\varepsilon}$. Comparing Eqs.(\ref{3.26}) with Eqs.(\ref{3.14},\ref{3.15},\ref{3.22},\ref{3.23}) we conclude that superconductor is equivalent to dielectric (in some effective sense, not in conductivity) with permittivity
\begin{equation}\label{3.27}
    \varepsilon=\frac{c^{2}}{\upsilon^{2}}\sim 10^{5},
\end{equation}
and \emph{the speed $\upsilon$ is the light speed in SC medium if there were no the skin-effect and Meissner effect}. We can see that this permittivity is giant compared to the permittivity of true dielectrics (maximum $\sim 10^{3}$ in segnetoelectrics). In vacuum $\upsilon=c$, that is $\varepsilon=1$ and $\widetilde{A}_{\mu}\equiv A_{\mu}$.

It should be noted that the dielectric permittivity is $\varepsilon=c^{2}/\upsilon^{2}$  in the long wave limit $q<1/\xi$ only. If frequency of electromagnetic wave is such that $\hbar\omega\geq 2|\Delta|$, then a photon can break a Cooper pair with transfer of its constituents in the free quasiparticle states. Hence in this area the strong absorption of the waves takes place. Thus, we can suppose the permittivity $\varepsilon$ is equal to $c^{2}/\upsilon^{2}=\mathrm{const}$ in a frequency interval $\hbar\omega<2|\Delta|$ only, at $\hbar\omega\gg 2|\Delta|$ we suppose $\varepsilon\rightarrow\varepsilon_{n}(\omega)$, where $\varepsilon_{n}(\omega)$ is the dielectric function of normal metal. As we transit into the normal phase, where $\Delta=0$, the permittivity $c^{2}/\upsilon^{2}$ losses sense, then $\varepsilon$ must be replaced by the dielectric function of normal metal $\varepsilon_{n}(\omega)$ with the following properties\cite{landau,tilley}:
\begin{equation}\label{3.56}
    \varepsilon_{n}(\omega\neq 0)\ll\frac{4\pi\sigma}{\omega},\quad\varepsilon_{n}(0)=\infty,
\end{equation}
where the conductivity $\sigma$ can be assumed constant at low frequencies. The case $\varepsilon(\omega=0)$ is special and it will be consider in next subsection.

In these terms, energy of the electric field $U_{\mathrm{el}}$ and density of the flux of electromagnetic energy $\mathbf{S}_{\mathrm{el}}$ are
\begin{eqnarray}
  U_{\mathrm{el}} &=& \frac{\mathbf{ED}}{8\pi}=\frac{\varepsilon E^{2}}{8\pi}=\frac{\widetilde{E}^{2}}{8\pi}\label{3.28} \\
  \mathbf{S}_{\mathrm{el}}&=&\frac{c}{4\pi}\mathbf{E}\times\mathbf{H}=\frac{\upsilon}{4\pi}\widetilde{\mathbf{E}}\times\mathbf{H}, \label{3.29}
\end{eqnarray}
so that $\mathrm{div}\mathbf{S}_{\mathrm{el}}=-\frac{\partial U_{\mathrm{el}}}{\partial t}$ (here, $U_{\mathrm{el}}$ should be understood as energy of the electromagnetic field $\frac{\widetilde{E}^{2}}{8\pi}+\frac{H^{2}}{8\pi}$). The electromagnetic field tensors in a dielectric are $F_{\mu\nu}=\left(\mathbf{E},\mathbf{H}\right)$ and $H^{\mu\nu}=\left(-\mathbf{D},\mathbf{H}\right)$ with an invariant $\frac{1}{2}F_{\mu\nu}H^{\mu\nu}=H^{2}-\mathbf{ED}=H^{2}-\varepsilon
E^{2}=H^{2}-\widetilde{E}^{2}$. In new representation the tensors can be written in a symmetrical form: $\widetilde{F}_{\mu\nu}=\left(\widetilde{\mathbf{E}},\mathbf{H}\right)$ and $\widetilde{F}^{\mu\nu}=\left(-\widetilde{\mathbf{E}},\mathbf{H}\right)$ with the corresponding invariant $\frac{1}{2}\widetilde{F}_{\mu\nu}\widetilde{F}^{\mu\nu}=H^{2}-\widetilde{E}^{2}=\frac{1}{2}F_{\mu\nu}H^{\mu\nu}$.

As a result of the above reasoning we can write the Lorentz-invariant gauge invariant Lagrangian:
\begin{eqnarray}\label{3.30}
    \mathcal{L}=\frac{\hbar^{2}}{4m}\left(\widetilde{\partial}_{\mu}+\frac{i2\widetilde{e}}{\upsilon\hbar}\widetilde{A}_{\mu}\right)\Psi
    \left(\widetilde{\partial}^{\mu}-\frac{i2\widetilde{e}}{\upsilon\hbar}\widetilde{A}^{\mu}\right)\Psi^{+}
    -a\left|\Psi\right|^{2}-\frac{b}{2}\left|\Psi\right|^{4}-\frac{1}{16\pi}\widetilde{F}_{\mu\nu}\widetilde{F}^{\mu\nu}
\end{eqnarray}
for the action
\begin{equation}\label{3.31}
    S=\frac{1}{\upsilon}\int\mathcal{L}(\Psi,\Psi^{+},\widetilde{A}_{\mu},\widetilde{A}^{\mu})\upsilon dtd^{3}r.
\end{equation}
The Lagrangian (\ref{3.30}) is a sum of energy of spatial-time inhomogeneity (the first term), potential energy of the field $\Psi$ and self-action (the second and third terms), and the last term is Lagrangian of the electromagnetic field $\widetilde{A}_{\mu}$ which is a sum of the external field and the self-consistent internal field.

\subsection{Anderson-Higgs mechanism}\label{electro2}

The modulus-phase representation (\ref{2.2}) can be considered as a local gauge $U(1)$ transformation \cite{sad}:
\begin{equation}\label{3.44a}
    \Psi=\left|\Psi\right|e^{i\frac{2\widetilde{e}}{\hbar\upsilon}\chi},
\end{equation}
so that the gauge field is transformed as
\begin{equation}\label{3.44b}
    \widetilde{A}_{\mu}'=\widetilde{A}_{\mu}+\widetilde{\partial}_{\mu}\chi.
\end{equation}
After transformations (\ref{3.44a},\ref{3.44b}) the Lagrangian (\ref{3.30}) takes a form:
\begin{eqnarray}\label{3.45}
    \mathcal{L}&=&\frac{\hbar^{2}}{4m}\left(\widetilde{\partial}_{\mu}+\frac{i2\widetilde{e}}{\hbar\upsilon}\widetilde{A}_{\mu}\right)\left|\Psi\right|
    \left(\widetilde{\partial}^{\mu}-\frac{i2\widetilde{e}}{\hbar\upsilon}\widetilde{A}^{\mu}\right)\left|\Psi\right|
    -a\left|\Psi\right|^{2}-\frac{b}{2}\left|\Psi\right|^{4}-\frac{1}{16\pi}\widetilde{F}_{\mu\nu}\widetilde{F}^{\mu\nu}\nonumber\\
    &\approx&\frac{\hbar^{2}}{4m}\widetilde{\partial}_{\mu}\phi\widetilde{\partial}^{\mu}\phi-2|a|\phi^{2}
    +\frac{1}{2}\frac{a^{2}}{b}+\frac{\hbar^{2}}{4m}\left(\frac{2\widetilde{e}}{\hbar\upsilon}\right)^{2}\Psi_{0}^{2}\widetilde{A}_{\mu}\widetilde{A}^{\mu}
    -\frac{1}{16\pi}\widetilde{F}_{\mu\nu}\widetilde{F}^{\mu\nu},
\end{eqnarray}
where we have  redesignated $\widetilde{A}_{\mu}'\rightarrow\widetilde{A}_{\mu}$ after the gauge transformation. We have neglected the nonlinear
term $\frac{\hbar^{2}}{2m}\left(\frac{2\widetilde{e}}{\hbar\upsilon}\right)^{2}\Psi_{0}\phi\widetilde{A}_{\mu}\widetilde{A}^{\mu}$ which describes coupling between a Higgs boson $\phi$ and two photons, since in the pure limit the Higgs-mode contribution is subleading to quasiparticles due to instability of the Higgs boson \cite{shim}. Comparing this Lagrangian with  Eq.(\ref{2.12}) we can see that the Goldstone boson $\theta$ is absorbed into the gauge field $\widetilde{A}_{\mu}$, i.e., Anderson-Higgs mechanism occurs. From this Lagrangian the equation for the field $\widetilde{A}_{\mu}$ can be obtained as
\begin{equation}\label{3.46}
    \widetilde{\partial}_{\mu}\frac{\partial \mathcal{L}}{\partial\left(\widetilde{\partial}_{\mu}\widetilde{A}_{\nu}\right)}-\frac{\partial
    \mathcal{L}}{\partial\widetilde{A}_{\nu}}=0\Rightarrow
\widetilde{\partial}_{\mu}\widetilde{F}^{\mu\nu}+\frac{1}{\lambda^{2}}\widetilde{A}^{\nu}=0,
\end{equation}
where we denoted
\begin{equation}\label{3.46a}
\lambda^{2}\equiv\frac{m\upsilon^{2}}{8\pi\widetilde{e}^{2}\Psi_{0}^{2}}=\frac{mc^{2}}{8\pi e^{2}\Psi_{0}^{2}}.
\end{equation}
Using the Lorentz gauge
$\widetilde{\partial}_{\mu}\widetilde{A}^{\mu}=0$, Eq.(\ref{3.46}) is reduced to
\begin{equation}\label{3.47}
    \widetilde{\partial}^{\mu}\widetilde{\partial}_{\mu}\widetilde{A}^{\nu}+\frac{1}{\lambda^{2}}\widetilde{A}^{\nu}=0
    \Rightarrow
    \begin{array}{c}
      \frac{1}{\upsilon^{2}}\frac{\partial^{2}\mathbf{A}}{\partial t^{2}}-\Delta\mathbf{A}+\frac{1}{\lambda^{2}}\mathbf{A}=0 \\
      \\
      \frac{1}{\upsilon^{2}}\frac{\partial^{2}\widetilde{\varphi}}{\partial t^{2}}-\Delta\widetilde{\varphi}+\frac{1}{\lambda^{2}}\widetilde{\varphi}=0 \\
    \end{array}.
\end{equation}
Taking the fields as harmonic modes $\mathbf{A}=\mathbf{A}_{0}e^{i(\mathbf{qr}-\omega t)}$ and $\widetilde{\varphi}=\widetilde{\varphi}_{0}e^{i(\mathbf{qr}-\omega t)}$ we obtain dispersion relation for photons in a superconductor:
\begin{equation}\label{3.48}
    \omega^{2}=\upsilon^{2}q^{2}+\frac{\upsilon^{2}}{\lambda^{2}}\Rightarrow E^{2}=\upsilon^{2}p^{2}+m_{A}^{2}\upsilon^{4},
\end{equation}
where
\begin{equation}\label{3.49}
   m_{A}=\frac{\hbar}{\lambda\upsilon}\propto\left(T_{c}-T\right)^{1/2}
\end{equation}
is mass of a photon, i.e., the Higgs mechanism takes place. It is noteworthy that the mass of a Higgs boson (\ref{2.19}) $\widetilde{m}=\frac{\sqrt{2}\hbar}{\xi\upsilon}$ and the mass of a photon (\ref{3.49}) are related as
\begin{equation}\label{3.50}
    \frac{\widetilde{m}(T)}{m_{A}(T)}=\sqrt{2}\kappa,
\end{equation}
where $\kappa=\lambda/\xi$ is GL parameter. That is for type-I superconductors $\widetilde{m}<m_{A}$, for type-II
superconductors $\widetilde{m}>m_{A}$.

The dispersion relation (\ref{3.48}) can be rewritten in a form:
\begin{equation}\label{3.48a}
    q^{2}=-\frac{1}{\lambda^{2}}+\frac{\omega^{2}}{\upsilon^{2}},
\end{equation}
that determines the penetration of electromagnetic field in a superconductor. Let us consider stationary case $\omega=0$, then $q^{2}=-\frac{1}{\lambda^{2}}$, hence $\mathbf{A}=\mathbf{A}_{0}e^{iqx}=\mathbf{A}_{0}e^{-x/\lambda}$. We can see that \emph{Anderson-Higgs mechanism manifests itself in that the electromagnetic (magnetic) field penetrates a superconductor in the depth} $\lambda$ (\ref{3.46a}), \emph{which is London penetration depth}. At the same time, for the nonstationary case $\omega\neq 0$ the penetration depth is obtained as $\lambda\left(1-\lambda^{2}\frac{\omega^{2}}{\upsilon^{2}}\right)^{-1/2}>\lambda$. If $\omega\geq\omega_{c}$, where
\begin{equation}\label{3.48b}
\omega_{c}=\frac{\upsilon}{\lambda}=\frac{m_{A}\upsilon^{2}}{\hbar}\propto\left(T_{c}-T\right)^{1/2},
\end{equation}
we have $q^2\geq 0$, that is electromagnetic field penetrates superconductor through entire its depth. Thus, \emph{the increase of the penetration depth with frequency and existence of the critical frequency} (\ref{3.48b}), \emph{when the depth becomes infinitely large, is principal result of the extended TDGL theory}. The London screening is caused by SC electrons with density $n_{\mathrm{s}}$. However in a superconductor the normal electrons with density $n_{\mathrm{n}}$ exist even at $T=0$ due to impurities \cite{sad1,levit}, at $T\neq 0$ the normal component is thermally excited quasiparticles. The normal component causes absorption of electromagnetic waves and the skin-effect. These processes smear the penetration effect, that will be considered in Sect.\ref{relax}. Moreover, if the frequency $\hbar\omega\geq 2|\Delta|$, then intensive absorption of the electromagnetic waves occurs due to the breaking of Cooper pairs, hence in order to observe the penetration effect it must be $\hbar\omega_{c}<2|\Delta|$. Using Eq.(\ref{2.17}) in a form $\hbar\upsilon\frac{\sqrt{2}}{\xi}=2|\Delta|$, where $\xi=\frac{\hbar}{\sqrt{4m|a|}}$ is coherence length, we can represent $\omega_{c}$ in another form:
\begin{equation}\label{3.48c}
\hbar\omega_{c}=2|\Delta|\frac{1}{\sqrt{2}\kappa}.
\end{equation}
Thus, the condition $\hbar\omega_{c}<2|\Delta|$ can be satisfied in type-II superconductors only (where $\kappa>1/\sqrt{2}$). It should be noted that this result has been obtained only for $s$-wave superconductors.

\begin{figure}[h]
\includegraphics[width=8cm]{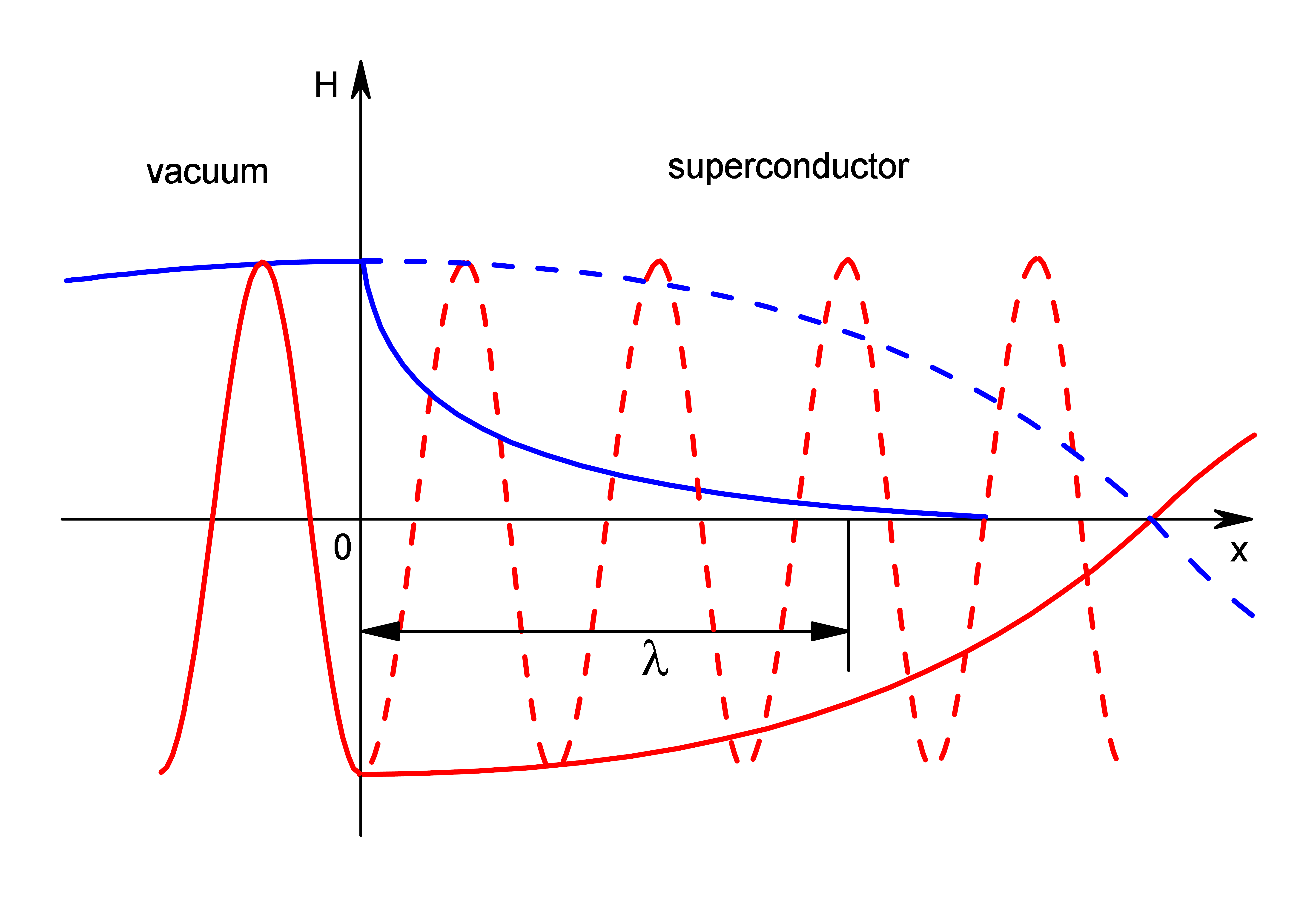}
\caption{long-wave field (blue line) $\Lambda\gg\lambda$ and short-wave field (red line) $\Lambda\ll\lambda$ fall on surface of a superconductor. The long-wave field can be screened by SC component in the London depth $\lambda$ (solid blue line). The short-wave oscillation changes its sign within the depth $\lambda$ (dash red line), hence the system cannot screen this field, because the screening supercurrent would have to turn over many times within the minimal length $\lambda$. Influence of the Anderson-Higgs mechanism on this wave is reduced to increasing of the wavelength (solid red line). The screening by the normal component (skin effect) is not considered here.}
\label{Fig6}
\end{figure}

The physical cause of the threshold $\omega_{c}$ is illustrated in Fig.\ref{Fig6} and it is as follows. Electromagnetic field $A_{\mu}$ is screened in the depth $\lambda$ by the induced supercurrent $\mathbf{j}_{s}=-\frac{c}{4\pi\lambda^{2}}\mathbf{A}$ according to Higgs mechanism. Thus, the system can respond to the external field $A_{\mu}$ in the \emph{minimal} length $\lambda$. For example, if we have a thin plate $d\ll\lambda$, then magnetic field penetrates it completely and does not affect its state \cite{schmidt,tinh}. Let, at first, the electromagnetic wave with wavelength $\Lambda=\frac{2\pi\upsilon}{\omega}\gg\lambda$ fall on a superconductor. Then the depth $\lambda$ is enough to screen this field. Now, let the wavelength be $\Lambda\ll\lambda$, then the field essentially changes within the length $\lambda$ - it changes the sign as illustrated in Fig.\ref{Fig6}. In this situation the SC system should screen the oppositely directed fields in the
length which is much less than $\lambda$, that cannot be done. Hence the fields with wavelength $\Lambda\lesssim\lambda$ (that is $\omega\geq\omega_{c}$) cannot be screened by the supercurrent. It should be noted that, according to our model, the speed of light in a superconductor is $\upsilon\ll c$, hence at given frequency $\omega$ the wavelength in the superconductor is much less than in vacuum: $\Lambda/2\pi=\frac{\upsilon}{\omega}\ll\frac{c}{\omega}$, precisely because of this property the screening effect can be observable: at the frequencies $\hbar\omega<2|\Delta|$ we can get into the interval $\Lambda\lesssim\lambda$. In turn, influence of the Anderson-Higgs mechanism on the wave with $\omega\geq\omega_{c}$ is reduced to increasing of the wavelength as $\Lambda\left(\omega\right)=\lambda\left(\frac{\omega^{2}}{\upsilon^{2}}\lambda^{2}-1\right)^{-1/2}$, so that $\Lambda\left(\omega_{c}\right)=\infty$.

If we suppose $q=0$ in Eq.(\ref{3.48}) then we obtain homogeneous oscillation (i.e., all points of a sample oscillate in a phase) with frequency $\omega_{c}$ (\ref{3.48b}). At $T=0$ we have
\begin{equation}\label{3.48d}
\omega_{c}(0)=\frac{\upsilon}{\lambda(0)}=\frac{\upsilon_{F}}{\sqrt{3}c}\sqrt{\frac{4\pi e^{2}n}{m}}\equiv\frac{\upsilon_{F}}{\sqrt{3}c}\omega_{p}\ll\omega_{p},
\end{equation}
where $\omega_{p}$ is the plasma frequency in metal. We can see this frequency is much lower than the ordinary plasma frequency. As have been demonstrated before these oscillations are consequence of Anderson-Higgs mechanism: Goldstone boson is absorbed into gauge field and the electromagnetic oscillation mode (\ref{3.47}) with spectrum (\ref{3.48}) appears. As will be demonstrated below Goldstone oscillations are not accompanied by longitudinal electric field (by oscillations of charge density) and they are eddy Meissner currents which generate the transverse field $\mathrm{div}\mathbf{A}=0$ only. Thus, these electromagnetic oscillations are \emph{eigen} oscillations of the SC system instead the phase oscillations. \emph{The critical frequency $\omega_{c}$ is a minimal limit of frequencies which can propagate through the system}, since for frequencies $\omega<\omega_{c}$ we have $q^{2}<0$.  If electromagnetic wave with frequency $\omega\geq\omega_{c}$ falls on superconductor then the wave is carried by these eigen oscillations hence the superconductor becomes transparent for such wave, at the same time for the wave with frequencies $\omega<\omega_{c}$ the carrier is absent hence such wave reflects from the surface of superconductor.

It should be noted that electromagnetic oscillations (\ref{3.47}) with spectrum (\ref{3.48}) and critical frequency $\omega_{c}$ (\ref{3.48b}) is analogous to propagation of electromagnetic wave (which can be presented as phase oscillations) in plane of Josephson junction \cite{tinh}. Such Josephson plasma frequency exists $\omega_{J}=\frac{\overline{c}}{\lambda_{J}}\ll\omega_{p}$ (where $\overline{c}\ll c$ is speed of the wave, $\lambda_{J}$ is magnetic penetration depth in the Josephson junction) which is the lowest limit of frequencies able to propagate in the junction plane and it has been observed in experiment \cite{dahm}. Analogy between  above-described electrodynamics of bulk superconductors and the phase wave in the plane of Josephson junction is presented in a Table \ref{tab1}.
\begin{table}[h]
\centering
\begin{tabular}{|c|c|c|c|c|}
  \hline\rule{0cm}{0.5cm}
   &bulk superconductor & Josephson junction at $\theta\ll\pi$  \\
  \hline\rule{0cm}{0.5cm}
   wave equation &$\Delta\mathbf{A}-\frac{1}{\upsilon^{2}}\frac{\partial^{2}\mathbf{A}}{\partial t^{2}}=\frac{1}{\lambda^{2}}\mathbf{A}$ & $\frac{\partial^{2}\theta}{\partial y^{2}}-\frac{1}{\overline{c}^{2}}\frac{\partial^{2}\theta}{\partial t^{2}}=\frac{1}{\lambda_{J}^{2}}\theta$     \\
  \hline\rule{0cm}{0.5cm}
   magnetic penetration depth & $\lambda=\sqrt{\frac{mc^{2}}{8\pi e^{2}|\Psi|^{2}}}$ & $\lambda_{J}=\sqrt{\frac{c\Phi_{0}}{8\pi^{2}J_{0}(2\lambda+d)}}
   \approx\sqrt{\frac{mc^{2}}{8\pi e^{2}|\Psi|^{2}\left(1+\frac{2\lambda}{d}\right)}}$    \\
  \hline\rule{0cm}{0.5cm}
   light speed &$\upsilon=\frac{c}{\sqrt{\varepsilon}}=\frac{\upsilon_{F}}{\sqrt{3}}\ll c$ & $\overline{c}=\frac{c}{\sqrt{\varepsilon_{J}}}\left(1+\frac{2\lambda}{d}\right)^{-\frac{1}{2}}\ll c$     \\
  \hline\rule{0cm}{0.5cm}
   critical frequency &$\omega_{c}=\frac{\upsilon}{\lambda}$&  $\omega_{J}=\frac{\overline{c}}{\lambda_{J}}$    \\
  \hline
\end{tabular}
\caption{Analogy between the electromagnetic wave in a bulk superconductor and the phase wave in the plane of Josephson junction \cite{tinh} when phase different is small $\theta\ll\pi$. Here, $J_{0}$ is a maximal current so that $J=J_{0}\sin\theta$ (we can suppose $J_{0}=\frac{e\hbar|\Psi|^{2}}{md}$), $\Phi_{0}=\frac{\pi\hbar c}{e}$ is the magnetic flux quantum, $d\ll\lambda$ is thickness of the junction, $\varepsilon_{J}$ is dielectric constant of material of the junction.}\label{tab1}
\centering
\end{table}

In order to obtain equations for the fields $\Psi$ and $\widetilde{A}_{\mu}$ we have to variate an action (\ref{3.31}):
\begin{eqnarray}
    D^{\mu}\frac{\partial\mathcal{L}}{\partial\left(D^{\mu}\Psi\right)^{+}}-\frac{\partial\mathcal{L}}{\partial\Psi^{+}}=0&\Rightarrow&
    \frac{\hbar^{2}}{4m}\left(\widetilde{\partial}^{\mu}+\frac{i2\widetilde{e}}{\hbar\upsilon}\widetilde{A}^{\mu}\right)
    \left(\widetilde{\partial}_{\mu}+\frac{i2\widetilde{e}}{\hbar\upsilon}\widetilde{A}_{\mu}\right)\Psi+a\Psi+b\left|\Psi\right|^{2}\Psi=0,\label{3.51a}\\
    \nonumber\\
    \widetilde{\partial}_{\nu}\frac{\partial\mathcal{L}}{\partial\left(\widetilde{\partial}_{\nu}\widetilde{A}_{\mu}\right)}-\frac{\partial   \mathcal{L}}{\partial\widetilde{A}_{\mu}}=0&\Rightarrow&    \widetilde{\partial}_{\nu}\widetilde{F}^{\mu\nu}=-\frac{i2\pi\widetilde{e}\hbar}{m\upsilon}\left[\Psi^{+}\left(\widetilde{\partial}^{\mu}
    +\frac{i2\widetilde{e}}{\upsilon\hbar}\widetilde{A}^{\mu}\right)\Psi
    -\Psi\left(\widetilde{\partial}^{\mu}-\frac{i2\widetilde{e}}{\upsilon\hbar}\widetilde{A}^{\mu}\right)\Psi^{+}\right],\label{3.51b}
\end{eqnarray}
where we have denoted $D^{\mu}\equiv\widetilde{\partial}^{\mu}+\frac{i2\widetilde{e}}{\hbar\upsilon}\widetilde{A}^{\mu}$. If we neglect space-time variation of $\Psi$, i.e., $\widetilde{\partial}^{\mu}\Psi=0$, then we obtain the field equation (\ref{3.46}). In Eq.(\ref{3.51b}) we can extract the current $\widetilde{j}^{\mu}$ using Maxwell equation $\widetilde{\partial}_{\nu}\widetilde{F}^{\mu\nu}=-\frac{4\pi}{\upsilon}\widetilde{j}^{\mu}$:
\begin{equation}\label{3.51c}
    \widetilde{j}^{\mu}=\frac{i\widetilde{e}\hbar}{2m}\left[\Psi^{+}\left(\widetilde{\partial}^{\mu}
    +\frac{i2\widetilde{e}}{\upsilon\hbar}\widetilde{A}^{\mu}\right)\Psi
    -\Psi\left(\widetilde{\partial}^{\mu}-\frac{i2\widetilde{e}}{\upsilon\hbar}\widetilde{A}^{\mu}\right)\Psi^{+}\right].
\end{equation}
The relevant boundary conditions are
\begin{equation}\label{3.52}
    \left(\widetilde{\partial}^{\mu}+\frac{i2\widetilde{e}}{\hbar\upsilon}\widetilde{A}^{\mu}\right)\Psi n_{\mu}=0,
\end{equation}
where $n^{\mu}=(n_{0},\mathbf{n})$ is a normal to some hypersurface. If the superconductor is bordered by vacuum or dielectric then the current $\mathbf{j}$ cannot flow in or out it \cite{schmidt}:
\begin{equation}\label{3.52a}
    \left(\nabla-\frac{i2e}{\hbar c}\mathbf{A}\right)\Psi\mathbf{n}=0,
\end{equation}
where $\mathbf{n}$ is a normal to the surface of superconductor. The vector $\mathbf{n}$ is determined by the surface of SC sample, i.e., this vector is given, at the same time the time component $n_{0}$ of the normal is arbitrary, then in order to satisfy the boundary condition (\ref{3.52}) we should set
\begin{equation}\label{3.52b}
    \frac{\partial\theta}{\partial t}+\frac{2\widetilde{e}}{\hbar}\widetilde{\varphi}=0,
\end{equation}
if $\frac{\partial|\Psi|}{\partial t}=0$. Relation (\ref{3.52b}) determines change of the phase $\theta$ in time in the presence of the scalar potential $\varphi$ that, in particular, gives the AC Josephson effect \cite{lifsh} if $\varphi$ is understood as the potential of a voltage source.

If to exclude the phase $\theta$ by means of gauge transformations (\ref{3.6}) with $\chi=\frac{\hbar\upsilon}{2\widetilde{e}}\theta$ then the current (\ref{3.51c}) can be represented in a form:
\begin{eqnarray}\label{3.53}
    \widetilde{j}^{\mu}\equiv(\upsilon\widetilde{\rho}_{s},\widetilde{\mathbf{j}}_{s})=\left[-\upsilon\frac{\widetilde{e}\hbar}{m\upsilon^{2}}|\Psi|^{2}
    \left(\frac{\partial\theta}{\partial t}+\frac{2\widetilde{e}}{\hbar}\widetilde{\varphi}\right),
    \frac{\widetilde{e}\hbar}{m}|\Psi|^{2}\left(\nabla\theta-\frac{2\widetilde{e}}{\upsilon\hbar}\mathbf{A}\right)\right]%\nonumber\\
    \rightarrow-\frac{\upsilon}{4\pi\lambda^{2}}\left(\widetilde{\varphi},\mathbf{A}\right)\equiv-\frac{\upsilon}{4\pi\lambda^{2}}\widetilde{A}^{\mu}.
\end{eqnarray}
From this equation we obtain the ordinary London equation
\begin{equation}\label{3.53a}
  \widetilde{\mathbf{j}}_{s}=-\frac{\upsilon}{4\pi\lambda^{2}}\mathbf{A}\Rightarrow\mathbf{j}_{s}=-\frac{c}{4\pi\lambda^{2}}\mathbf{A}
\end{equation}
and the London-like equation for charge $\rho_{s}$ and scalar potential $\varphi$:
\begin{equation}\label{3.53b}
  \widetilde{\rho}_{s}=-\frac{1}{4\pi\lambda^{2}}\widetilde{\varphi}\Rightarrow\rho_{s}=-\frac{\varepsilon}{4\pi\lambda^{2}}\varphi.
\end{equation}
Using Eq.(\ref{3.53}) we can turn the continuity condition into the Lorentz gauge:
\begin{equation}\label{3.53d}
  \frac{\partial\widetilde{\rho}}{\partial t}+\textrm{div}\widetilde{\mathbf{j}}=0\Rightarrow\frac{1}{\upsilon}\frac{\partial\widetilde{\varphi}}{\partial t}+\mathrm{div}\mathbf{A}=0.
\end{equation}
which is another manifestation of Anderson-Higgs mechanism. However we must apply the boundary condition (\ref{3.52b}) in 4D London equation (\ref{3.53}). Then we obtain $\rho_{s}=0$, hence, from electrostatic "London equation" (\ref{3.53b}), we have $\varphi=0$ (in the London gauge). This means that \emph{inside superconductor we have $\varphi=\mathrm{const}$, i.e., the potential electric field is absent $\mathbf{E}=-\nabla\varphi=0$ like in normal metals}. From Lorentz gauge $\frac{1}{\upsilon}\frac{\partial\widetilde{\varphi}}{\partial t}+\textrm{div}\mathbf{A}=0$ we obtain $\mathrm{div}\mathbf{A}=0$, hence transverse fields take place only.

It should be noted that the obtained results essentially differ from the results of the theory of gauge-invariant response of superconductors to external electromagnetic field presented in \cite{ars}. In this model the Coulomb interaction "pushes" the frequency of oscillations of the phase $\theta$ from acoustic spectrum to plasma frequency $\omega_{p}^{2}=4\pi ne^{2}/m$. The Goldstone mode is related to the appearance of oscillations in the order parameter phase, necessarily causing oscillations of the current $\mathbf{j}=\frac{e\hbar}{m}|\Psi|^{2}\nabla\theta$, which, in turn, give rise to oscillations in the electron number density. Any change in the charge density in metals generates strong longitudinal electric fields, which results in oscillations at the plasma frequency. Thus, Goldstone mode becomes unobservable in itself since it turns to plasma oscillations. However in our model the unobservability of Goldstone mode is explained with the Anderson-Higgs mechanism: the phase $\theta$ is absorbed into the gauge field $\widetilde{A}^{\mu}$. This makes impossible to accompany the phase oscillations by charge density oscillations. Indeed, according to the continuity condition $\frac{\partial\rho}{\partial t}+\mathrm{div}\mathbf{j}=0$, presence of such current that $\mathrm{div}\mathbf{j}\neq0$ changes the charge density $\rho$. Using equation for Goldstone mode (\ref{2.14}) we obtain $\textrm{div}\mathbf{j}=\frac{e\hbar}{m}|\Psi|^{2}\Delta\theta=\frac{e\hbar}{m}|\Psi|^{2}\frac{1}{\upsilon^{2}}\frac{\partial}{\partial t}\frac{\partial\theta}{\partial t}$. However we must use the gauge invariant expression $\frac{\partial\theta}{\partial t}+\frac{2\widetilde{e}}{\hbar}\widetilde{\varphi}$ instead $\frac{\partial\theta}{\partial t}$. Then the continuity condition takes a form:
\begin{equation}\label{3.54}
    \frac{\partial\widetilde{\rho}}{\partial t}+\frac{\widetilde{e}\hbar}{m}|\Psi|^{2}\frac{1}{\upsilon^{2}}\frac{\partial}{\partial t}\left(\frac{\partial\theta}{\partial t}+\frac{2\widetilde{e}}{\hbar}\widetilde{\varphi}\right)=0 \Rightarrow\frac{\partial\widetilde{\rho}}{\partial t}=0,
\end{equation}
where we have used the boundary condition (\ref{3.52b}). Thus, \emph{the phase oscillations are not accompanied by oscillations of the charge density} and the condition $\mathrm{div}\mathbf{j}=0$ takes place.

From other hand the current generates magnetic field as $\mathrm{curl}\mathbf{H}=\frac{4\pi}{c}\mathbf{j}$, thus the continuity condition can be written as follows:
\begin{equation}\label{3.55}
    \frac{\partial\widetilde{\rho}}{\partial t}+\frac{\widetilde{e}\hbar}{m}|\Psi|^{2}\mathrm{div}\left(\nabla\theta-\frac{2\widetilde{e}}{\upsilon\hbar}\mathbf{A}\right)=0
\rightarrow\frac{\partial\widetilde{\rho}}{\partial t}-\frac{\widetilde{e}\hbar}{m}\frac{2\widetilde{e}}{\upsilon\hbar}|\Psi|^{2}\mathrm{div}\mathbf{A} \Rightarrow\mathrm{div}\mathbf{A}=0,
\end{equation}
where we have excluded the phase $\theta$ by gauge transformation $\mathbf{A}=\mathbf{A}'+\nabla\chi$ with $\chi=\frac{\hbar\upsilon}{2\widetilde{e}}\theta$ and we have used Eq.(\ref{3.54}). Thus, \emph{Goldstone oscillations generate the transverse field $\mathrm{div}\mathbf{A}=0$ only} as result of the boundary condition (\ref{3.52b}), thereby remaining themselves as low-energy excitations $\hbar\omega(q\rightarrow0)\rightarrow0$ (acoustic spectrum $\omega=\upsilon q$) unlike plasmons $\hbar\omega_{p}\gg|\Delta|$. For this reason in the London equation $\mathbf{j}=-\frac{c}{4\pi\lambda^{2}}\mathbf{A}$ the transverse field $\mathrm{div}\mathbf{A}=0$ takes place only, since this excitation corresponds to Goldstone mode with the lowest energy $\hbar\omega(q\rightarrow0)\rightarrow0$, that ensures the closed current $\mathrm{div}\mathbf{j}=0$. Thereby a superconductor is characterized with the dielectric permittivity $\varepsilon=\frac{c^{2}}{\upsilon^{2}}$ for the induced electric field $\mathbf{E}=-\frac{1}{c}\frac{\partial\mathbf{A}}{\partial t}$ only (for frequencies $0<\hbar\omega<2|\Delta|$) and the speed $\upsilon$ is the light speed in SC medium. For electrostatic field $\mathbf{E}=-\nabla\varphi$ the permittivity is $\varepsilon(\omega=0,\mathbf{q}=0)=\infty$ like in metals. This fact explains experimental results in \cite{bert3} which illustrate that the external electric field does not affect significantly the SC state unlike predictions of the covariant extension of GL theory \cite{bert1,bert2}, since in our model a superconductor screens electrostatic field by metallic mechanism in the microscopic Thomas-Fermi length, unlike the screening of magnetic fields by SC electrons in London depth. Thus, unlike models \cite{bert1,bert2,bert3,hirs1}, electrostatics and magnetostatics have different form despite Lorentz covariance of the model. It should be noted that \emph{the fixing of the transverse gauge} $\mathrm{div}\mathbf{A}=0$ \emph{is result of the spontaneously broken gauge symmetry in SC phase}.

Thus, from Eq.(\ref{3.54}) we can see that the oscillations of phase (Goldstone mode) cannot change the charge density $\rho$, hence the phase oscillations are not affiliated with the plasmons. Indeed, the phase oscillations $\theta(\mathbf{r},t)$ are oscillations of the order parameter $|\Psi|e^{i\theta}$, i.e., are specific for the SC state. At the same time, the plasma oscillations exist unchanged both in SC phase and in the normal metal phase, i.e., they are unrelated to the SC ordering and are not specific for the SC state. Therefore if the plasma oscillations were the phase oscillations of the order parameter, then this would be affected them at the transition point $T_{c}$ necessarily, however the spectrum of the plasma oscillations does not depend on temperature.

Since inside superconductor the electrostatic field is absent $\mathbf{E}=0$ as consequence of boundary condition (\ref{3.52b}), then in order to screen the field the electron liquid must become spatial inhomogeneous in the Thomas-Fermi length $\lambda_{\mathrm{TF}}$ (screening charge is created in this region). However the SC phase is characterized by spatial scales: London depth $\lambda(T)$ and coherence length $\xi(T)$ which are $\lambda,\xi\gg\lambda_{\mathrm{TF}}$ even at $T=0$. Analogously, there is a time scale $\tau\sim\hbar/|\Delta|$, which is such that $|\Delta|\ll\hbar\omega_{p}$, where $\omega_{p}$ is the plasma frequency. At variations over shorter distances $l\ll\xi$ and time intervals $\delta t\ll\tau$ the SC system does not have reserve of the free energy $\frac{a^{2}}{2b}\sim\frac{\hbar^{2}}{m}\frac{|\Psi|^{2}}{\xi^{2}}\sim\frac{\hbar^{2}}{m\upsilon^{2}}\frac{|\Psi|^{2}}{\tau^{2}}$ (where the scales $\xi$ and $\tau$ are related as $\xi/\tau\sim\upsilon$) to compensate the spatial-time variation energies $\sim\frac{\hbar^{2}}{m}\frac{1}{l^{2}}|\Psi|^{2}$, $\sim\frac{\hbar^{2}}{m\upsilon^{2}}\frac{1}{(\delta t)^{2}}|\Psi|^{2}$. Thus, variation of order parameter is impossible on distances $\sim\lambda_{\mathrm{TF}}$ and time intervals $\sim 1/\omega_{p}$. Therefore within these spatial-time scales (TF length and plasma frequency) the SC order parameter $\Psi$ (or $\Delta$) losses any sense and a superconductor should behave like ordinary metal. Thus, the plasma (Thomas-Fermi) screening and the plasma oscillations exist in metal independently on its state.

Let us differentiate Eq.(\ref{3.47}) with respect to time and apply the operation $\mathrm{curl}$ supposing $\widetilde{\varphi}=0$ and $\mathrm{div}\mathbf{A}=0$, then we obtain:
\begin{eqnarray}
    &\frac{1}{\upsilon^{2}}\frac{\partial^{2}\widetilde{\mathbf{E}}}{\partial t^{2}}-\Delta\widetilde{\mathbf{E}}+\frac{1}{\lambda^{2}}\widetilde{\mathbf{E}}=0 \label{3.55a}\\
    &\frac{1}{\upsilon^{2}}\frac{\partial^{2}\mathbf{H}}{\partial t^{2}}-\Delta\mathbf{H}+\frac{1}{\lambda^{2}}\mathbf{H}=0, \label{3.55b}
\end{eqnarray}
where we have used $\widetilde{\mathbf{E}}=-\frac{1}{\upsilon}\frac{\partial\mathbf{A}}{\partial t}$ and $\mathbf{H}=\mathrm{curl}\mathbf{A}$. Using Maxwell equations $\mathrm{curl}\widetilde{\mathbf{E}}=-\frac{1}{\upsilon}\frac{\partial\mathbf{H}}{\partial t}$ and $\mathrm{curl}\mathbf{H}= \frac{1}{\upsilon}\frac{\partial\widetilde{\mathbf{E}}}{\partial t}+\frac{4\pi}{\upsilon}\widetilde{\mathbf{j}}$ we obtain accordingly:
\begin{eqnarray}
    &\frac{\partial \mathbf{j}_{s}}{\partial t}=\frac{c^{2}}{4\pi\lambda^{2}}\mathbf{E} \label{3.56a}\\
    &\mathrm{curl}\mathbf{j}_{s}=-\frac{c}{4\pi\lambda^{2}}\mathbf{H}, \label{3.56b}
\end{eqnarray}
which are the first and the second London equations. Since Goldstone oscillations generate the transverse field only $\mathbf{A}\equiv\mathbf{A}_{\perp}$ (here $\mathrm{div}\mathbf{A}_{\perp}=0$) then the electric field $\mathbf{E}$ is transverse too: $\mathbf{E}_{\perp}=-\frac{1}{c}\frac{\partial\mathbf{A}_{\perp}}{\partial t}$. However Eqs.(\ref{3.56a},\ref{3.56b}) are not equivalent to Eqs.(\ref{3.55a},\ref{3.55b}) (or to Eq.(\ref{3.47})) since we have extracted the current $\mathbf{j}$ from the field equations by Maxwell equations (\ref{3.14},\ref{3.15},\ref{3.22},\ref{3.23}) which have \emph{unbroken gauge symmetry}. The current is Goldstone mode which are absorbed into the gauge field $\mathbf{A}$ as we have seen before. Thus, the extraction of $\mathbf{j}$ violates Anderson-Higgs mechanism. As a result we obtain the first London equation which is the second Newton law for SC electrons (charges are accelerated by the electric field $\mathbf{E}$ but the friction is absent), that is Eq.(\ref{3.56a}) is equation for ideal conductor. As a result we lose the term $\frac{1}{\upsilon^{2}}\frac{\partial^{2}\mathbf{A}}{\partial t^{2}}$ which plays role at relatively high frequencies $\omega\sim\omega_{c}$, and the frequency dependence is caused by the complex conductivity $\sigma_{s}=-i\frac{c^{2}}{4\pi\lambda^{2}\omega}$ only that provides the low-frequency regime such that the penetration depth does not depend on frequency: $L(\omega)=\lambda$ \cite{tinh}. Thus, the field equations (\ref{3.47},\ref{3.55a},\ref{3.55b}) are more general than London equations (\ref{3.56a},\ref{3.56b}). In the ordinary GL theory the extraction of the current $\mathbf{j}_{s}$ by Maxwell (Poisson) equation $\mathrm{curl}\mathbf{H}=-\Delta\mathbf{A}=\frac{4\pi}{c}\widetilde{\mathbf{j}}$ from the field equation
\begin{equation}\label{3.57}
  \Delta\mathbf{A}=\frac{1}{\lambda^{2}}\mathbf{A}\Rightarrow\mathbf{j}_{s}=-\frac{c}{4\pi\lambda^{2}}\mathbf{A}
\end{equation}
does not lead to the losses in dynamics because GL theory is stationary (equilibrium) theory. At the same time, as it has been demonstrated in Appendix \ref{london}, the first London equation (\ref{3.56a}) is a consequence of the second London equation (\ref{3.56b}). This confusion is caused by the violation of Anderson-Higgs mechanism in the following way. The second equation has a form without the current as $\Delta\mathbf{H}=\frac{1}{\lambda^{2}}\mathbf{H}$ which is a result of minimization of free energy of a superconductor $\frac{1}{8\pi}\int\left[\mathbf{H}^{2}+\lambda^{2}(\mathrm{curl}\mathbf{H})^{2}\right]dV$. As described above, the applying of Maxwell equation $\mathrm{curl}\mathbf{H}=\frac{4\pi}{\upsilon}\widetilde{\mathbf{j}}$ extracts the current $\mathbf{j}$ which had to be absorbed into the gauge field $\mathbf{A}$. This violation results in the appearance of equation for ideal conductor (the first London equation) from the equation for superconductor (the second London equation).

\section{Goldstone modes in a two-band superconductor}\label{two}

In previous section we could observe Anderson-Higgs mechanism: the Goldstone boson $\theta$ is absorbed into the gauge field $\widetilde{A}_{\mu}$ - Eq.(\ref{3.45}). Thus, the Goldstone mode is not observable in single-band superconductors. Let us consider this problem in two-band superconductors. In presence of two order parameters in a bulk isotropic s-wave superconductor the GL free energy functional (at $\mathbf{A}=0$) can be written as
\cite{asker1,asker2,doh}:
\begin{eqnarray}\label{4.1}
    F&=&\int d^{3}r[\frac{\hbar^{2}}{4m_{1}}\left|\nabla\Psi_{1}\right|^{2}+\frac{\hbar^{2}}{4m_{2}}\left|\nabla\Psi_{2}\right|^{2}\nonumber\\
    &+&a_{1}\left|\Psi_{1}\right|^{2}+a_{2}\left|\Psi_{2}\right|^{2}+\frac{b_{1}}{2}\left|\Psi_{1}\right|^{4}+\frac{b_{2}}{2}\left|\Psi_{2}\right|^{4}
    +\epsilon\left(\Psi_{1}^{+}\Psi_{2}+\Psi_{1}\Psi_{2}^{+}\right)],
\end{eqnarray}
where $m_{1,2}$ denote the effective mass of carriers in the correspond band, the coefficients $a_{1,2}$ are given as $a_{i}=\gamma_{i}(T-T_{ci})$ where $\gamma_{i}$ are some constants, the coefficients $b_{1,2}$ are independent on temperature, the quantity $\epsilon$ describes interband mixing of the two order parameters (proximity effect). If we switch off the interband interaction $\epsilon=0$, then we will have two independent superconductors with different critical temperatures $T_{c1}$ and $T_{c2}$ because the intraband interactions can be different. Interaction between gradients of the order parameters (drag effect), which is considered in \cite{grig}, is omitted for easing of consideration.

Minimization of the free energy functional with respect to the order parameters, if $\nabla \Psi_{1,2}=0$, gives
\begin{equation}\label{4.2}
\left\{\begin{array}{c}
  a_{1}\Psi_{1}+\epsilon\Psi_{2}+b_{1}\Psi_{1}^{3}=0 \\
  a_{2}\Psi_{2}+\epsilon\Psi_{1}+b_{2}\Psi_{2}^{3}=0 \\
\end{array}\right\}.
\end{equation}
Near critical temperature $T_{c}$ we have $\Psi_{1,2}^{3}\rightarrow 0$, hence we can find the critical temperature as a solvability condition of the linearized
system (\ref{4.2}):
\begin{equation}\label{4.3}
a_{1}a_{2}-\epsilon^{2}=\gamma_{1}\gamma_{2}(T_{c}-T_{c1})(T_{c}-T_{c2})-\epsilon^{2}=0.
\end{equation}
Solving this equation we find $T_{c}>T_{c1},T_{c2}$, moreover, the solution does not depend on the sign of $\epsilon$. The sign determines the phase difference of the order parameters $|\Psi_{1}|e^{i\theta_{1}}$ and $|\Psi_{2}|e^{i\theta_{2}}$:
\begin{equation}\label{4.4}
    \begin{array}{cc}
      \cos(\theta_{1}-\theta_{2})=1 & \mathrm{if}\quad\epsilon<0  \\
      \cos(\theta_{1}-\theta_{2})=-1 & \mathrm{if}\quad\epsilon>0 \\
    \end{array},
\end{equation}
that follows from Eq.(\ref{4.2}). The case $\epsilon<0$ corresponds to attractive interband interaction, the case $\epsilon>0$ corresponds to repulsive interband interaction.

According to the method described in Sect.\ref{normal} the two-component scalar fields $\Psi_{1,2}(\mathbf{r},t)$ should minimize an action $S$ in the Minkowski space:
\begin{equation}\label{4.5}
    S=\frac{1}{\upsilon}\int\mathcal{L}(\Psi_{1},\Psi_{2},\Psi^{+}_{1},\Psi^{+}_{2})\upsilon dtd^{3}r.
\end{equation}
The Lagrangian $\mathcal{L}$ is built by generalizing the density of free energy in Eq.(\ref{4.1}) to the "relativistic" invariant form by substitution of covariant and contravariant differential operators (\ref{2.4}) instead the gradient operator:
\begin{eqnarray}\label{4.6}
    \mathcal{L}&=&\frac{\hbar^{2}}{4m_{1}}\left(\widetilde{\partial}_{\mu}\Psi_{1}\right)\left(\widetilde{\partial}^{\mu}\Psi_{1}^{+}\right)
    +\frac{\hbar^{2}}{4m_{2}}\left(\widetilde{\partial}_{\mu}\Psi_{2}\right)\left(\widetilde{\partial}^{\mu}\Psi_{2}^{+}\right)\nonumber\\
    &-&a_{1}\left|\Psi_{1}\right|^{2}-\frac{b_{1}}{2}\left|\Psi_{1}\right|^{4}-a_{2}\left|\Psi_{2}\right|^{2}-\frac{b_{2}}{2}\left|\Psi_{2}\right|^{4}
    -\epsilon\left(\Psi_{1}^{+}\Psi_{2}+\Psi_{1}\Psi_{2}^{+}\right),
\end{eqnarray}
where for $\Psi_{1}$ and $\Psi_{2}$ with the masses $m_{1}$ and $m_{2}$ accordingly the same speed $\upsilon$ is used. According to \cite{grig} the theory of a two-band superconductor can be reduced to GL theory of a single-band superconductor for equilibrium values of the orders parameters (time independent). In this model the orders parameters are related as $\Psi_{2}=C(T)\Psi_{1}$ at $T\rightarrow T_{c},T>T_{c1},T_{c2}$, where the coefficient $C$ is
\begin{equation}\label{4.7}
    \begin{array}{cc}
      C=\sqrt{\frac{a_{1}}{a_{2}}}, & \mathrm{if} \quad\epsilon<0 \\
      C=-\sqrt{\frac{a_{1}}{a_{2}}}, & \mathrm{if} \quad\epsilon>0 \\
    \end{array}.
\end{equation}
Then the relation (\ref{2.17}) can be written in a form:
\begin{equation}\label{4.8}
    \sqrt{8|A|M\upsilon^{2}}=2|\Delta|,
\end{equation}
where $A=a_{1}+a_{2}C^{2}+2\epsilon C$, $B=b_{1}+b_{2}C^{4}$, $M^{-1}=\frac{1}{m_{1}}+\frac{C^{2}}{m_{2}}$ and $|\Delta|\propto\Psi_{0}=\sqrt{-A/B}$ is an effective gap. Obviously, $\upsilon\sim v_{F1},v_{F2}$ like in Eq.(\ref{2.17}).

Let us consider movement of the phases only. Using the modulus-phase representation and assuming $|\Psi_{1}|=\mathrm{const}$ and $|\Psi_{2}|=\mathrm{const}$ Lagrangian (\ref{4.6}) takes a form:
\begin{eqnarray}\label{4.10}
    \mathcal{L}&=&\frac{\hbar^{2}}{4m_{1}}|\Psi_{1}|^{2}\left(\widetilde{\partial}_{\mu}\theta_{1}\right)\left(\widetilde{\partial}^{\mu}\theta_{1}\right)
    +\frac{\hbar^{2}}{4m_{2}}|\Psi_{2}|^{2}\left(\widetilde{\partial}_{\mu}\theta_{2}\right)\left(\widetilde{\partial}^{\mu}\theta_{2}\right)
    -2|\Psi_{1}||\Psi_{2}|\epsilon\cos\left(\theta_{1}-\theta_{2}\right)+\mathcal{L}\left(|\Psi_{1}|,|\Psi_{2}|\right).
\end{eqnarray}
Corresponding Lagrange equations are
\begin{eqnarray}
  \frac{\hbar^{2}}{4m_{1}}|\Psi_{1}|^{2}\widetilde{\partial}_{\mu}\widetilde{\partial}^{\mu}\theta_{1}
  -|\Psi_{1}||\Psi_{2}|\epsilon\sin\left(\theta_{1}-\theta_{2}\right) &=& 0\label{4.11a}\\
  \frac{\hbar^{2}}{4m_{2}}|\Psi_{2}|^{2}\widetilde{\partial}_{\mu}\widetilde{\partial}^{\mu}\theta_{2}
  +|\Psi_{1}||\Psi_{2}|\epsilon\sin\left(\theta_{1}-\theta_{2}\right) &=& 0.\label{4.11b}
\end{eqnarray}
The phases can be written in a form of oscillations:
\begin{equation}\label{4.12}
    \begin{array}{c}
      \theta_{1}=\theta_{1}^{0}+Ae^{i(\mathbf{qr}-\omega t)} \\
      \theta_{2}=\theta_{2}^{0}+Be^{i(\mathbf{qr}-\omega t)} \\
    \end{array},
\end{equation}
where equilibrium phases $\theta_{1,2}^{0}$ satisfy the relation (\ref{4.4}). Substituting the phases in Eqs.(\ref{4.11a},\ref{4.11b}) and linearizing (using the relation (\ref{4.4}) in a form $\epsilon\cos(\theta_{1}^{0}-\theta_{2}^{0})=-|\epsilon|$ and $\sin(\theta_{1}^{0}-\theta_{2}^{0})=0$) we obtain the following dispersion relations:
\begin{equation}\label{4.13}
    \omega^{2}=q^{2}\upsilon^{2},
\end{equation}
wherein $A=B$, and
\begin{equation}\label{4.14}
    (\hbar\omega)^{2}=4|\epsilon|\frac{|\Psi_{1}|^{2}m_{2}+|\Psi_{2}|^{2}m_{1}}{|\Psi_{1}||\Psi_{2}|}\upsilon^{2}+(\hbar q)^{2}\upsilon^{2},
\end{equation}
wherein
\begin{equation}\label{4.15}
    \frac{A}{B}=-\frac{m_{1}}{m_{2}}\frac{|\Psi_{2}|^{2}}{|\Psi_{1}|^{2}}.
\end{equation}
For symmetrical bands $m_{1}=m_{2}\equiv m$ and
$|\Psi_{1}|=|\Psi_{2}|$ we obtain
\begin{equation}\label{4.16}
    (\hbar\omega)^{2}=8|\epsilon|m\upsilon^{2}+(\hbar q)^{2}\upsilon^{2},\quad A=-B.
\end{equation}
Thus, we can see that in two-band superconductors there are two modes of the phase oscillations: the common mode oscillations with spectrum (\ref{4.13}) like Goldstone mode (\ref{2.16}) in single-band superconductors, and the oscillations of the relative phase between two SC condensates (\ref{4.14},\ref{4.16}) which can be identified as Leggett's mode \cite{legg,shar,yanag}. In a two band superconductor a current (flow) takes the following form:
\begin{equation}\label{4.17}
\mathbf{j}=e\hbar\left(\frac{|\Psi_{1}|^{2}}{m_{1}}\nabla\theta_{1}+\frac{|\Psi_{2}|^{2}}{m_{2}}\nabla\theta_{2}\right)
=ie^{i(\mathbf{qr}-\omega t)}e\hbar\left(\frac{|\Psi_{1}|^{2}}{m_{1}}A+\frac{|\Psi_{2}|^{2}}{m_{2}}B\right)\mathbf{q},
\end{equation}
from where we can see that for Leggett's mode (\ref{4.15}) the current is $\mathbf{j}=0$. Thus, unlike the Goldstone mode (\ref{4.13}) which is the eddy current, Leggett's oscillations are not accompanied by any currents.

Let us consider the Anderson-Higgs mechanism in a two-band superconductor. The gauge invariant form ($U(1)\times U(1)$ symmetry) of Lagrangian (\ref{4.6}) is
\cite{asker1,asker2,doh,grig}:
\begin{eqnarray}\label{4.18}
    \mathcal{L}&=&\frac{\hbar^{2}}{4m_{1}}\left(\widetilde{\partial}_{\mu}+\frac{i2\widetilde{e}}{\upsilon\hbar}\widetilde{A}_{\mu}\right)\Psi_{1}
    \left(\widetilde{\partial}^{\mu}-\frac{i2\widetilde{e}}{\upsilon\hbar}\widetilde{A}^{\mu}\right)\Psi_{1}^{+}
    +\frac{\hbar^{2}}{4m_{2}}\left(\widetilde{\partial}_{\mu}+\frac{i2\widetilde{e}}{\upsilon\hbar}\widetilde{A}_{\mu}\right)\Psi_{2}
    \left(\widetilde{\partial}^{\mu}-\frac{i2\widetilde{e}}{\upsilon\hbar}\widetilde{A}^{\mu}\right)\Psi_{2}^{+}\nonumber\\
    &-&a_{1}\left|\Psi_{1}\right|^{2}-\frac{b_{1}}{2}\left|\Psi_{1}\right|^{4}-a_{2}\left|\Psi_{2}\right|^{2}-\frac{b_{2}}{2}\left|\Psi_{2}\right|^{4}
    -\epsilon\left(\Psi_{1}^{+}\Psi_{2}+\Psi_{1}\Psi_{2}^{+}\right)-\frac{1}{16\pi}\widetilde{F}_{\mu\nu}\widetilde{F}^{\mu\nu}.
\end{eqnarray}
As in previous consideration the modulus-phase representation (\ref{3.44a}) can be considered as local gauge $U(1)$ transformation:
\begin{equation}\label{4.19}
    \Psi_{1}=\left|\Psi_{1}\right|e^{i\frac{2\widetilde{e}}{\hbar\upsilon}\chi_{1}},\quad
    \Psi_{2}=\left|\Psi_{2}\right|e^{i\frac{2\widetilde{e}}{\hbar\upsilon}\chi_{2}}.
\end{equation}
Then the gauge field should be transformed as
\begin{equation}\label{4.20}
    \widetilde{A}_{\mu}'=\widetilde{A}_{\mu}+\alpha\widetilde{\partial}_{\mu}\chi_{1}+\beta\widetilde{\partial}_{\mu}\chi_{2},
\end{equation}
where
\begin{equation}\label{4.21}
\alpha=\frac{\frac{|\Psi_{1}|^{2}}{m_{1}}}{\frac{|\Psi_{1}|^{2}}{m_{1}}+\frac{|\Psi_{2}|^{2}}{m_{2}}},\quad
\beta=\frac{\frac{|\Psi_{2}|^{2}}{m_{2}}}{\frac{|\Psi_{1}|^{2}}{m_{1}}+\frac{|\Psi_{2}|^{2}}{m_{2}}},
\end{equation}
so that
\begin{equation}\label{4.22}
    \alpha+\beta=1,\quad \frac{|\Psi_{1}|^{2}}{m_{1}}\beta=\frac{|\Psi_{2}|^{2}}{m_{2}}\alpha.
\end{equation}
The transformation (\ref{4.20}) excludes the phases $\theta_{1}$ and $\theta_ {2}$ from Lagrangian (\ref{4.18}) individually leaving only their difference:
\begin{eqnarray}\label{4.23}
   \mathcal{L}&=&\frac{\hbar^{2}}{4m_{1}}\left(\widetilde{\partial}_{\mu}+\frac{i2\widetilde{e}}{\upsilon\hbar}\widetilde{A}_{\mu}\right)|\Psi_{1}|
    \left(\widetilde{\partial}_{\mu}-\frac{i2\widetilde{e}}{\upsilon\hbar}\widetilde{A}_{\mu}\right)|\Psi_{1}|
    +\frac{\hbar^{2}}{4m_{2}}\left(\widetilde{\partial}_{\mu}+\frac{i2\widetilde{e}}{\upsilon\hbar}\widetilde{A}_{\mu}\right)|\Psi_{2}|
    \left(\widetilde{\partial}_{\mu}-\frac{i2\widetilde{e}}{\upsilon\hbar}\widetilde{A}_{\mu}\right)|\Psi_{2}|\nonumber\\
    &+&\frac{\hbar^{2}}{4}\frac{|\Psi_{1}|^{2}|\Psi_{2}|^{2}}{|\Psi_{1}|^{2}m_{2}+|\Psi_{2}|^{2}m_{1}}
    \widetilde{\partial}_{\mu}\left(\theta_{1}-\theta_{2}\right)\widetilde{\partial}^{\mu}\left(\theta_{1}-\theta_{2}\right)
    -2\epsilon|\Psi_{1}||\Psi_{2}|\cos\left(\theta_{1}-\theta_{2}\right)
    +\mathcal{L}\left(|\Psi_{1}|,|\Psi_{2}|,\widetilde{F}_{\mu\nu}\widetilde{F}^{\mu\nu}\right).
\end{eqnarray}
Thus, the gauge field $\widetilde{A}_{\mu}$ absorbs the Goldstone bosons $\theta_{1,2}$ so that the Lagrangian becomes dependent on
the difference $\theta_{1}-\theta_{2}$ only. The equation for $\theta_{1}-\theta_{2}$ has a form:
\begin{equation}\label{4.24}
    \widetilde{\partial}_{\mu}\widetilde{\partial}^{\mu}\left(\theta_{1}-\theta_{2}\right)
    -\frac{4}{\hbar^{2}}\frac{|\Psi_{1}|^{2}m_{2}+|\Psi_{2}|^{2}m_{1}}{|\Psi_{1}||\Psi_{2}|}\epsilon\sin\left(\theta_{1}-\theta_{2}\right)=0.
\end{equation}
This equation is similar to the sin-Gordon equation, but the coefficient $\epsilon$ is a function of difference of the equilibrium phases
$\left|\theta_{1}^{0}-\theta_{2}^{0}\right|=0,\pi$ according to Eq.(\ref{4.4}). Eq.(\ref{4.24}) can be linearized using $\epsilon\cos(\theta_{1}^{0}-\theta_{2}^{0})=-|\epsilon|$, $\sin(\theta_{1}^{0}-\theta_{2}^{0})=0$ and small oscillations (\ref{4.12}), that gives the following spectrum:
\begin{eqnarray}
    (\hbar\omega)^{2}=4|\epsilon|\frac{|\Psi_{1}|^{2}m_{2}+|\Psi_{2}|^{2}m_{1}}{|\Psi_{1}||\Psi_{2}|}\upsilon^{2}+(\hbar q)^{2}\upsilon^{2},\nonumber
\end{eqnarray}
which coincides with Leggett's mode spectrum (\ref{4.14}). Thus, we can see that in two-band superconductors the common mode oscillations with the spectrum (\ref{4.13}) are absorbed into the gauge field $\widetilde{A}_{\mu}$ like in single-band superconductors. At the same time, the oscillations of the relative phase between two SC condensates (Leggett's mode) "survive" due to that these oscillations are not accompanied by current $\mathbf{j}=0$ - Eqs.(\ref{4.15},\ref{4.17}). Hence Leggett's mode can be observable, that is confirmed in experiment \cite{blum}.

\section{Damping and relaxation}\label{relax}

In the previous sections we have considered eigen harmonic oscillations of the order parameter - Higgs mode and Goldstone modes absorbed into the gauge field except Leggett's mode (in multi-band systems). At the same time, movement of the normal component is accompanied with friction and generation of heat by Joule-Lenz law $Q=\mathbf{j}_{n}^{2}/\sigma$, where $\mathbf{j}_{n}=en_{\mathrm{n}}\mathbf{v}_{n}$ is normal current, $n_{\mathrm{n}}=n-n_{\mathrm{s}}=n-2|\Psi|^{2}$ is density of the normal component, $\mathbf{v}_{n}$ is its speed, $\sigma$ is conductivity. Thus, to support the normal movement some electric field $\mathbf{E}$ must act, the current and the field are connected with Ohm's law: $\mathbf{j}_{n}=\sigma\mathbf{E}$. Then, instead the harmonic oscillations, we will have situations illustrated in Fig.\ref{Fig1}b - relaxation, Fig.\ref{Fig1}c - damping oscillations, Fig.\ref{Fig1}d - forced (undamped) oscillations under the action of external field.

The energy dissipation is accounted with the Rayleigh dissipation function $R$ which determines speed of change of the energy of the system $W$ as $dW/dt=-2R$, that is $Q=2R$. As a rule, the dissipative force (friction) is proportional to generalized velocity $\dot{q}$: $F=-k\dot{q}$ ($k$ is a friction coefficient), then $R=\frac{1}{2}k\dot{q}^{2}$ and $F=-\frac{dR}{d\dot{q}}$. Corresponding equation of motion is
\begin{equation}\label{5.1}
\frac{d}{dt}\frac{\partial L}{\partial \dot{q}}-\frac{\partial L}{\partial q}+\frac{\partial R}{\partial \dot{q}}=0.
\end{equation}
Let us consider several important examples using Lagrangian (\ref{3.30}).

\subsection{Wave skin-effect in superconductors}

Let monochromatic electromagnetic wave $\widetilde{A}^{\mu}=(0,\mathbf{A})$ (in a gauge $\varphi=0$, $\textrm{div}\mathbf{A}=0$) with frequency $\hbar\omega<2|\Delta|$ fall on a superconductor perpendicularly to its surface. The wave induces eddy currents both superconducting $\mathbf{j}_{s}$ and normal $\mathbf{j}_{n}$. For normal electrons an equation of motion is $m\dot{\mathbf{v}}=e\mathbf{E}(t)-\frac{m}{\tau_{ph}}\mathbf{v}$, where $\mathbf{E}(t)=-\frac{1}{c}\frac{\partial\mathbf{A}}{\partial t}$ is the electric field, $\tau_{ph}$ is the mean free path time of electrons ($\tau_{ph}^{-1}$ is frequency of collisions of electrons with phonons or with the lattice defects). Then equation for the normal current is
 \begin{equation}\label{5.2a}
   \frac{m}{en_{\mathrm{n}}}\frac{d\mathbf{j}_{n}}{dt}+\frac{m}{en_{\mathrm{n}}}\frac{\mathbf{j}_{n}}{\tau_{E}}=e\mathbf{E}(t).
 \end{equation}
For the harmonic field $\mathbf{E}=\mathbf{E}_{0}e^{i(\mathbf{qr}-\omega t)}$ and current $\mathbf{j}_{n}=\mathbf{j}_{n0}e^{i(\mathbf{qr}-\omega t)}$ we have the Ohm's law in a form $\mathbf{j}_{n}=\sigma\frac{1+i\omega\tau_{ph}}{1+(\omega\tau_{ph})^{2}}\mathbf{E}$, where $\sigma=\frac{\tau_{ph}}{m}e^{2}n_{\textrm{n}}$ is conductivity of the normal component. We will consider regime of normal skin-effect only, that is when $\omega\tau_{ph}\ll 1$. The Joule-Lenz heat is $Q=\sigma_{\omega}\mathbf{E}^{2}$, where $\sigma_{\omega}=\frac{\sigma}{1+(\omega\tau_{ph})^{2}}$. Respectively, the Rayleigh dissipation function is
\begin{eqnarray}\label{5.2}
    R&=&\frac{1}{2}\sigma_{\omega}\left(1+i\omega\tau_{ph}\right)\mathbf{E}^{2}
    =\frac{1}{2}\widetilde{\sigma}_{\omega}\left(1+i\omega\tau_{ph}\right)\widetilde{\mathbf{E}}^{2}.
\end{eqnarray}
Here, $\widetilde{\mathbf{E}}\equiv-\frac{1}{\upsilon}\frac{\partial\mathbf{A}}{\partial t}\equiv-\widetilde{\partial}^{0}\widetilde{A}^{\mu}=\widetilde{\partial}_{0}\widetilde{A}_{\mu}$, so that
$\sigma\mathbf{E}^{2}=\widetilde{\sigma}\widetilde{\mathbf{E}}^{2}$ where $\widetilde{\sigma}=\frac{\upsilon^{2}}{c^{2}}\sigma$.
Unlike the Joule-Lenz law the Rayleigh dissipation function has both an active part and a reactive part determined by the term $i\omega\tau_{ph}$ for  agreement with the results of London theory \cite{schmidt}. The active part determines dissipation of the electromagnetic energy, the reactive part determines the phase shift of the current $\mathbf{j}_{n}$ relatively to the field $\mathbf{E}$ due to inertia of the system. Then an analog of Eq.(\ref{5.1}) using Eq.(\ref{3.46}) can be written:
\begin{equation}\label{5.3}
    \widetilde{\partial}_{\mu}\frac{\partial\mathcal{L}}{\partial\left(\widetilde{\partial}_{\mu}\widetilde{A}_{\nu}\right)}-\frac{\partial \mathcal{L}}{\partial\widetilde{A}_{\nu}}+\frac{1}{\upsilon}\frac{\partial R}{\partial\left(\widetilde{\partial}_{0}\widetilde{A}_{\nu}\right)}=0.
\end{equation}
Obviously, this equation is not Lorentz covariant due to the dissipative term. Dissipation distinguishes a time direction, i.e., violates the time symmetry $t\leftrightarrow -t$ which is symmetry for the Lorentz boost. Using the gauge $\varphi=0$, $\textrm{div}\mathbf{A}=0$ Eq.(\ref{5.3}) is reduced to
\begin{equation}\label{5.4}
    \frac{1}{\upsilon^{2}}\frac{\partial^{2}\mathbf{A}}{\partial t^{2}}-\Delta\mathbf{A}+\frac{1}{\lambda^{2}}\mathbf{A}
    +\frac{4\pi\widetilde{\sigma}_{\omega}\left(1+i\omega\tau_{ph}\right)}{\upsilon^{2}}\frac{\partial\mathbf{A}}{\partial t}=0.
\end{equation}
Taking the field as harmonic mode $\mathbf{A}=\mathbf{A}_{0}e^{i(\mathbf{qr}-\omega t)}$ we obtain a dispersion relation for photons in a superconductor:
\begin{equation}\label{5.5}
    \omega^{2}=\upsilon^{2}q^{2}+\frac{\upsilon^{2}}{\lambda^{2}_{\omega}}-i4\pi\widetilde{\sigma}_{\omega}\omega
    \Rightarrow  q^{2}=\frac{\omega^{2}}{\upsilon^{2}}-\frac{1}{\lambda^{2}_{\omega}}+i\frac{4\pi}{c^{2}}\sigma_{\omega}\omega.
\end{equation}
Here, we have denoted:
\begin{equation}\label{5.4a1}
\frac{1}{\lambda_{\omega}^{2}}\equiv\frac{1}{\lambda^{2}}+\frac{4\pi}{c^{2}}\sigma_{\omega}\omega^{2}\tau_{ph}
=\frac{1}{\lambda^{2}}\left(1+\frac{n_{\mathrm{n}}}{n_{\mathrm{s}}}\frac{(\omega\tau_{ph})^{2}}{1+(\omega\tau_{ph})^{2}}\right).
\end{equation}
Obviously, $\lambda_{\omega}\leq\lambda$, at $T=0$ in pure metal $n_{\mathrm{n}}=0$ takes place \cite{sad1,levit} hence we have $\lambda_{\omega}=\lambda$ in this case. The obtained expression (\ref{5.5}) differs from the result of London theory $q^{2}=-\frac{1}{\lambda^{2}_{\omega}}+i\frac{4\pi}{c^{2}}\sigma_{\omega}\omega$ by a term $\frac{\omega^{2}}{\upsilon^{2}}$. However, at small frequencies $\hbar\omega\ll|\Delta|$, $\frac{\omega^{2}}{\upsilon^{2}}\ll\frac{1}{\lambda^{2}}$ this term can be omitted, main contribution is given by the Meissner effect $-\frac{1}{\lambda^{2}}$ and by the skin-effect of the normal component $i\frac{4\pi}{c^{2}}\sigma_{\omega}\omega$. At frequencies $\omega\sim\omega_{c}=\frac{\upsilon}{\lambda}$ the term $\frac{\omega^{2}}{\upsilon^{2}}$ becomes important, that is discussed in Sect.\ref{electro}. The term $\frac{\omega^{2}}{\upsilon^{2}}$ has the following nature. The Maxwell equations for the electromagnetic field in normal metal are
\begin{equation}\label{5.5a}
    \left\{\begin{array}{c}
      \textrm{curl}\mathbf{H} = \frac{4\pi}{c}\sigma\mathbf{E}+\frac{1}{c}\frac{\partial \mathbf{D}}{\partial t} \\
      \\
      \textrm{curl}\mathbf{E} = -\frac{1}{c}\frac{\partial \mathbf{H}}{\partial t} \\
    \end{array}\right\}
\Rightarrow\Delta\mathbf{H}=\frac{4\pi\sigma}{c}\frac{\partial\mathbf{H}}{\partial t}
+\frac{\varepsilon_{n}(\omega)}{c^{2}}\frac{\partial^{2}\mathbf{H}}{\partial t^{2}},
\end{equation}
where $\mathbf{D}=\varepsilon_{n}\mathbf{E}$ has been used. Taking the field as harmonic mode $\mathbf{H}=\mathbf{H}_{0}e^{i(\mathbf{qr}-\omega t)}$ we obtain the dispersion relation for photons in metal:
\begin{equation}\label{5.5b}
q^{2}=i\frac{4\pi}{c^{2}}\sigma\omega+\frac{\varepsilon_{n}(\omega)}{c^{2}}\omega^{2}.
\end{equation}
In good metals $\frac{4\pi\sigma}{\omega}\gg\varepsilon_{n}(\omega)$ (if $\omega\neq 0$), then the second term in Eq.(\ref{5.5b}) can be
neglected \cite{landau,tilley}. Hence we obtain usual expression for the skin-effect: $q^{2}=i\frac{4\pi}{c^{2}}\sigma\omega$. In superconductors the conductivity is determined with the normal component $\sigma=\frac{\tau_{ph}}{m}e^{2}n_{\mathrm{n}}$ which is $\sigma\rightarrow 0$ in pure system at small temperatures $T\ll T_{c}$, so that $\frac{\sigma}{\omega}\lesssim\varepsilon$ can be. Thus, for SC materials the induction term $\frac{1}{c}\frac{\partial\mathbf{D}}{\partial t}$ is as important as the ohmic term $\sigma\mathbf{E}$, unlike normal metals where the induction term can be neglected for quasistationary fields.

It should be noted that the accounting of the complex conductivity from the first London equation $\mathbf{j}_{s}=-i\frac{c^{2}}{4\pi\lambda^{2}\omega}\mathbf{E}$ (i.e., $\sigma_{s}=-i\frac{c^{2}}{4\pi\lambda^{2}\omega}$) is unnecessary because the induced electric field $-\frac{1}{c}\frac{\partial\mathbf{A}}{\partial t}$ drives the eddy currents which are Goldstone oscillations, which, in turn, are absorbed into the gauge field $\mathbf{A}$. Thus, this complex conductivity has already hidden in the nondissipative part of the field equation (\ref{5.4}). As we have seen before, the London equations  (\ref{3.56a},\ref{3.56b}) are not equivalent to Eqs.(\ref{3.47},\ref{3.55a},\ref{3.55b}) since the extraction of current $\mathbf{j}$ violates Anderson-Higgs mechanism. As a result of the extraction we obtain the first London  equation (\ref{3.56a}) which is the second Newton law for SC electrons, i.e., it is equation for an ideal conductor. Since we lose a term $\frac{1}{\upsilon^{2}}\frac{\partial^{2}\widetilde{\mathbf{A}}}{\partial t^{2}}$ then the frequency dependence is caused by the complex conductivity $\sigma_{s}$ only, that provides the low-frequency regime such that the penetration depth does not depend on frequency: $L(\omega)=\lambda$ \cite{tinh} (if we exclude the wave skin-effect due to normal electrons where $L(\omega)=\sqrt{\frac{c^{2}}{2\pi\sigma\omega}}$).

\begin{figure}[h]
\includegraphics[width=8.5cm]{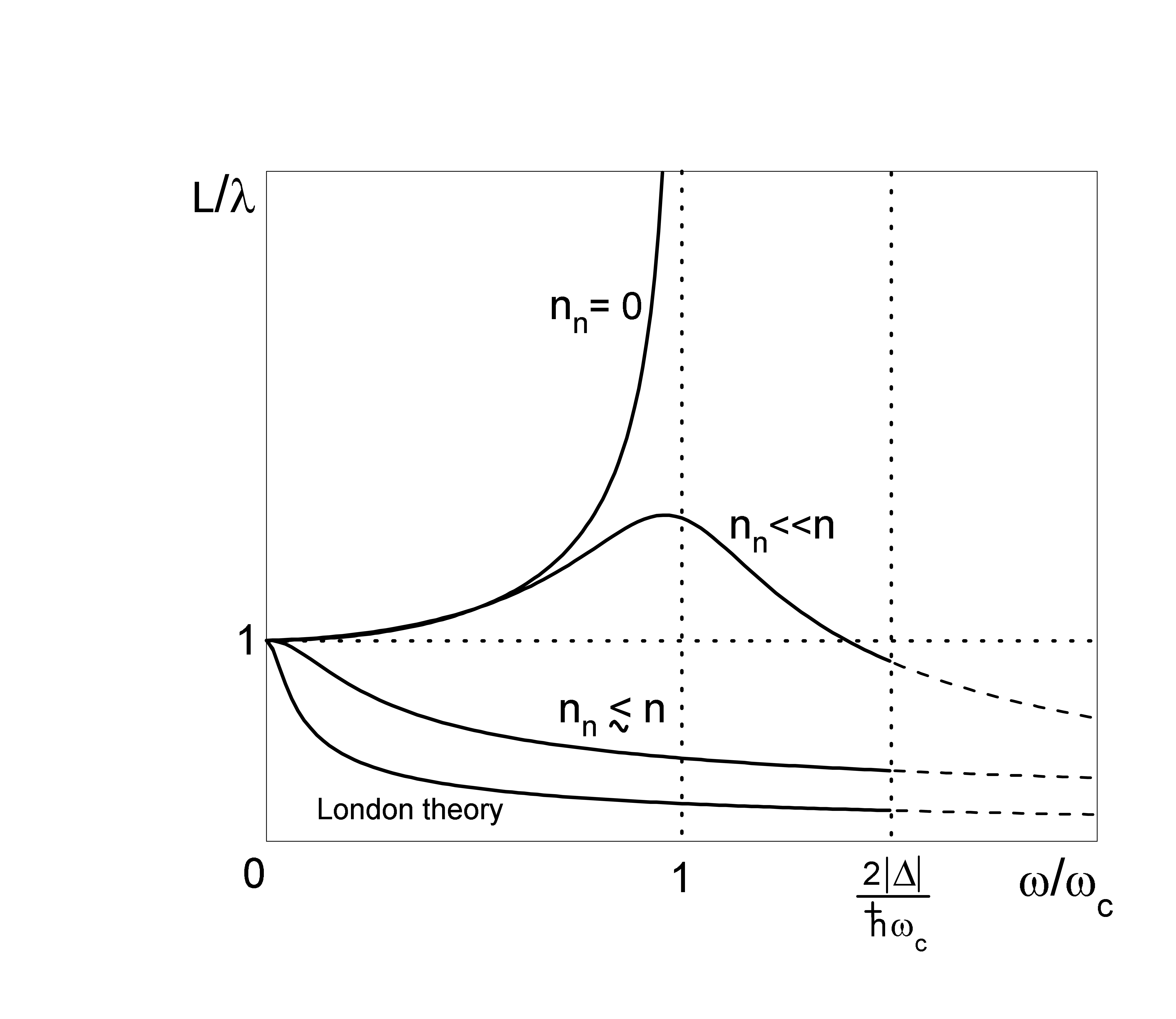}
\caption{Schematic image of dependence of the penetration depth $L$ on the frequency $\omega$ for different densities of the normal component $n_{\mathrm{n}}$ and for the London theory. At $\omega\rightarrow 0$ the penetration depth is equal to the London depth $\lambda$ for all regimes. If $n_{\mathrm{n}}=0$ then for $\omega\geq\omega_{c}$ SC material becomes transparent i.e., $L\rightarrow\infty$, however at frequencies $\hbar\omega\geq 2|\Delta|$ strong absorption takes place (the dashed lines). For large normal density (low SC density) $n_{\mathrm{n}}\lesssim n$ the result for $L$ is close to the London theory ($L$ weakly depends on frequency). It should be noted that for different normal densities $n_{\mathrm{n}}$ we have different parameters $\omega_{c}$, $|\Delta|$ and $\lambda$.} \label{Fig7}
\end{figure}

From Eq.(\ref{5.5}) we obtain:
\begin{equation}\label{5.6}
q=\sqrt[4]{\frac{1}{\lambda^{4}_{\omega}}\left(\frac{\omega^{2}}{\upsilon^{2}}\lambda^{2}_{\omega}-1\right)^{2}+\left(\frac{4\pi}{c^{2}}\sigma_{\omega}\omega\right)^{2}}
\left(\cos\frac{\varphi}{2}+i\sin\frac{\varphi}{2}\right),
\end{equation}
where
\begin{equation}\label{5.6a}
\varphi=\textrm{arccot}\frac{\textrm{Re}\left(q^{2}\right)}{\textrm{Im}\left(q^{2}\right)}=\textrm{arccot}\frac{-\frac{1}{\lambda^{2}_{\omega}}\left(1-\frac{\omega^{2}}{\upsilon^{2}}
\lambda^{2}_{\omega}\right)}{\frac{4\pi}{c^{2}}\sigma_{\omega}\omega},
\end{equation}
that corresponds to attenuation of the wave $\mathbf{A}=\mathbf{A}_{0}e^{i(qx-\omega t)}$ in the depth of the superconductor which occupies half-space $x>0$. Then substituting $q$ (\ref{5.6}) in this field $\mathbf{A}$ we obtain the penetration depth:
\begin{equation}\label{5.6b}
L=\frac{\lambda_{\omega}}{\sqrt[4]{\left(\frac{\omega^{2}}{\upsilon^{2}}\lambda^{2}_{\omega}-1\right)^{2}
+\left(\frac{4\pi}{c^{2}}\lambda^{2}_{\omega}\sigma_{\omega}\omega\right)^{2}}}\frac{1}{\sin\frac{\varphi}{2}},
\end{equation}
in a sense $\mathbf{A}=\mathbf{A}_{0}e^{-x/L}e^{i\left(\mathrm{Re}(q)x-\omega t\right)}$. Let us consider the following limit cases:
\begin{enumerate}
\item $\omega=0$. Then $\varphi=\pi$ and from Eq.(\ref{5.6b}) we have $L=\lambda$. i.e., low-frequency field is screened like static field.

\item The frequency is equal to the critical frequency (\ref{3.48b}): $\omega=\omega_{c}\equiv\frac{\upsilon}{\lambda_{\omega}}$ (it should be noted that at $n_{\mathrm{n}}\rightarrow 0$ we have $\lambda_{\omega}=\lambda$). In this case we have $\varphi=\frac{\pi}{2}$, hence
    \begin{equation}\label{5.6c}
    L=\frac{1}{\sqrt{\frac{2\pi e^{2}}{mc^{2}}n_{\mathrm{n}}\frac{\omega_{c}\tau_{ph}}{1+(\omega_{c}\tau_{ph})^{2}}}}.
    \end{equation}
    We can see that at $n_{\mathrm{n}}\rightarrow 0$ we have result of Sect.\ref{electro}, where the penetration depth becomes infinitely large $L\gg\lambda$. Formally we can suppose $\tau_{ph}=0$, then $L\rightarrow\infty$ too, however the condition $\tau_{ph}=0$ means full blocking of movement of normal electrons i.e., the substance ceases to be conductor.

\item $\omega>\omega_{c}$. Then from Eq.(\ref{5.6a}) we can see $0\leq\varphi<\frac{\pi}{2}$. If $n_{\mathrm{n}}\rightarrow 0$ then $\varphi\rightarrow0$, hence from Eq.(\ref{5.6b}) we have $L\rightarrow\infty$, i.e., the superconductor becomes transparent for such electromagnetic waves.

\item $\omega<\omega_{c}$. Then from Eq.(\ref{5.6a}) we can see $\frac{\pi}{2}<\varphi\leq\pi$. If even $n_{\mathrm{n}}\rightarrow 0$ then $\varphi\rightarrow\pi$, hence from Eq.(\ref{5.6b}) we have $L<\infty$, i.e., the superconductor reflects such electromagnetic waves.

\item $n_{\mathrm{n}}=n$ hence $n_{\mathrm{s}}=0$, $\lambda=\infty$, $\frac{c^{2}}{\upsilon^{2}}\rightarrow\varepsilon_{n}(\omega)\ll\frac{4\pi\sigma}{\omega}$. Then from Eq.(\ref{5.6a}) we can see $\varphi\rightarrow\frac{\pi}{2}$, and then, we obtain $L(\omega)=\sqrt{\frac{c^{2}}{2\pi\sigma\omega}}$, i.e., the wave skin-effect in normal metal takes place.
\end{enumerate}
At the same time, according to the London theory:
\begin{equation}\label{5.6d}
q^{2}=-\frac{1}{\lambda^{2}_{\omega}}+i\frac{4\pi}{c^{2}}\sigma_{\omega}\omega\Rightarrow
L_{\mathrm{London}}=\frac{\lambda_{\omega}}{\sqrt[4]{1+\left(\frac{4\pi}{c^{2}}\lambda^{2}_{\omega}\sigma_{\omega}\omega\right)^{2}}}\frac{1}{\sin\frac{\varphi}{2}},
\quad
\varphi=\textrm{arccot}\frac{-\frac{1}{\lambda^{2}_{\omega}}}{\frac{4\pi}{c^{2}}\sigma_{\omega}\omega}.
\end{equation}
We can see that at $\omega\rightarrow 0$ we have $L_{\mathrm{London}}\rightarrow\lambda$ like in Eq.(\ref{5.6b}). At the same time, even if $n_{\mathrm{n}}=0$ (hence $\sigma=0$), then we have $L_{\mathrm{London}}(\omega)=\lambda$ also, at that for any frequency.

The total result is shown in Fig.\ref{Fig7} schematically. We can see that at $\omega\rightarrow 0$ both $L_{\mathrm{London}}\rightarrow\lambda$ and $L\rightarrow\lambda$. At large frequencies $\omega\sim\omega_{c}$ the results of the extended TDGL theory and the London theory can be essentially different. So, at $n_{\mathrm{n}}\rightarrow n$ (i.e., $n_{\mathrm{s}}\rightarrow 0$) the penetration depth $L$ is close to $L_{\mathrm{London}}$. However at $n_{\mathrm{n}}\rightarrow 0$ (i.e., $n_{\mathrm{s}}\rightarrow n$) we obtain $L\gg\lambda$ inside an interval $\omega_{c}\leq\omega<2|\Delta|/\hbar$ (only for type-II superconductors - Eq.(\ref{3.48c})), that corresponds to result of Sect.\ref{electro} where superconductor becomes transparent for electromagnetic waves with frequencies from this interval. Photons with frequency $\hbar\omega\geq2|\Delta|$ break Cooper pairs, hence intensive absorption of the electromagnetic waves takes place. \emph{Thus, at $T=0$ pure (in order to ensure the condition $n_{\mathrm{n}}\rightarrow 0$) type-II (in order to ensure $\hbar\omega_{c}<2|\Delta|$) superconductor should be transparent for electromagnetic waves with frequencies $\omega_{c}\leq\omega<2|\Delta|/\hbar$, that can be subject for experimental verification}. However we can see from Eq.(\ref{5.6c}) the penetration depth $L$ is very sensitive to the normal density $n_{\mathrm{n}}$, in addition, for clear observation of this effect the substance should not absorb in this frequency range in itself, therefore observation of this effect can be difficult. It should be noted that this problem has been considered only for s-wave superconductors.

Results of calculation of the frequency interval $\nu_{c}(0)<\nu<\frac{2|\Delta(0)|}{h}$ (here $\nu=2\pi\omega$, $h=2\pi\hbar$) for pure elemental type-II superconductors niobium \texttt{Nb}, technetium \texttt{Tc} and vanadium \texttt{V} at $T=0$ are presented in a Table \ref{tab2}. Parameters $T_{c}$ and $\kappa$ has been taken from \cite{niobium,technetium,vanadium}. We can see that the transparency interval is in a terahertz range.
\begin{table}[h]
\centering
\begin{tabular}{|c|c|c|c|c|c|}
  \hline\rule{0cm}{0.5cm}
   & $T_{c}$, K & $2|\Delta(0)|=3.52T_{c}$, K & $\kappa$ & $\nu_{c}=\frac{k_{B}}{h}\frac{2|\Delta(0)|}{\sqrt{2}\kappa}$, THz & $\frac{k_{B}}{h}2|\Delta(0)|$, THz\\
  \hline\rule{0cm}{0.5cm}
  \texttt{Nb} & 9.25 & 32.56 & $1.43$ & 0.34 & 0.68 \\
  \hline\rule{0cm}{0.5cm}
  \texttt{Tc} & 7.77 & 27.35 & 0.92  & 0.44 & 0.57 \\
  \hline\rule{0cm}{0.5cm}
  \texttt{V}  & 5.43 & 19.11 & 0.85 & 0.33 & 0.40 \\
  \hline
\end{tabular}
\caption{Transparency intervals $\nu_{c}(0)\leq\nu<\frac{k_{B}}{h}2|\Delta(0)|$ for pure elemental type-II superconductors niobium, technetium and vanadium in THz. The energy gap $|\Delta(0)|$ is measured in Kelvins, $\kappa=\frac{\lambda}{\xi}=\frac{H_{c2}}{\sqrt{2}H_{cm}}$ is a GL parameter.}\label{tab2}
\centering
\end{table}

\subsection{Eigen electromagnetic oscillations}

In the Sect.\ref{electro2} we have demonstrated that Goldstone boson is absorbed into gauge field, and instead the phase oscillations the electromagnetic oscillation mode (\ref{3.47}) with spectrum (\ref{3.48}) appears. These electromagnetic oscillations are specific for SC system, and they are eddy Meissner current. The transparency of superconductor for electromagnetic wave with frequency $\omega\geq\omega_{c}$ is caused by the eigen electromagnetic oscillation of SC medium with minimal frequency $\omega_{c}$. However, at $T>0$ and for dirty superconductors for $T=0$ even \cite{levit} the normal electrons are present. Induced electric field $\mathbf{E}=-\frac{1}{c}\frac{\partial\mathbf{A}}{\partial t}$ in such oscillations causes movement of the normal electrons according to Ohm's law $\mathbf{j}_{n}=\sigma\mathbf{E}$ and dissipation in a form of Joule-Lenz heat $Q=\mathbf{j}_{n}^{2}/\sigma=\sigma\mathbf{E}^{2}$ occurs. Let us consider a homogeneous mode, i.e., $q=0$ is supposed in Eq.(\ref{5.5}), and $\lambda_{\omega}=\lambda$, $\sigma_{\omega}=\sigma$ are supposed for simplicity. Then we obtain an equation for frequency:
\begin{equation}\label{5.7}
    \omega^{2}+i\frac{4\pi\upsilon^{2}}{c^{2}}\sigma\omega-\frac{\upsilon^{2}}{\lambda^{2}}=0,
\end{equation}
whose solution is
\begin{equation}\label{5.8}
    2\omega=-i\frac{4\pi\upsilon^{2}}{c^{2}}\sigma\pm\sqrt{\frac{4\upsilon^{2}}{\lambda^{2}}-\left(\frac{4\pi\upsilon^{2}}{c^{2}}\sigma\right)^{2}}.
\end{equation}
This solution determines oscillations of electromagnetic field $\mathbf{A}=\mathbf{A}_{0}e^{-i\omega t}=\mathbf{A}_{0}e^{-t/\tau}e^{-i\mathrm{Re}(\omega) t}$, where $1/\tau=-\mathrm{Im}(\omega)$ is the decay time. We can see that the oscillations are sensitive to the normal density $n_{\mathrm{n}}$ via conductivity $\sigma=\frac{\tau_{ph}}{m}e^{2}n_{\textrm{n}}$. Let us consider the following limit cases:

\begin{enumerate}
\item $T\rightarrow T_{c}$. Then $\lambda(T)\rightarrow\infty$ and $n_{\textrm{n}}\approx n$, hence  $\omega=-i\frac{4\pi\upsilon^{2}}{c^{2}}\sigma$, that is $\tau=\frac{c^{2}}{4\pi\upsilon^{2}\sigma}$ and $\mathrm{Re}(\omega)=0$. Thus, monotonic attenuation of the electromagnetic excitation occurs.

\item $n_{\textrm{n}}=0$, that takes place at $T=0$ in pure material \cite{levit}. Then $\omega=\frac{\upsilon}{\lambda}$ which coincides with the critical frequency $\omega_{c}$ (\ref{3.48b}), and $\mathrm{Im}(\omega)=0$. Thus, harmonic oscillations of the electromagnetic field in superconductor occur.
\end{enumerate}

Obviously, if $\hbar\omega_{c}\geq2|\Delta|$, then a quant of these oscillations breaks Cooper pair, hence intensive absorption of these oscillations takes place. As we can see from Eq.(\ref{5.8}) the frequency $\omega$ is sensitive to the normal density $n_{\mathrm{n}}$ in the sense that the normal electrons caused attenuation of the oscillations. The crossover between overdamped (at high $T$) and underdamped (low $T$) regimes can be found from Eq.(\ref{5.8}) in a form:
\begin{equation}\label{5.9}
    \frac{c}{\lambda(T)}=\frac{2\pi\upsilon}{c}\sigma(T).
\end{equation}
If take $\tau_{ph}\sim 10^{-10}\mathrm{c}$, $n\sim 10^{22}\mathrm{cm}^{-3}$, $\upsilon\sim 10^{8}\mathrm{cm/c}$, $\lambda=\lambda(0)\left(1-\left(\frac{T}{T_{c}}\right)^{4}\right)^{-1/2}$ and $n_{\mathrm{n}}=n\left(\frac{T}{T_{c}}\right)^{4}$, then we can evaluate the temperature of crossover: $\frac{T}{T_{c}}\sim 0.5$. However these oscillations are bulk oscillations, and, as we could see in previous subsection, propagation of fluctuations of electromagnetic field (with $q\neq 0$) is blocked by the skin-effect if $n_{\mathrm{n}}\neq 0$. That is any field, by which such eigen oscillations are tried to excite, will be reflected from surface of a superconductor. Thus, experimental observation of such eigen electromagnetic oscillations is problematically. Only if normal electrons are absent $n_{\mathrm{n}}=0$, such fluctuations can show themselves by the transparency of superconductor for electromagnetic wave with frequency $\omega_{c}<\omega<2|\Delta|/\hbar$.

Similar frequency (\ref{5.8}) and crossover (\ref{5.9}) have been obtained in \cite{hirs1}. However in our model these oscillations are Goldstone mode absorbed into the gauge field, and at that $\frac{\partial\rho}{\partial t}=0$ (hence $\mathrm{div}\mathbf{j}_{s}=0$), $\mathrm{div}\mathbf{A}=0$. At the same time, in model \cite{hirs1} the analogous oscillations are oscillations of charge density $\rho$ and longitudinal electric field $\mathrm{div}\mathbf{E}=4\pi\rho$ respectively. In our model at $T>T_{c}$ the eigen electromagnetic oscillations do not turn into relaxation of charge density with relaxation time $\frac{1}{4\pi\sigma}$. According to the rule $\frac{c^{2}}{\upsilon^{2}}\rightarrow\varepsilon_{n}(\omega)$  (either supposing $q=0$ in Eq.(\ref{5.5b})) we have $\tau(T>T_{c})=\frac{\varepsilon_{n}(\omega)}{4\pi\sigma}$, which is relaxation time for \emph{eddy} current in metal.

\subsection{Relaxation of a fluctuation}

\begin{figure}[h]
\includegraphics[width=8.5cm]{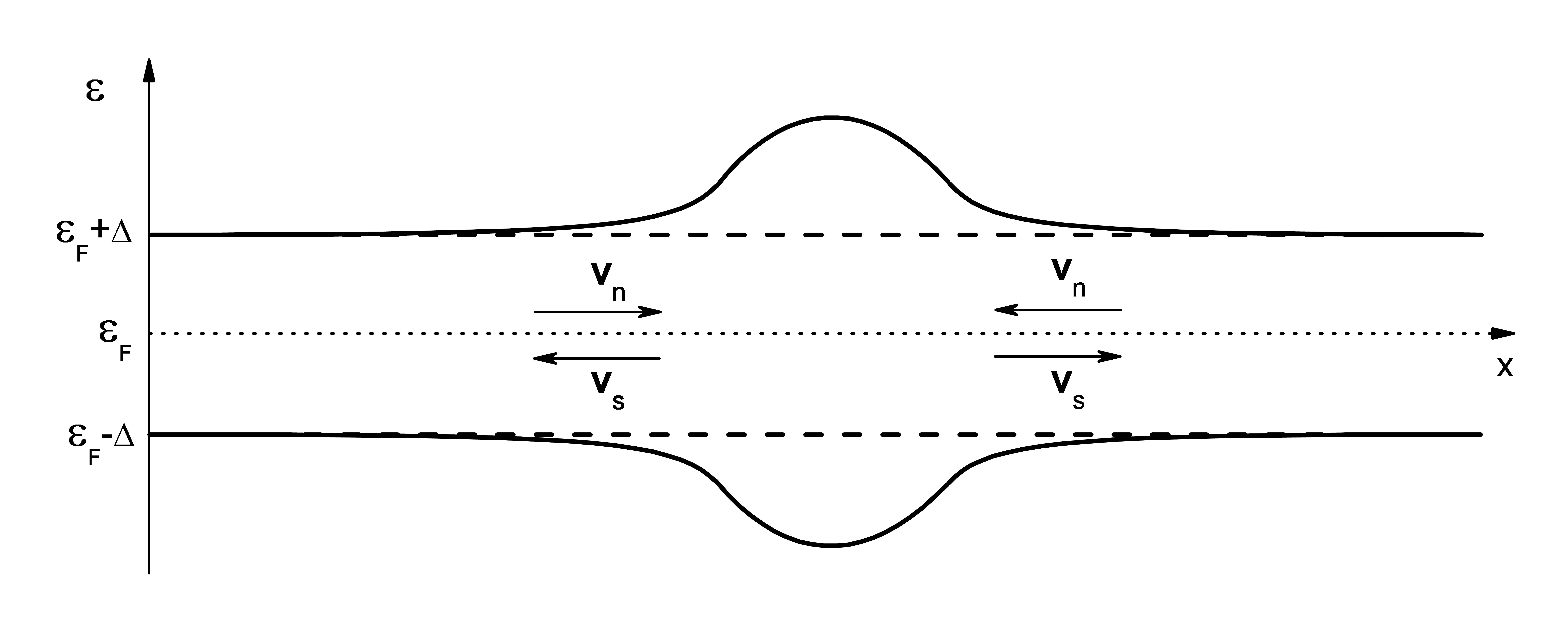}
\caption{relaxation of a fluctuation of modulus of the order parameter $|\Psi|\propto|\Delta|$. The fluctuation is accompanied by changes of density of SC electrons $n_{\mathrm{s}}=2\left|\Psi\right|^{2}$ and normal electrons $n_{\mathrm{n}}$ so that $n_{\mathrm{s}}+n_{\mathrm{n}}=n=\mathrm{const}$. The relaxation is carried out by counterflows of SC and normal components so that $n_{\mathrm{s}}\mathbf{v}_{s}+n_{\mathrm{n}}\mathbf{v}_{n}=0$. The dissipation is caused by the friction $\mathbf{f}=-\frac{m}{\tau_{ph}}\mathbf{v}_{n}$ of the normal component.}\label{Fig8}
\end{figure}

Let a fluctuation is formed in some region so that $|\Delta(x)|>|\Delta|$, where $|\Delta|$ is the equilibrium value - Fig.\ref{Fig8}, but $n=n_{\mathrm{n}}+n_{\mathrm{s}}=\mathrm{const}$. Let us consider relaxation of this bubble which can be both damped oscillation and monotonous relaxation to the equilibrium - Fig.\ref{Fig1}b,c. After the fluctuation is formed the system tends to the equilibrium: the flow of the normal component is directed to the bubble, at the same time the flow of the superfluid component is directed from the bubble so that the total flow is $\mathbf{j}=n_{\mathrm{s}}\mathbf{v}_{s}+n_{\mathrm{n}}\mathbf{v}_{n}=0$. Then $\textrm{div}\mathbf{j}_{s}=\textrm{div}(n_{\mathrm{s}}\mathbf{v}_{s})
=-\textrm{div}(n_{\mathrm{n}}\mathbf{v}_{n})\approx-\frac{1}{\xi}n_{\mathrm{n}}\mathbf{v}_{n}$, because the changes of superfluid and normal components occur in the coherence length $\xi(T)\propto(T_{c}-T)^{-1/2}$. The density of the superfluid component is $n_{\mathrm{s}}=2|\Psi|^{2}$. The normal motion is accompanied with friction $\mathbf{f}=-\frac{m}{\tau_{ph}}\mathbf{v}_{n}$, corresponding Rayleigh dissipation function is $R=\frac{m}{2\tau_{ph}}n_{\mathrm{n}}\mathrm{v}_{n}^{2}=\frac{1}{2}\frac{\sigma}{e^{2}}\mathrm{v}_{n}^{2}$. Using the continuity equation $\frac{\partial n_{\mathrm{s}}}{\partial t}=-\textrm{div}\mathbf{j}_{s}$ and $n_{\mathrm{n}}\approx n$ the dissipation function takes a form:
\begin{equation}\label{5.14}
    R=\frac{\sigma}{e^{2}}\frac{2\xi^{2}}{n^{2}}\left(\Psi\frac{\partial\Psi^{+}}{\partial t}+\Psi^{+}\frac{\partial\Psi}{\partial t}\right)^{2}
    =\frac{\sigma}{e^{2}}\frac{2\xi^{2}\upsilon^{2}}{n^{2}}\left(\Psi\widetilde{\partial}_{0}\Psi^{+}+\Psi^{+}\widetilde{\partial}_{0}\Psi\right)^2.
\end{equation}
Then equation for the field $\Psi$ is
\begin{equation}\label{5.15}
    \widetilde{\partial}^{\mu}\frac{\partial\mathcal{L}}{\partial\left(\widetilde{\partial}^{\mu}\Psi^{+}\right)}-\frac{\partial
    \mathcal{L}}{\partial\Psi^{+}}+\frac{1}{\upsilon}\frac{\partial R}{\partial\left(\widetilde{\partial}_{0}\Psi^{+}\right)}=0
    \Rightarrow\frac{\hbar^{2}}{4m}\widetilde{\partial}_{\mu}\widetilde{\partial}^{\mu}\Psi+a\Psi+b\left|\Psi\right|^{2}\Psi
    +\frac{\sigma}{e^{2}}\frac{4\xi^{2}\upsilon}{n^{2}}\left(\Psi\widetilde{\partial}_{0}\Psi^{+}+\Psi^{+}\widetilde{\partial}_{0}\Psi\right)\Psi=0.
\end{equation}
Analogously to previously considered problem about the skin-effect this equation is not Lorentz covariant due to the dissipative term. Using the modulus-phase representation (\ref{2.2}) and linearizing with help $|\Psi|=\sqrt{-\frac{a}{b}}+\phi\equiv\Psi_{0}+\phi$ (where $|\phi|\ll\Psi_{0}$) we obtain an equation for small deviations of the superfluid density $\phi$ and for the phase $\theta$ accordingly:
\begin{eqnarray}
&&\frac{\hbar^{2}}{4m}\widetilde{\partial}_{\mu}\widetilde{\partial}^{\mu}\phi+2|a|\phi
+\frac{\sigma}{e^{2}}\frac{8\xi^{2}\upsilon}{n^{2}}\Psi_{0}^{2}\widetilde{\partial}_{0}\phi=0,\label{5.16a}\\
&&\widetilde{\partial}_{\mu}\widetilde{\partial}^{\mu}\theta=0.\label{5.16b}
\end{eqnarray}
The oscillations of phase remains without damping in this approximation, since, as has been demonstrated in Sect.\ref{normal}, the Goldstone mode is the eddy supercurrent. For harmonic oscillations $\phi(\mathbf{r},t)=\phi_{0}e^{i(\mathbf{qr}-\omega t)}$ we obtain their dispersion relation in a form:
\begin{equation}\label{5.17}
\omega^{2}-q^{2}\upsilon^{2}-\omega_{0}^{2}+i2\gamma\omega=0,
\end{equation}
where
$\omega_{0}^{2}=\frac{8|a|m\upsilon^{2}}{\hbar^{2}}=\frac{4|\Delta|^{2}}{\hbar^{2}}$ is a characteristic frequency of the system and
$\gamma=\frac{\sigma}{e^{2}}\frac{16\xi^{2}m\upsilon^{2}}{\hbar^{2}n^{2}}\Psi_{0}^{2}$ is an attenuation constant. We can see that
$\omega_{0}\propto(T_{c}-T)^{1/2}$, at the same time $\gamma\propto\xi^{2}\Psi_{0}^{2}=\mathrm{const}$ at $T\rightarrow T_{c}$, i.e., $\omega_{0}\ll\gamma$. Thus, the regime of overdamped oscillation takes place - the field $\Psi(\mathbf{r},t)$ is monotonically relaxing to the thermodynamically steady state $\Psi_{0}$. Then evolution of the fluctuation is $\phi(t)\propto e^{-t/\tau}$, where
\begin{equation}\label{5.18}
    \tau=\frac{2\gamma}{\omega_{0}^{2}}\propto(T_{c}-T)^{-1}
\end{equation}
is the relaxation time for a homogeneous ($q=0$) mode in the limit $\omega_{0}\ll\gamma$. Another solution with $\tau=1/2\gamma$ can be omitted because this mode decays much faster. The relaxation time is lifetime of the fluctuation: the closer temperature to $T_{c}$ the larger size of a fluctuation $\xi(T\rightarrow T_{c})\rightarrow\infty$ and it lives longer $\tau(T\rightarrow T_{c})\rightarrow\infty$. The new phase is a fluctuation of infinite size (fills the entire system) and infinite lifetime. The relaxation time (\ref{5.18}) corresponds to the reduced equation describing the relaxation only:
\begin{equation}\label{5.19}
    q^{2}\upsilon^{2}+\omega_{0}^{2}-i2\gamma\omega=0\Rightarrow-\frac{\hbar^{2}}{4m}\Delta\Psi+a\Psi+b\left|\Psi\right|^{2}\Psi    +2\gamma\frac{\hbar^{2}}{4m\upsilon^{2}}\frac{\partial\Psi}{\partial t}=0.
\end{equation}\enlargethispage{\baselineskip}
This equation can be made dimensionless using a dimensionless order parameter $\psi=\Psi/\Psi_{0}$:
\begin{eqnarray}\label{5.20}
    \tau\frac{\partial\psi}{\partial t}=\xi^{2}\Delta\psi+\psi-\left|\psi\right|^{2}\psi,
\end{eqnarray}
where $\xi(T)=\sqrt{\frac{\hbar^{2}}{4m|a(T)|}}$ is temperature-dependent coherence length and $\tau(T)=\frac{\hbar^{2}\gamma}{4m|a(T)|\upsilon^{2}}=\frac{\xi^{2}(T)\gamma}{\upsilon^{2}}$ is the temperature-dependent relaxation time. Thus, due to the strong damping at $T\rightarrow T_{c}$ the equation (\ref{5.15}) is reduced to Eq.(\ref{5.20}) which is analogous to TDGL equation (\ref{1.3}). Thus, \emph{the TDGL equation is a limit case of the extended TDGL theory proposed here}. We can see that in consequence of the strong damping the monotonous relaxation of fluctuation with the relaxation time (\ref{5.18}) takes place. Moreover, this means strong damping of the free Higgs mode, so that observation of these oscillations is problematical at $T\rightarrow T_{c}$. On the other hand at $T\rightarrow 0$ we can have such situation that $\omega_{0}>\gamma$ then the Higgs mode can be more clearly observable.

\section{Results}\label{concl}

In this work we have formulated the extended TDGL theory which is generalization of the GL theory for the nonstationary regimes: damped eigen oscillations (including relaxation) and forced oscillations of the order parameter $\Psi(\mathbf{r},t)$ under the action of external field. In this theory, instead the GL functional (\ref{1.1}), we propose action (\ref{3.31}) with Lorentz invariant Lagrangian (\ref{3.30}) for the complex scalar field $\Psi=|\Psi|e^{i\theta}$ and the gauge field $A^{\mu}=(\varphi, \mathbf{A})$ in some 4D Minkowski space $\{\upsilon t,\mathbf{r}\}$, where speed $\upsilon$ is determined with dynamical properties of the system. At the same time, the dynamics of conduction electrons remains non-relativistic. For accounting of movement of the normal component, which is accompanied by friction, we have used approach with the Rayleigh dissipation function which determines speed of the dissipation. This makes the theory is not Lorentz covariant since dissipation distinguishes a time direction, i.e., violates the time symmetry $t\leftrightarrow -t$ which is symmetry of the Lorentz boost. Our results are follows:

1) The SC system has two types of collective excitations: with an energy gap (quasi-relativistic spectrum) $E^{2}=\widetilde{m}^{2}\upsilon^{4}+p^{2}\upsilon^{2}$ (where $\widetilde{m}$ is the mass of a Higgs boson, so that $\widetilde{m}\upsilon^{2}=2|\Delta|$) - free Higgs mode, and with acoustic (ultrarelativistic) spectrum $E=p\upsilon$ - Goldstone mode, which are oscillations of the order parameter $\Psi(t,\mathbf{r})$. The light speed $\upsilon$ is determined with dynamical properties of the system (\ref{2.17}), and it is much less than the vacuum light speed: $\upsilon=v_{F}/\sqrt{3}\ll c$. The Higgs mode is oscillation of modulus of the order parameter $|\Psi(t,\mathbf{r})|$ and it can be considered as sound in the gas of above-condensate quasiparticles. It should be noted that the free Higgs mode in a pure superconductor is unstable due to both strong damping of these oscillations at $T\rightarrow T_{c}$, so that aperiodic relaxation takes place and decay into above-condensate quasiparticles since $E(q)\geq 2|\Delta|$. At the same time, various manifestations of Higgs mechanism plays important role in the dynamics of SC system. The Goldstone mode is oscillations of the phase $\theta$ which are eddy current and they are absorbed into the gauge field $A^{\mu}$ according to Anderson-Higgs mechanism.

2) In two-band superconductors the Goldstone mode splits into two branches: common mode oscillations $\nabla\theta_{1}=\nabla\theta_{2}$ with the acoustic spectrum, and the oscillations of the relative phase $\theta_{1}-\theta_{2}$ between two SC condensates (for symmetrical condensates we have $\nabla\theta_{1}=-\nabla\theta_{2}$) with energy gap in spectrum determined by interband coupling  - Leggett's mode. The common mode oscillations are absorbed into the gauge field $A_{\mu}$ like in single-band superconductors, at the same time Leggett's mode "survives" due to these oscillations are not accompanied by current. Hence Leggett's mode can be observable that is confirmed in experiment \cite{blum}.

3) From the gauge invariance of Lagrangian (\ref{3.30}) it follows that superconductor is equivalent to dielectric (in some effective sense, not in conductivity) with permittivity $\varepsilon=\frac{c^{2}}{\upsilon^{2}}\sim 10^{5}$. At the same time, inside superconductor the potential electric field is absent $\mathbf{E}=-\nabla\varphi=0$ as consequence of boundary condition (\ref{3.52b}). Hence the dielectric permittivity $\varepsilon=\frac{c^{2}}{\upsilon^{2}}$ is correct for the induced electric field $\mathbf{E}=-\frac{1}{c}\frac{\partial\mathbf{A}}{\partial t}$ only, and the speed $\upsilon$ is the speed of light in SC medium if there were no skin-effect and Meissner effect. For electrostatic field $\mathbf{E}=-\nabla\varphi$ the permittivity is $\varepsilon(\omega=0,\mathbf{q}=0)=\infty$ like in metals.  This fact explains experimental results in \cite{bert3} which illustrate that the external electric field does not affect significantly the SC state unlike predictions of the covariant extension of GL theory \cite{bert1,bert2}, since in our model a superconductor screens electrostatic field by metallic mechanism in the microscopic Thomas-Fermi length, unlike the screening of magnetic fields by SC electrons in the London depth. Thus, unlike the models \cite{bert1,bert2,bert3,hirs1}, electrostatics and magnetostatics take different forms despite Lorentz covariance of the model. It should be noted a photon with frequency $\hbar\omega\geq 2|\Delta|$ can break a Cooper pair with transfer of its constituents in the free quasiparticle states. Hence in this frequency region the strong absorption of the waves takes place. Thus, the permittivity $\varepsilon$ is equal to $c^{2}/\upsilon^{2}$ only when $0<\hbar\omega<2|\Delta|$. At $\hbar\omega\gg 2|\Delta|$ we can suppose $\varepsilon=\varepsilon_{n}(\omega)$, where $\varepsilon_{n}(\omega)$ is the dielectric function of normal metal.

4) Unlike popular opinion \cite{ars} the Goldstone mode cannot be associated with the plasmon mode. We have demonstrated - Eqs.(\ref{3.54},\ref{3.55}), that Goldstone oscillations cannot be accompanied by oscillations of charge density, they generate the transverse field $\textrm{div}\mathbf{A}=0$ only and they are currents for which $\textrm{div}\mathbf{j}=0$ as result of the boundary condition (\ref{3.52b}). This is expressed in that the Anderson-Higgs mechanism takes place: the oscillations of the phase $\theta$ are absorbed into the gauge field $A^{\mu}$, hence the pure Goldstone oscillations become unobservable. It should be noted that the fixing of the transverse gauge $\textrm{div}\mathbf{A}=0$ is result of the spontaneously broken gauge symmetry in SC phase. Variation of order parameter is impossible in distances $\sim\lambda_{\mathrm{TF}}$ and time intervals $\sim 1/\omega_{p}$, since the SC system does not have reserve of the free energy to compensate the spatial-time variation energies. Therefore on these spatial-time scales (TF length and plasma frequency) the SC order parameter $\Psi$ (or $\Delta$) losses any sense and a superconductor should behave like ordinary metal. Thus, the plasma (Thomas-Fermi) screening and the plasma oscillations exist in metal independently on its state.

5) As result of interaction of the gauge field $A^{\mu}$ with the scalar field $\Psi$ with spontaneously broken $U(1)$ symmetry a photon obtains mass $m_{A}=\frac{\hbar}{\lambda\upsilon}$ in a superconductor, i.e., the Anderson-Higgs mechanism takes place, which manifests itself as the penetration depth $L(0)=\lambda$. The penetration depth depends on frequency: \emph{the depth $L(\omega)$ increases with frequency and such frequency exists $\omega_{c}=\frac{\upsilon}{\lambda}$, that the depth becomes infinitely large: $L(\omega\geq\omega_{c})=\infty$}. This is principal result of the extended TDGL theory. It should be noted that $\hbar\omega_{c}<2|\Delta|$ for type-II superconductors only. However the normal component causes absorption of electromagnetic waves and the wave skin-effect occurs, as a result the penetration depth $L$ is very sensitive to the normal density $n_{\mathrm{n}}$, that is \emph{for observation of the effect $L(\omega\geq\omega_{c})\rightarrow\infty$ the normal electrons should be absent} $n_{\mathrm{n}}=0$. Thus, we have shown that \emph{at $T=0$ pure type-II superconductors (in understanding of a monocrystalline sample without defects and impurities) should become transparent for electromagnetic waves with frequency $\omega$ from interval $\omega_{c}\leq\omega<2|\Delta|/\hbar$, that can be subject for experimental verification}, that illustrated in Fig.\ref{Fig7}. Thereby observation of this effect can be difficult. At the other hand, when Goldstone mode is absorbed into gauge field, the electromagnetic oscillation mode (\ref{3.47}) with spectrum (\ref{3.48}) appears instead the phase oscillations. These electromagnetic oscillations are \emph{eigen} oscillations of the SC system instead phase oscillations and they are eddy Meissner current. The critical frequency $\omega_{c}$ is a minimal limit of frequencies which can propagate through the system. And besides this frequency is much lower than the ordinary plasma frequency: $\omega_{c}(T=0)=\frac{\upsilon}{c}\omega_{p}\ll\omega_{p}$. If electromagnetic wave with frequency $\omega\geq\omega_{c}$ falls on superconductor then the wave is carried by these eigen oscillations hence the superconductor becomes transparent for such wave, at the same time for the wave with frequencies $\omega<\omega_{c}$ the carrier is absent hence such wave reflects from the surface of superconductor. Moreover, the analogy between above-described electrodynamics of bulk superconductors and the phase wave in a plane of Josephson junction exists that is presented in a Table \ref{tab1}. Results of calculation of the transparency intervals $\nu_{c}\leq\nu<\frac{2|\Delta|}{h}$ for pure elemental type-II superconductors niobium \texttt{Nb}, technetium \texttt{Tc} and vanadium \texttt{V} at $T=0$ are presented in a Table \ref{tab2}. The interval is in terahertz range.

6) The London electrodynamics is a low frequency limit of the extended TDGL theory, i.e., the term $\frac{1}{\upsilon^{2}}\frac{\partial^{2}\mathbf{A}}{\partial t^{2}}$ can be neglected in the equation for the field (\ref{5.4}), but the Rayleigh dissipation function (\ref{5.2}), which gives the term $\sim\sigma\frac{\partial\mathbf{A}}{\partial t}$, has to be accounted. At high frequencies $\omega\sim\omega_{c}$, we should use the extended TDGL theory. At the same time, at the large normal density (low SC density) $n_{\mathrm{n}}\rightarrow n$ the results are close to the London theory at high frequencies too. Moreover, the TDGL equation (\ref{1.3}) is a limit case of the extended TDGL theory at $T\rightarrow T_{c}$: due to strong damping of oscillations of $|\Psi|$ the monotonous relaxation of a fluctuation to the equilibrium state takes place, the process is described with Eq.(\ref{5.20}) which is analogous to the TDGL equation.

In conclusion, it should be noted that the extended TDGL theory has a form of a scalar field theory with spontaneously broken gauge symmetry interacting with a gauge field. However, the boundary condition (\ref{3.52}) violates this analogy due to the fact that, unlike fields, a superconductor is finite system with some surface on which corresponding boundary conditions must be set by determining of current through the surface. Consequently, electrostatics and magnetostatics take different forms despite Lorentz covariance of the model. In addition, it should be noted, that GL theory is two-liquid approximation (the total electron density is sum of the SC density and the normal density $n=n_{\mathrm{s}}+n_{\mathrm{n}}$). At the same time, in dynamics and electrodynamics of superconductors the coherency factors $u_{\mathbf{k}'}u_{\mathbf{k}}\pm v_{\mathbf{k}'}v_{\mathbf{k}}$ and $u_{\mathbf{k}'}v_{\mathbf{k}}\pm v_{\mathbf{k}'}u_{\mathbf{k}}$ play some role \cite{tinh}: they give corrections at temperature slightly less than $T_{c}$, i.e., when there are many quasiparticles. At low temperatures $T\ll T_{c}$, there are few quasiparticles, hence the affect of the coherency factors is negligible.

\appendix
\section{"Derivation" of the second London equation from the first London equation and vice versa}\label{london}

Following \cite{svir} let us write the Newton equation for SC electrons (they do not experience friction) in the electric field $\mathbf{E}$: $m\frac{d\mathbf{v}}{dt}=e\mathbf{E}$, which can be given a form:
\begin{equation}\label{A1}
  \frac{d\mathbf{j}_{s}}{dt}=\frac{n_{\mathrm{s}}e^{2}}{m}\mathbf{E},
\end{equation}
where $\mathbf{j}_{s}=n_{\mathrm{s}}e\mathbf{v}$ is supercurrent, $n_{\mathrm{s}}$ is density of SC electrons. Eq.(\ref{A1}) is the first London equation (\ref{1.5a}).  Making the operation $\mathrm{curl}$ for both sides of the equation and taking into account the Maxwell equation $\mathrm{curl}\mathbf{E}=-\frac{1}{c}\frac{\partial \mathbf{H}}{\partial t}$ we obtain:
\begin{equation}\label{A2}
  \frac{d}{dt}\left(\mathrm{curl}\mathbf{j}_{s}\right)=-\frac{n_{\mathrm{s}}e^{2}}{mc}\frac{\partial \mathbf{H}}{\partial t}.
\end{equation}
Integrating over time we obtain:
\begin{equation}\label{A3}
  \textrm{curl}\mathbf{j}_{s}=-\frac{n_{\mathrm{s}}e^{2}}{mc}\left(\mathbf{H}-\mathbf{H}_{0}\right),
\end{equation}
where $\mathbf{H}_{0}$ - a constant of integration which does not depend on time, but it can be function of coordinates. Thus, by setting the field $\mathbf{H}_{0}$ we set the initial condition, i.e., \emph{configuration of the field $\mathbf{H}$ is determined by both response of the medium and the initial field}. Supposing $\mathbf{H}_{0}=0$ (a sample is introduced into magnetic field) we obtain the second London equation (\ref{1.5b}). However supposing $\mathbf{H}_{0}\neq 0$ (a sample in the normal state is in the magnetic field $\mathbf{H}_{0}$, then it is cooled below the transition temperature $T_{c}$), we obtain the freezing of the magnetic field inside the sample. Thus, the field $\mathbf{H}_{0}$ is a constant of motion, that characterizes an ideal conductor: $\mathrm{curl}\mathbf{E}=-\frac{1}{c}\frac{\partial\mathbf{B}}{\partial t}\Rightarrow \mathbf{B}=\mathrm{const}$ since we have inside the sample $\mathbf{E}=\rho\mathbf{j}=0$ due to ideal conductivity $\rho=0$. Therefore Eq.(\ref{A3}) does not describe thermodynamically steady state, this equation describes the ideal conductor which pushes out or freezes magnetic field due to the electromagnetic induction and Lenz's rule.

From other hand, following \cite{tinh} we can differentiate the second London equation in a form $\mathbf{j}_{s}=-\frac{c}{4\pi\lambda^{2}}\mathbf{A}$, where $\mathrm{div}\mathbf{A}=0$ (i.e., $\mathbf{A}=\mathbf{A}_{\perp}$), with respect to time:
\begin{equation}\label{A4}
  \frac{\partial\mathbf{j}_{s}}{\partial t}=-\frac{c}{4\pi\lambda^{2}}\frac{\partial\mathbf{A}}{\partial t}=
  \frac{c^{2}}{4\pi\lambda^{2}}\mathbf{E}=\frac{n_{s}e^{2}}{m}\mathbf{E}.
\end{equation}
Thus, we obtain the first London equation (\ref{A1}) which is the second Newton law for SC electrons, i.e., it is equation for an ideal conductor. At the same time, the second London equation is result of minimization of the free energy functional, i.e., it is equation for a superconductor. This contradiction is resolved in Subsection \ref{electro2} within the Anderson-Higgs mechanism. It should be noticed that in Eq.(\ref{A4}) the field $\mathbf{E}$ is transverse field $\mathbf{E}=\mathbf{E}_{\perp}=-\frac{1}{c}\frac{\partial\mathbf{A}_{\perp}}{\partial t}$ only.  At the same time, in Eq.(\ref{A1}) the electric field can be longitudinal $-\nabla\varphi-\frac{1}{c}\frac{\partial\mathbf{A}_{||}}{\partial t}$ also (here $\mathrm{div}\mathbf{A}_{||}\neq 0$). Thus, the first London equation (\ref{A1}) cannot be obtain from the second London equation $\mathbf{j}_{s}=-\frac{c}{4\pi\lambda^{2}}\mathbf{A}_{\perp}$ completely.

\acknowledgments

This research was supported by theme grant of Department of physics and astronomy of NAS of Ukraine: "Mathematical models of nonequilibrium processes in open systems" 0120U100857 and by grant of National Research Foundation of Ukraine "Models of nonequilibrium processes in colloidal systems" 2020.02/0220.

\end{document}